\documentclass[sn-mathphys]{sn-jnl}

\usepackage{mathptmx} 
\usepackage{eurosym}
\usepackage{threeparttable} 
\usepackage{subcaption}
\usepackage{url}
\usepackage{xurl} 
\usepackage{xr} 
\usepackage[acronym, automake, nonumberlist]{glossaries-extra}
\usepackage{lineno}
\modulolinenumbers[5]

\usepackage{lineno,hyperref}
\modulolinenumbers[1]
\usepackage{amsmath}
\usepackage[T1]{fontenc}
\usepackage{siunitx}
\usepackage{eurosym}
\usepackage[europeanresistors,americaninductors]{circuitikz}
\usepackage{adjustbox}
\usepackage{xspace}
\usepackage{caption}
\usepackage{booktabs}
\usepackage{tabularx}
\usepackage{threeparttable}
\usepackage{multicol, multirow}
\usepackage{float}
\usepackage{graphicx,dblfloatfix}
\usepackage{csvsimple}
\usepackage{amssymb}
\usepackage{pifont}
\usepackage{tikzsymbols}
\usepackage{textcomp}
\usepackage{etoolbox}
\usepackage{longtable}
\usepackage{placeins}
\usepackage{float}

\newcommand{\kw}{\ \officialeuro /kW}

\newcommand{\kwe}{\ \officialeuro /kW$_{\text{elec}}$\ }
\newcommand{\kwee}{\ \officialeuro /kW$_{\text{elec}}$}

\newcommand{\budget}[1]{+{#1}$^\text{o}$C budget}
\newcommand{\ce}[1]{+{#1}$^\text{o}$C}

\newcommand*\sref[1]{%
S\ref{#1}}
\newcommand{\kg}{\ \officialeuro/kg$_{\text{H}_2}$}

\makeglossaries
\newacronym{mip}{MIP}{Mixed Integer Problem}
\newacronym{milp}{MILP}{Mixed Integer Linear Problem}
\newacronym{tes}{TES}{Thermal energy storage (in form of water tanks)}
\newacronym{chp}{CHP}{Combined heat and power plant}
\newacronym{cop}{COP}{Coefficient of Performance}
\newacronym{bdew}{BDEW}{Bundesverband der Energie- und Wasserwirtschaft}
\newacronym{PV}{PV}{Solar photovoltaics}
\newacronym{ghg}{GHG}{Greenhouse gas}
\newacronym{EU}{EU}{European Union}
\newacronym{dea}{DEA}{Danish Energy Agency}
\newacronym{ice}{ICE}{Internal combustion engine}
\newacronym{helmeth}{HELMETH}{Integrated High-Temperature Electrolysis and Methanation for Effective Power to Gas Conversion}
\newacronym{pem}{PEM}{Polymer electrolyte membrane electrolysis}
\newacronym{jrc}{JRC}{Joint Research Center}
\newacronym{idees}{IDEES}{Integrated Database of the European Energy System}
\newacronym{ecmwf}{ECMWF}{European Centre for Medium-Range Weather Forecasts}
\newacronym{ocgt}{OCGT}{Open cyclic gas turbine}
\newacronym{ccgt}{CCGT}{Closed cyclic gas turbine}
\newacronym{v2g}{V2G}{Vehicle to Grid}
\newacronym{pypsa}{PyPSA}{Python for Power System Analysis}
\newacronym{tyndp}{TYNDP}{Ten-year network development plan}
\newacronym{entso-e}{ENTSO-E}{European network for transmission system operators electricity}
\newacronym{nuts}{NUTS}{Nomenclature of Territorial Units for Statistics}
\newacronym{covid}{COVID-19}{Coronavirus disease 2019}
\newacronym{entsog}{ENTSO-G}{European Network of Transmission System Operators for Gas}
\newacronym{dsm}{DSM}{Demand side management}
\newacronym{hvdc}{HVDC}{High voltage direct current}
\newacronym{phs}{PHS}{Pumped hydro storage}
\newacronym{dac}{DAC}{Direct air capture}
\newacronym{sng}{SNG}{Synthetic natural gas}
\newacronym{ise}{Fraunhofer ISE}{Fraunhofer Institute for Solar Energy Systems}
\newacronym{iea}{IEA}{International Energy Agency}
\newacronym{soec}{SOEC}{Solid oxide electrolysis cell}
\newacronym{diw}{DIW}{German Institute for Economic Research (Deutsches Institut f\"ur Wirtschaftsforschung)}
\newacronym{SMR}{SMR}{Steam methane reforming}
\newacronym{fom}{FOM}{Fixed operation and maintenance costs}
\newacronym{vom}{VOM}{Variable operation and maintenance costs}
\newacronym{KKT}{KKT}{Karush-Kuhn-Tucker}
\newacronym{hdd}{HDD}{Heating degree days [d/a]}
\newacronym{tabula}{TABULA}{Typology Approach for Building Stock Energy Assessment}
\newacronym{EN}{EN}{European Norm}
\newacronym{ISO}{ISO}{International Organization for Standardization}
\newacronym{esm}{ESM}{Energy System Model}
\newacronym{genie}{GENIE}{Global ENergy system with Internalized Experience curves}
\newacronym{irena}{IRENA}{International Renewable Energy Agency}
\newacronym{sos}{SOS2}{Special ordered set of type 2}
\newacronym{aec}{AEC}{Alkaline electrolysis cells}
\newacronym{iam}{IAM}{Integrated Assessment Model}
\newacronym{wacc}{WACC}{Weighted average cost of capital}
\newacronym{scc}{SCC}{Social cost of carbon}
\newacronym{ccs}{CCS}{Carbon capture and storage}
\newacronym{etc}{ETC}{Energy Transitions Commission }

\jyear{2021}%

\theoremstyle{thmstyleone}%
\theoremstyle{thmstyletwo}%

\theoremstyle{thmstylethree}%

\unnumbered

\begin{document}

\title[Endogenous learning for green hydrogen in a sector-coupled energy model for Europe]{Endogenous learning for green hydrogen in a sector-coupled energy model for Europe}


\author*[1,2]{\fnm{Elisabeth} \sur{Zeyen}}\email{e.zeyen@tu-berlin.de}

\author[3,4]{\fnm{Marta} \sur{Victoria}}\email{t.brown@tu-berlin.de}

\author[1,2]{\fnm{Tom} \sur{Brown}}\email{mvp@mpe.au.dk}

\affil*[1]{\orgdiv{Department of Digital Transformation in Energy Systems, Faculty of Process Engineering}, \orgname{TU Berlin}, \orgaddress{\street{Einsteinufer 25 (TA 8)}, \city{Berlin}, \postcode{10587}, \state{Berlin}, \country{Germany}}}

\affil[2]{\orgdiv{Institute for Automation and Applied Informatics (IAI)}, \orgname{Karlsruhe Institute of Technology (KIT)}, \orgaddress{\street{Forschungszentrum 449}, \city{Eggenstein-Leopoldshafen}, \postcode{76344}, \state{Baden-Württemberg}, \country{Germany}}}

\affil[3]{\orgdiv{Department of Mechanical and Production Engineering}, \orgname{Aarhus University}, \orgaddress{\street{Inge Lehmanns Gade 10}, \city{Aarhus}, \postcode{8000}, \country{Denmark}}}

\affil[4]{\orgdiv{Novo Nordisk Foundation CO2 Research Center}, \orgaddress{\street{Gustav Wieds Vej 10}, \city{Aarhus}, \postcode{8000},  \country{Denmark}}}

\abstract{\textbf{Many studies have shown that hydrogen could play a large role in the energy transition for hard-to-electrify sectors, but previous modelling has not included the necessary features to assess its role. They have either left out important sectors of hydrogen demand, ignored the temporal variability in the system or neglected the dynamics of learning effects. We address these limitations and consider learning-by-doing for the full green hydrogen production chain with different climate targets in a detailed European sector-coupled model. Here, we show that in the next 10 years a faster scale-up of electrolysis and renewable capacities than envisaged by the EU in the REPowerEU Plan is cost-optimal in order to reach the \ce{1.5} target. This reduces the costs for hydrogen production to 1.26\ \officialeuro/kg by 2050. Hydrogen production switches from grey to green hydrogen, omitting the option of blue hydrogen. If electrolysis costs are modelled without dynamic learning-by-doing, then the electrolysis scale-up is significantly delayed, while total system costs are overestimated by up to 13\% and the levelised cost of hydrogen is overestimated by 67\%.}}




\maketitle
\thispagestyle{empty}
\clearpage
\setcounter{page}{1}
\section{Introduction}\label{main}
\begin{figure}[h]
	\centering
	\includegraphics[width=1.0\linewidth]{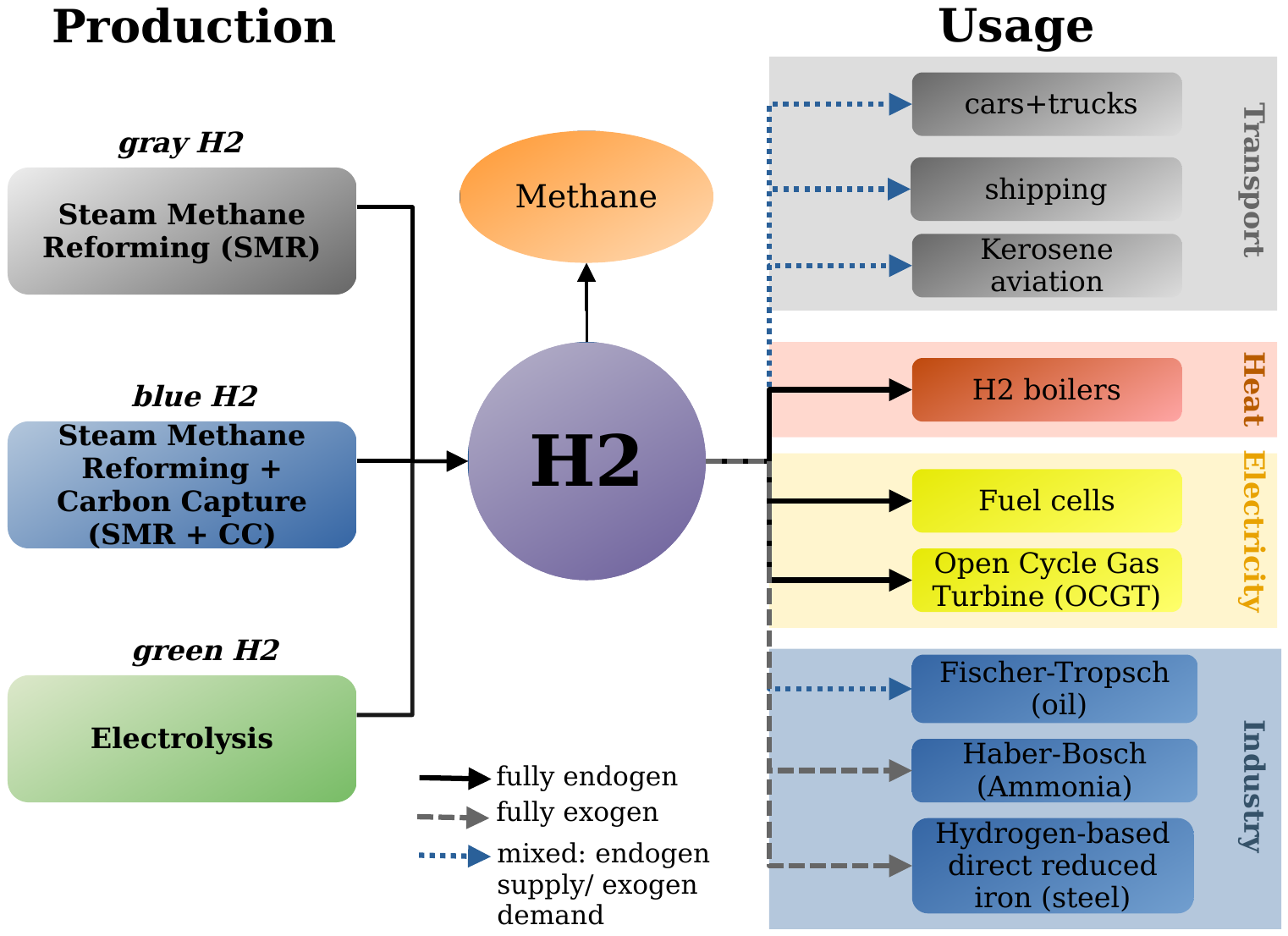}
	\caption{Overview of hydrogen usage and production in the model. In most sectors, both the production and demand of hydrogen is endogenous. With the exception of the production of ammonia and steel in industry, for these sectors fixed hydrogen demands are specified for the respective reference year. For the transport sector, e.g. the share of internal combustion engine (\gls{ice}) vehicles is predefined for each year (\textit{exogenous demand}), whether the fuel is of fossil or synthetic origin is optimised (\textit{endogenous supply}).}
	\label{fig:h2usage}
\end{figure}

 Today hydrogen plays a minor role in European final energy consumption with a 2\% share \cite{euhydrogen}, most of which is produced from natural gas with associated CO$_2$ emissions. In the future, however, hydrogen produced with low carbon emissions is likely to play an increasing role in the energy mix as decarbonisation progresses with shares of up to 24\% of the final energy demand by 2050 \cite{h2_roadmap}. Applications could include the production of green steel \cite{greensteel} or synthetic fuels \cite{noauthor_hydrogen_2016, emonts_2019}. There are several ways in which hydrogen can be produced: (i) from fossil fuels such as coal or natural gas, (ii) from fossil fuels in combination with CO$_2$ capture, or (iii) via electrolysis. The production costs for hydrogen depend in the first two cases on the price of fossil fuels, and in the second case on the cost of electricity. In order to reduce dependence on fossil fuels with high gas prices and to accelerate the energy transition, the European Commission aims to boost the deployment of hydrogen electrolysis. This objective has been concretised in the REPowerEU plan \cite{repowereu} published in May 2022, which has set a goal to produce 10 million tonnes of hydrogen with renewable electricity and to import further 10 million tonnes to the EU by 2030. \\
\\Hydrogen produced from water electrolysis using renewable electricity, so-called green hydrogen, is considered to be crucial to decarbonise hard-to-electrify sectors of the energy system. Since it is an immature technology, it is highly likely that costs and efficiencies will improve as production scales up, a phenomenon known as `learning-by-doing'. In this paper, we provide the first model that considers learning-by-doing for the complete green hydrogen production chain in a detailed energy system model. We include learning-by-doing for both renewable generators and electrolysis in a fully sector-coupled model with high temporal resolution. \\
\\ Several studies have been carried out on how future hydrogen electrolysis investment costs develop over time or with capacity deployment. However, many are based either on predefined production or capacity developments \cite{boehm2019, oxford2021, irena2020} or on expert surveys \cite{schmidt2017}. Way et al. \cite{oxford2021} explore various cost distributions for the world's energy system up to year 2070 for multiple technologies via a Monte Carlo approach.  Through the different investment cost trajectories they cover uncertainties in the cost developments, but the path for installed capacities are exogenously defined and energy dispatch is not modelled.\\
\\ The concept of the experience curve, which means that costs and efficiencies improve as production increases, is not a new idea. Already in 1936, Wright \cite{wright1936} described cost reductions related to aircraft production in a mathematical unit cost model. Since then, global cumulative production volume or overall installed capacity is used to quantify experience, attaining a good match with reality in many different technologies. However, experience curves are often neglected in energy system models, since they result in non-linear, non-convex optimisation problems which adds complexity. Modelling the endogenous cost curves leads to a better representation of the dynamics of new technologies \cite{victoria2021}, but requires a significantly higher computational effort. This is especially important when modelling with perfect foresight over a large time horizon. If the cost and efficiency improvements are given exogenously to the model, the model can `wait' with investments until they are profitable, while in reality, costs would only dynamically decrease as investments take place. Nevertheless, in many models, technology trends including cost reduction or efficiency improvement, are only considered as exogenously assumptions. An example of how these exogenous assumptions overestimate future investment costs is solar photovoltaic (PV). The investment costs of solar have decreased rapidly due to successful policy support. A comparison by Krey et al. \cite{krey2019} of the cost assumptions in 15 Integrated Assessment Models (\gls{iam}) shows that the cost of PV in 2020 have already fallen below the model expectations for 2050. Way et al. \cite{oxford2021} show that the progressive cost projections for solar, onshore wind, batteries and polymer electrolyte membrane (\gls{pem}) electrolysis from several \gls{iam}s and the International Energy Agency (\gls{iea}) are high compared to historical developments or even above costs in 2020. Several other studies \cite{creutzig2017, shiraki2020, mohn2020, xiao2021} criticise that the link between cost reduction and capacity installation is not well represented in models and that exogenously-set constraints such as floor costs or excessively low annual growth rates lead to an underestimation of cost reductions. This highlights the need to model investment costs endogenously without assuming excessive high floor costs, extremely constraining restrictions on maximum annual expansion rates or maximum penetration of renewable energies.\\
\\ Endogenous cost reductions are currently used in some \gls{iam}s \cite{remind, witch, image, times} which consider learning in multiple sectors and global developments. But because of the low temporal resolution in these models, they cannot represent the challenges of an energy system with a high share of variable renewable generation. In addition, most \gls{iam}s apply the Hotelling rule \cite{Hotelling1931} to determine the CO$_2$ prices, which with endogenous learning is not always applicable \cite{hof2021}. Bottom-up techno-economical Energy System Models (\gls{esm}) can better capture the temporal variability of the energy system. However, studies applying endogenous learning only deal with the power sector  \cite{Mattson1997, mattson, messner1997, barreto2001,heuberger2017} or focus on a single country \cite{felling2021}. \\
\\In this study, we explore for the first time how learning-by-doing on the full hydrogen production chain interacts with a fully sector-coupled energy model. We apply endogenous learning for electrolysis and renewable energy in the European model PyPSA-Eur-Sec \cite{githubpypsa} with full sector coupling and enough temporal resolution to capture renewable variability. The period between 2020-2050 is investigated with seven investment periods and perfect foresight over the whole modelling horizon. We do not specify the maximum annual expansion rates of renewable capacities nor CO$_2$ emission targets for every single year. These additional conditions often lead to a lower computational effort but also predetermine the transition paths. In this way, we determine the cost-optimal annual expansion rates without making any assumptions about expansion limits. Such limits have been estimated to be artificially low in many studies compared to actual capacity developments. We want to address the following two research questions:

\begin{itemize}
	\item What are the possible cost developments of green hydrogen production in Europe under the assumption of different CO$_2$ budgets without fixed capacity deployment projections?
	\item How do different methods of modelling cost reduction influence the results?
\end{itemize}
We consider three different competing options for producing hydrogen: (i) grey hydrogen (via steam methane reforming (\gls{SMR})), (ii) blue hydrogen (\gls{SMR} + carbon capture, capture rate 90\%) and (iii) green hydrogen (via electrolysis). Hydrogen can be used for methanation, for heating (hydrogen boilers), electricity (fuel cells and retrofitted open cyclic gas turbines (\gls{ocgt})), in the industry and in the transport sector.  The synthetic gas from methanation can be used in the heating sector (gas boilers or combined heat and power plants (\gls{chp}s)), for industry processes or in the power sector (\gls{ocgt}s or closed cyclic gas turbines \gls{ccgt}s). The demand pathway for hydrogen in parts of industry and transport is exogenously defined, while in all others sectors, hydrogen competes with other ways of supplying demand. This means both demand for hydrogen (e.g. where it competes with heat pumps in the heating sector) and the supply side (e.g. if the hydrogen is produced via electrolysis or \gls{SMR} with or without carbon capture), are part of the optimisation. \gls{ocgt} and gas boilers for heating can be retrofitted to run flexibly with natural gas or hydrogen (see Figure \ref{fig:h2usage}). \\
\\ There are various technologies for hydrogen electrolysis. In this study, cost and efficiency assumptions of alkaline electrolysis cells (\gls{aec}) are used since they are currently the most common electrolysers available on the market. We only consider large plants (above 100 MW) to avoid the scaling effects of very small plants. These costs consist of the equipment and installation costs. They include the expenses for the stack, power electronics, gas conditioning, balance of the plant and labour. The annual operational and maintenance costs are estimated in \cite{cost_dea} based on current projects to be 2\% of the investment costs. The cost of replacing the stack is not included in the fixed operational and maintenance cost (\gls{fom}) since it is assumed that the stack does not need to be replaced within the technical lifetime. The investment costs do not include costs for water purification, transformer costs or  connection fees to the transmission system. \gls{aec} is currently the dominant technology, but there are other types of electrolysis, such as \gls{pem} or solid oxide electrolysis cell (\gls{soec}), which may play a more prominent role in the future. We analyse the impact of higher initial investment costs (comparable to the investment cost of \gls{pem}) in the Supplementary Material. All costs are given in real 2015 Euro.
\subsection{Scenarios - Endogenous learning with different climate targets}
We minimise total system costs for the period 2020-2050 with perfect foresight and investments in new capacity in a five-years interval for three different CO$_2$ budgets. The investment costs of renewable generation capacity (solar, onshore and offshore wind) and hydrogen electrolysis are endogenous and thus a result of the optimisation in the form of a piecewise-linearised one-factor experience curve. This results in a mixed integer linear problem \gls{milp} which is further described in the Methods Section. Endogenous global learning of the renewable generation capacities for solar PV, onshore and offshore wind is applied.  Global learning assumes that global capacity grows proportionally to European capacity. The relative factor corresponds to today's share. For example, if 1 GW of solar PV is newly built in Europe and the share of European solar PV capacity is 22\%, the global capacity grows by about 4.5 GW. Further details about the initial investment cost assumptions of renewable generation are given in the Supplementary Material. For hydrogen electrolysis local learning is applied, since it cannot be assumed that global expansion will keep up with Europe. This means investment costs reduce only based on installed capacities within Europe. Learning rates are varied by $\pm 10\%$ for all learning technologies to understand the robustness of the results. Only the extreme cases are covered, i.e. a very optimistic scenario in which all default learning rates are higher by +10\% and a pessimistic case in which all learning rates are -10\% compared to the base case. The cost and efficiency assumptions for all other technologies are exogenously given depending on their build year. 
\begin{figure}[t]
	\centering
	\includegraphics[width=0.8\linewidth]{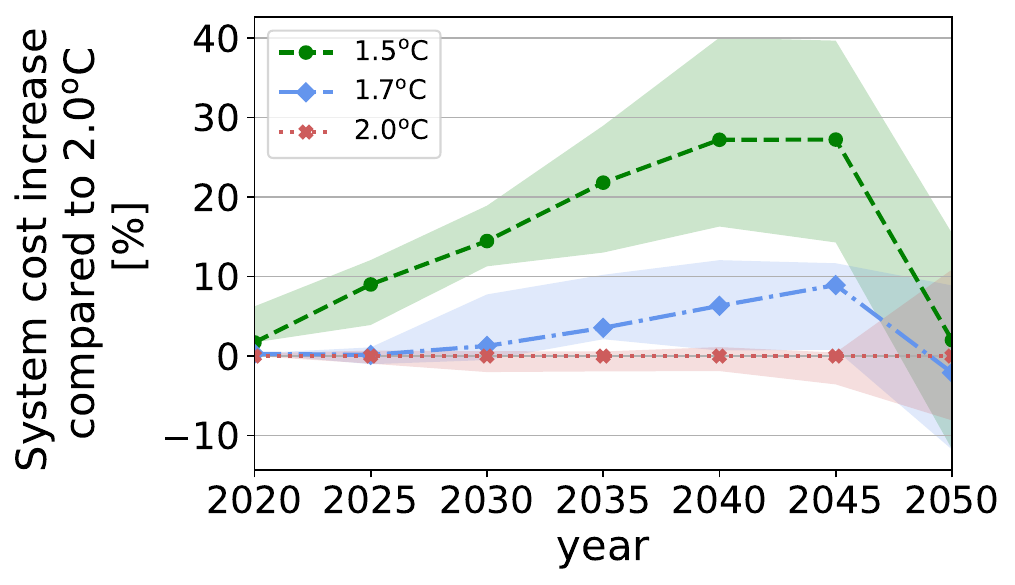}
	\caption{Total annualised system costs compared to the endogenous \budget{2} scenario without estimated costs of climate change damage. The costs of individual years are shown here, however, the entire period from 2020-2050 is solved in one single optimisation. The shaded areas represent the sensitivity analysis to $\pm$10\% learning rates for different technologies.}
	\label{fig:total_cost}
\end{figure}
	\begin{figure}
	\centering
	\includegraphics[width=0.8\linewidth]{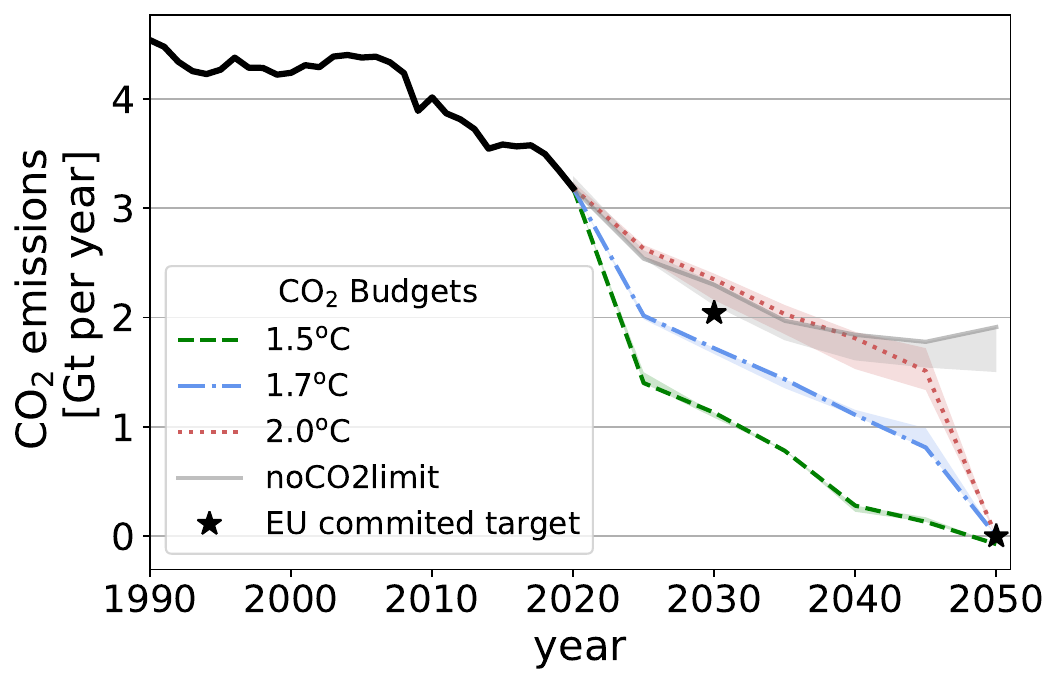}
	\caption{Annual CO$_2$ emissions. Three different budgets are assumed for Europe deducted from the global budget assuming equal per capita share (i) budget 25.7 Gt (\ce{1.5}), (ii) 45 Gt (\ce{1.7}), (iii) 73.9 Gt (\ce{2}). Further, carbon neutrality is required in all scenarios by 2050.  Hashed line shows a scenario with the default learning rate presented in Table \ref{tab:learning_rates}. Contour indicates scenarios with $\pm 10\%$ variation of the learning rate for all technologies with endogenous learning. We compare the budget scenarios with a fourth scenario (\textit{noCO2limit}) in which we do not specify any CO$_2$ constraints.}
	\label{fig:co2emissions}
\end{figure}
\section{Results}
\subsection{Total system costs to achieve the carbon targets}\label{sec:results_budgets}
To stay within the \budget{1.5}, the direct total annualised system costs (ignoring any costs of climate change damages) are up to 27\% higher than the \budget{2.0} (see Figure \ref{fig:total_cost}). Cost are higher in the \ce{1.5} scenario because (i) existing assets with high CO$_2$ emissions are  phased out before the end of their lifetime, (ii) large-scale investments are made before the costs are reduced by learning, and (iii) major parts of the oil demand are already supplied by synthetic fuels in 2030, which are more expensive to produce. However, total annualised system costs of the scenarios in 2050 are similar and vary only by 2\% between the \ce{1.5} and \ce{2.0} budget since the investment costs decline. The annualised system costs of the \ce{1.7} scenario in 2050 are 2\% lower compared to the \ce{2.0} scenarios. The reason is that investments in low-carbon infrastructure with the \ce{1.7} budget happen earlier, while for the \ce{2.0} scenarios a major infrastructure transformation with associated higher costs is needed in 2050 to meet the condition of climate neutrality.  The European Union target of 55\% greenhouse gas reduction in 2030 compared to 1990 \cite{fitfor55}, applied to CO$_2$, is within the \ce{1.7} and \ce{2.0} scenarios (see Figure \ref{fig:co2emissions}). If estimated costs of climate change damage with a social cost of carbon (\gls{scc}) of 120\ \officialeuro \ per tonne CO$_2$ are added, the total system costs are even slightly higher in the \ce{2.0} scenarios compared to the \ce{1.5} and \ce{1.7} scenario in the period 2025-2045 due to the higher CO$_2$ emissions (see Figure \sref{fig:total_cost_scc} in the Supplementary Material). A previous study \cite{victoria2021speed} deals in more detail with the impact of \gls{scc} on the total system costs. \\
\begin{figure*}
	\centering
	\includegraphics[width=12.cm]{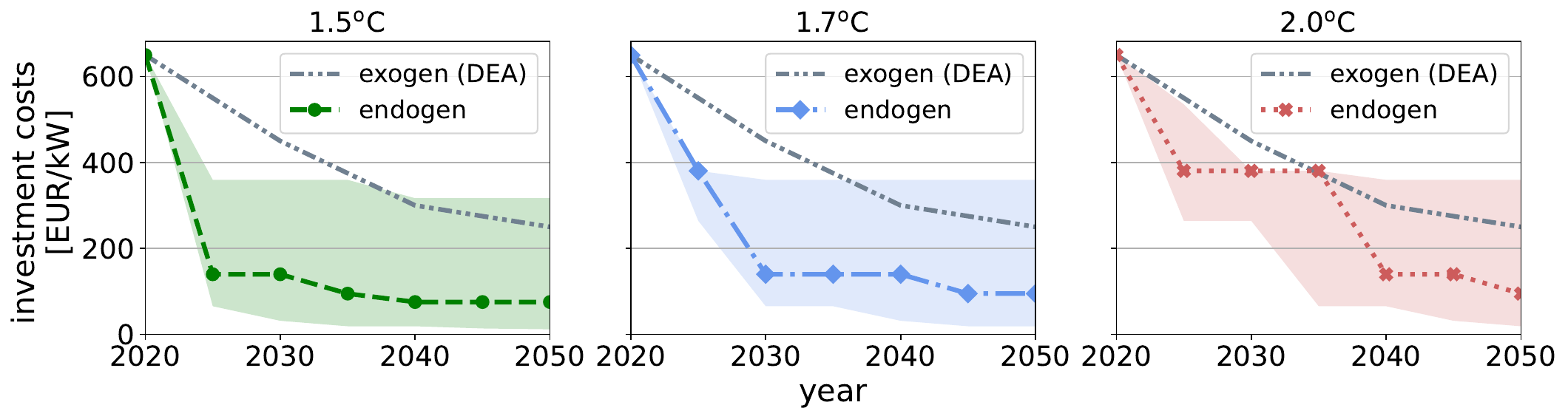}
	\caption{Investment costs of electrolysis for different carbon budgets. Hashed line shows a scenario with the default learning rate presented in Table \ref{tab:learning_rates}. Contour indicates scenarios with $\pm 10\%$ variation of the learning rate for all technologies which undergo learning. Investment costs of the other endogenous learning technologies are shown in the Supplementary Material Figure \sref{fig:investment_cost_renewables}.}%
	\label{fig:investment_cost_h2}
\end{figure*}
%
\subsection{Electrolysis investment costs}
 The investment costs of electrolysis decline strongly for all three assumed CO$_2$ budgets. For the base learning rate assumptions (see Table \ref{tab:learning_rates}), investment costs are for every year and budget below the cost estimates of Danish Energy Agency (\gls{dea}) (see Figure \ref{fig:investment_cost_h2}).  The tighter the budget, the more renewable generation and electrolysis capacities are employed and the stronger is the cost decline. All scenarios lead to a significant reduction in the investment costs of electrolysis, ranging from 140 to 380\kwe already in 2030. The largest cost reductions for the \ce{1.5} scenario occur until 2025, for the \ce{1.7} budget until 2030 while the investment costs in the \ce{2.0} scenarios decline most strongly in later years between 2035 and 2040. In 2050, investment costs for electrolysis decrease down to 75 and 95\kwe (compared to \gls{dea} 250\kwee) for \ce{1.5} and \ce{2} respectively (see Figure \ref{fig:investment_cost_h2}). \\
 \\ To examine the robustness of the results, the learning rates are varied by $\pm$10\% in additional scenarios. In the scenarios with the most pessimistic assumptions (all learning rates  are 10\% less than the base case), the electrolysis investment costs range between 317 and 360 \kwe in 2050.\\
%
\subsection{Installed electrolysis capacities and hydrogen usage}
 The \ce{1.5} scenario sees cost-optimal deployment of 435 GW of electrolysis which produce 36 million tonnes of hydrogen in 2030. This is clearly not realistic given the necessary scale-up of production facilities that would be required. In the scenario corresponding to a temperature increase of \ce{1.7}, 60 GW of electrolysis are installed by 2030 which produce 4 million tonnes of hydrogen. Under the \ce{2} scenario, 4 GW of electrolysis are deployed and 0.1 million tonnes of hydrogen are produced (see Figure \ref{fig:cap_h2}). The European Commission's REPowerEU plan \cite{repowereu} targets to produce 10 million tonnes of hydrogen within Europe, which corresponds to electrolysis capacities of 140 GW$_{\text{el}}$ assuming average utilisation factors of 43\% and 70\% conversion efficiency \cite{electrolyser_summit_joint}, and importing further 10 million tonnes of hydrogen from neighbouring countries by 2030. This target is in between our \budget{1.5} and \budget{1.7} result, given that we do not consider hydrogen imports from outside Europe. \\
\\ The production of hydrogen switches from \gls{SMR} (grey hydrogen) to  electrolysis (green hydrogen) in all scenarios, while \gls{SMR} in combination with carbon capture (blue hydrogen) is not installed in any scenario. To further explore the option of \gls{SMR} with carbon capture, we run sensitivity scenarios for blue hydrogen production depending on capture rate, investment costs and available CO$_2$ sequestration potential (see Supplementary Material). Blue hydrogen is only produced at scale under certain optimistic conditions. For example, with our base assumptions for CO$_2$ storage potential (200 Mt$_{\text{CO}_2}$/a) and capture rate (90\%), blue hydrogen has a share of 8\% in total production with \gls{SMR} investment costs of 286 EUR/kW$_{\text{H}2}$ (50\% of our reference assumption). A large sequestration potential of 2000 Mt$_{\text{CO}_2}$ per year with base assumptions on investment costs (572 EUR/kW$_{\text{H}2}$) and capture rate (90\%) leads to a  production of blue hydrogen with a share of 19\% of total production. With more optimistic assumptions on capture rate (100\%) and CO$_2$ storage potential (2000 Mt$_{\text{CO}_2}$/a), most of the hydrogen is produced as blue hydrogen from investment costs below 286 EUR/kW$_{\text{H}2}$. \\
\\ The timing of the transition from grey to green hydrogen production differs between the scenarios. In the scenario with a \budget{1.5}, green hydrogen is already produced in 2025, while in the scenario with a \budget{2} hydrogen is supplied by \gls{SMR} until 2040-2045 (see Figure \sref{fig:eb_H2} in the Supplementary Material for a breakdown of hydrogen supply and usage). For all budgets, hydrogen demand increases from about 110 TWh$_{\text{H}_2}$/a in 2020 to up to 4000 TWh$_{\text{H}_2}$/a in 2050. In the \ce{1.5} scenario, the hydrogen is primarily used to produce synthetic fuels (1800 TWh$_{\text{H}_2}$ in 2050) and synthetic methane (700 TWh$_{\text{H}_2}$/ 560 TWh$_{\text{CH}_4}$ in 2050). The produced synthetic methane is mainly used in industry processes (424 TWh$_{\text{CH}_4}$) but also in \gls{ocgt}s (135 TWh$_{\text{CH}_4}$). There is an option to convert the \gls{ocgt}s to run on hydrogen, but this option is not used. A smaller fraction of the hydrogen is used in the industry (500 TWh$_{\text{H}_2}$ in 2050), to power fuel cells (440 TWh$_{\text{H}_2}$ in 2050) and for shipping (200 TWh$_{\text{H}_2}$ in 2050). The option of retrofitting existing natural gas boilers for operation with hydrogen is not exploited. The natural gas boilers are largely replaced by heat pumps until 2040. \\
\\ Hydrogen is stored in salt caverns and the energy capacity ranges between 278-349 TWh from the \ce{2.0} to the \ce{1.5} scenario. These storage capacities are comparable to existing natural gas storage of 1075 TWh, once adjusted for the lower volumetric energy density of hydrogen, and below the European technical potential of 84.8 PWh$_{\text{H}_2}$ \cite{caglayan2020}. No costs are assumed in the main results for the hydrogen transport. A sensitivity analysis of a scenario with a higher spatial resolution and corresponding costs for electricity and hydrogen infrastructure, as well as an annual breakdown of hydrogen supply and usage and installed capacities is shown in the Supplementary Material. In scenarios with a higher spatial resolution, the hydrogen and electricity grids account for a share of 0.1-7.6\% of the total system costs, higher electrolysis capacities are installed and total system cost increase by 12-16\%. The trade-offs of an electricity and hydrogen grid in a decarbonised European energy system are discussed in more detail in a further publication \cite{neumann2022}. \\
\subsection{Renewable generation costs and capacities}
 In order to achieve net-zero emissions in 2050, renewable generation capacity must be strongly expanded in all scenarios to at least 3.2 TW solar, 1.7 TW onshore wind and 175 GW offshore wind. The timing of the capacity build out differs across the individual scenarios: with an ambitious budget, there is a strong expansion between 2030-2040, while in the other scenarios the strongest expansion takes place in later years from 2040 (\ce{1.7}) or 2045 (\ce{2.0}). The REPowerEU plan \cite{repowereu}, with targeted \gls{PV} capacities of 750 GW$_{\text{DC}}$ \cite{eusolarstrategy} ( which corresponds to 600 GW$_{\text{AC}}$) by 2030, lies within our results of \ce{1.7} with 764 GW and the \ce{2} with 422 GW. However, the wind expansion targets of REPowerEU with 480 GW \cite{euwindpower} by 2030, lag behind our findings with at least 555 GW in the \ce{2} scenario by 2030. \\
\\ In scenarios with base learning rate assumptions and endogenous learning, investment costs (without grid connection costs) for renewable generation capacity in 2050 range between 171-237\kw \ for solar \gls{PV},  818-900\kw \ for onshore wind and 1327\kw \ for offshore wind from the tight \ce{1.5} to the \budget{2} (see Figure \sref{fig:investment_cost_renewables} in the Supplementary Material). These are below the \gls{dea}'s cost projections. In particular, the costs for solar \gls{PV} decrease significantly and are 43\% lower in 2050 with ambitious climate targets than the \gls{dea} estimates of 300\kw. One should be aware that  grid connection costs are added to these investment costs in the model that do not undergo any learning. For offshore wind, grid connection costs depend on the location and connection type, while for solar and onshore wind additional grid connection costs of 133\kw \ are added. \\
%
\subsection{\budget{1.5} hard to accomplish}
 The build-out rates of renewables and electrolysis in scenarios with a \budget{1.5} are significantly higher than the historical record, which emphasises the challenge to stay within this budget. For example, the average annual build out rate is 77 GW for solar PV in the \ce{1.5} scenario between 2020-2030. This corresponds to almost a tripling of the historical maximum annual expansion rates in Europe \cite{growth_limit_solar}. The same applies to onshore and offshore wind, with average annual build-out rates of 97 GW and 16 GW between 2020-2030 in our scenario, which are roughly five times the historical maximum annual build-out rates \cite{growth_limit_onwind, growth_limit_offwind}. Between 2030-2040 even higher build rates are required of up to 200 GW solar capacity per year while the annual expansion of onshore wind decreases to 58 GW. A sensitivity analysis, in which the maximum annual expansion rates of renewables are limited, can be found in the Supplementary Material. Limiting maximum annual expansion rates leads to higher costs because more offshore wind and nuclear power plants are deployed instead of less costly onshore wind and solar. Hydrogen production is lower because methanation is no longer favoured. The option of producing blue hydrogen is not used. One should be aware, that we do not assume any negative emissions after 2050, nor any additional sufficiency measures, nor do we allow a fully endogenous transition path for the transport sector. These are all factors which could reduce the necessary build out rates and make the \ce{1.5} scenarios more achievable. \\
\begin{figure}[t]
	\centering
	{{\includegraphics[width=9.cm]{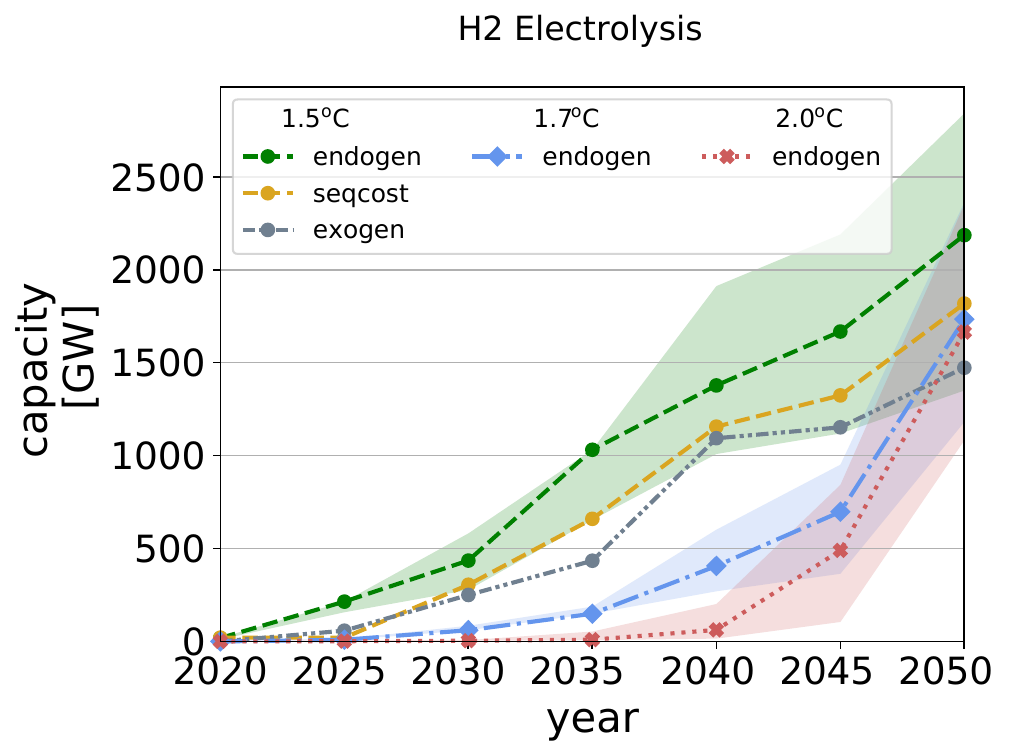} }}%
	\caption{Electrolysis capacity of the three different budgets. For the \ce{1.5} scenarios also the three different methods are compared.  The sequential cost (\textit{seqcost}) and the exogenous (\textit{exogen}) method results in later investments and underestimations of installed electrolysis capacities compared to the endogenous method (\textit{endogen}).}
	\label{fig:cap_h2}
\end{figure}
\subsection{Scenario without any carbon target}
 In one further scenario, no constraints on CO$_2$ emissions are assumed in order to investigate what a cost-optimal system without a carbon target looks like. As in the previous scenarios, the investment costs of renewables and green hydrogen are subject to endogenous experience curves. Compared to the current system, the share of renewable hydrogen increases even without climate targets. CO$_2$ emissions reduce from historical levels of 3.2 Gt/a in 2020 to 1.9 Gt/a in 2050 (see Figure \ref{fig:co2emissions}). Emissions in 2030 are 2.3 Gt/a, which is only slightly above the EU's 2030 target.  Hydrogen is largely produced via \gls{SMR}. In 2050, 23\% of the hydrogen demand in the CO$_2$-unconstrained scenario is covered by green hydrogen via electrolysis. CO$_2$ emissions stagnate from 2040 onward even with very high learning rates, whereby a large part of the emissions is generated by the use of fossil gas in the heating and power sector, as well as emissions from aviation and feedstocks for the petrochemical industry. It should be noted that our CO$_2$-unconstrained scenario does not continue historical trends, but allows for the transformation of energy sectors if it is cost-effective. For example, even without a set CO$_2$ limit, the comparatively expensive coal-fired power generation is phased out and \gls{ice} vehicles are replaced by electric vehicles and hydrogen-powered trucks, as in the other scenarios with a CO$_2$ budget. Sectors in which decarbonisation is not cost-optimal, are not transformed and continue to generate emissions, such as production of feedstocks for the petrochemical industry. 

\FloatBarrier
\subsection{Comparing different methods of modelling technology learning}\label{sec:results_methods}
\begin{figure}
	\centering
	{{\includegraphics[width=8.3cm]{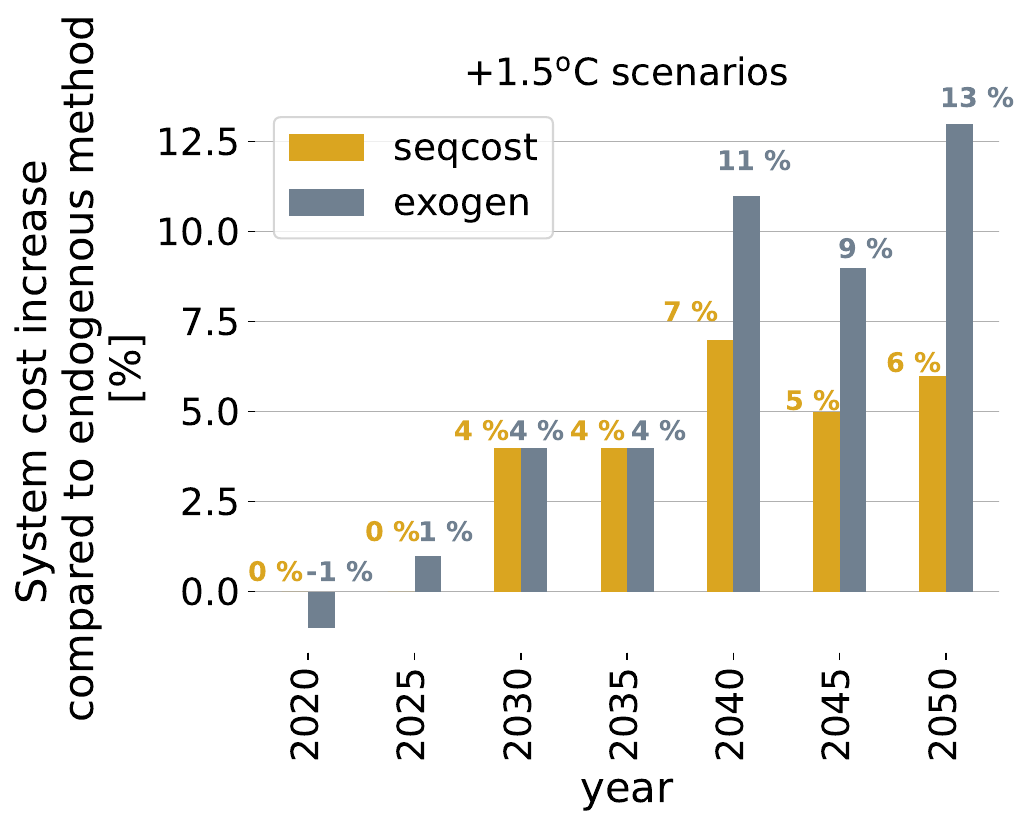} }}%
	\caption{Difference in total system costs for \ce{1.5} scenarios compared to the endogenous method with base learning rate. Exogenous (\textit{exogen}) and sequential (\textit{seqcost}) method result in  higher total annualised system costs compared to the endogenous scenarios. The comparison for the \ce{1.7} and \ce{2.0} scenarios and the impact of different learning rates can be found in the Supplementary Material \sref{fig:cost_bar_methods}.}%
	\label{fig:cap_cost_h2}
\end{figure}

In this section, we examine the difference that modelling learning-by-doing dynamically makes to the results. We compare three different methods of modelling investment cost reduction typically used in \gls{esm}, including (i) the endogenous method where investment costs are adjusted according to the installed capacity, (ii) the standard exogenous method with given fixed cost trajectories for each technology and investment year and (iii) the sequential method. In the sequential method, the optimisation problem is first solved using the exogenous cost assumptions, then the investment costs are updated depending on the optimised installed capacities. The same experience curves are assumed as in the endogenous method. The process of solving and updating investment costs is iteratively repeated until the difference of investment costs between two iterations is below a threshold. The threshold is set to a maximal mean square difference between optimised investment costs of the current and the previous iteration of 5\%. The sequential method is a linear problem that requires less computational resources than the \gls{milp} of the endogenous method, but unlike the exogenous method, the investment costs are adjusted based on the installed capacities. 
\paragraph{Impact on the total system costs} The cost difference of not using endogenous learning-by-doing is shown in Figure \ref{fig:cap_cost_h2}. In contrast to the exogenous method, the investment costs of the endogenous and sequential method depend on the installed capacities. For the \ce{1.5} budget, the sequential and exogenous methods result in up to 7\% and 13\% higher annualised total system costs compared to the endogenous method. The cost difference between the \ce{1.5} and the \ce{2.0} budget is smaller with the endogenous method compared to the sequential and exogenous method, because although larger capacities of renewables and electrolysis are needed at an earlier point in time, the costs also decrease more due to the faster scaling up and the foresight of the endogenous method of potential investment cost reductions. The endogenous method therefore allows a better comparison of costs of scenarios with different infrastructure needs or transition speeds.
\paragraph{Timing of investments}  The endogenous method leads to an earlier deployment of the learning technologies since it is the only one of the three methods that has the foresight of how far costs can decrease. As a result, investments are made early in order to reduce investment costs. With the sequential and exogenous methods, investment is delayed and the model `waits' until costs fall. For example in \ce{1.5} scenarios in 2025, 214 GW of electrolysis are installed in the endogenous scenarios and only 20 GW electrolysis in both the sequential and exogenous cost scenarios. With the endogenous method, the hydrogen is used for the synthesis of fuels and methane earlier and to a larger extent. In scenarios corresponding to a \budget{1.7} about 1800 TWh$_{\text{H}_2}$ are produced in 2045 mainly via electrolysis, while only half of this amount is produced in the exogenous case.  The larger volume of hydrogen is used in the endogenous case for synthetic fuels while in the exogenous case they have a fossil origin. This is particularly noteworthy since most \gls{esm} assume exogenous cost reductions and thereby underestimate initial investments in early years (see Figure \sref{fig:eb_H2} in the Supplementary Material).  The foresight of the endogenous method is particularly important when modelling the dynamics of emerging technologies for which strong cost reductions are possible. The endogenous method can consider the potential cost development during optimisation and make investment decisions based on this. With the sequential and exogenous method, the investment decisions are strongly subject to the initial assumed cost projection. 
\paragraph{Hydrogen prices}  The prices for hydrogen are an output of the linear optimisation and are determined by the dual variables. For the exogenous and sequential method, these are obtained directly from the optimisation. In the endogenous scenarios, the investment costs and capacities of the learning technologies determined from the optimisation are assumed fixed and the optimisation is rerun as a linear problem. The exogenous method overestimates the cost of hydrogen by up to 67\% in 2030 and 38\% in 2050 compared to the endogenous scenarios.  For example, in endogenous scenarios with a \budget{1.5} hydrogen costs drop to 1.32\kg in 2030, while in the sequential and exogenous scenarios they are 1.95\kg\ and 2.22\kg\ respectively. In 2050, the costs in the endogenous scenarios reach 1.26\kg and the exogenous ones 1.73\kg. 
\paragraph{System composition} The system composition in 2050 is primarily influenced by the assumed CO$_2$ budget and not by the modelling method (see Figure \ref{fig:cap_h2}). However, our results show deviations between the methods. The exogenous method leads to the lowest installed electrolysis capacities in 2050 followed by the sequential method with -33\% and -17\% respectively compared to the endogenous method in the \ce{1.5} scenarios. This can be explained by the lower investment costs for electrolysis obtained in the endogenous scenarios (75\ \officialeuro /kW$_{\text{elec}}$ in 2050 with a \budget{1.5}), which are below the sequential optimised investment costs of 79\ \officialeuro /kW$_{\text{elec}}$ and the exogenous assumptions of 250\ \officialeuro /kW$_{\text{elec}}$. The produced volume of hydrogen differs between the methods. For example, in scenarios corresponding to a \budget{1.5}, 24\% and 30\% less hydrogen is produced with the sequential and exogenous methods in 2030. With the endogenous method, more hydrogen is used for methanation and for re-electrification in fuel cells compared to the other two methods.\\
%
\paragraph{Global learning in regional energy systems}
Technology learning depends on global developments while most \gls{esm} model single regions or continents. Cost in Europe, for example for solar \gls{PV}, would reduce if solar panels are employed at a large scale in China. In this study, we have assumed that global renewable capacities increase in proportion to the growth in Europe. For electrolysis, we assume local learning. The exogenous method can provide a better description of cost developments for technologies for which large expansion rates outside the modelled region are expected. One could circumvent this problem with the sequential and endogenous method by splitting up the technology learning into two factors describing regional and global learning. The regional factor depends on the installed capacities in the regional model, while the global factor describes an exogenous cost decrease dependent on expected global development. The split into two factors was outside the scope of this study.
\paragraph{Computational resources}
The exogenous and sequential method require significantly less memory and computing time compared to the endogenous method. For example, using the commercial solver Gurobi \cite{gurobi} with 12 threads, the scenarios with endogenous learning need about 21 hours to solve and require 30 GB RAM, while the sequential models solve in less than 1 hour with only 3 GB RAM and the exogenous method solves in 1 minute with only 3 GB RAM. The sequential and the exogenous method require lower computational effort and thus offer the possibility to calculate e.g. with higher spatial resolution. The solution time and memory requirements for the endogenous method increase with the number of modelled learning technologies. In contrast, the computational effort of the sequential method is independent of the number of learning technologies, since the costs are only updated after optimisation. This allows to consider cost reductions for all technologies depending on the installed capacity without increasing the computing time.

\paragraph{Choice of method for modelling learning effects}
The chosen method of modelling technology learning should depend on the particular research question. The sequential and exogenous method are advantageous if the available computational resources are small or a greater level of detail (e.g. higher spatial resolution) is important. The number of technologies that are subject to learning can be increased with the sequential method without additional computational effort. The exogenous method can represent developments outside the modelled region. If the cost reductions are expected primarily through local learning within the modelled region, the sequential and endogenous methods are appropriate. The endogenous method is favourable if the timing of the investment is analysed, costs of different climate target scenarios are compared or trade-offs between emerging technologies are considered. Technologies with a large learning potential can be better modelled with the endogenous method, since small additional capacities already lead to strong cost reductions.
 \subsection{Comparison to other studies}
 In the following, we compare our findings concerning the cost of green hydrogen production to other studies. We have converted all cost assumptions into 2015-Euro (using the exchange rate of 1 USD = 0.876 Euro in 2020), assuming an annual inflation rate of 2\%. Our results of 1.32-1.99\kg in 2030 with the endogenous method are in good agreement with studies by \gls{irena} \cite{irena_g7}, Energy Transitions Commission (\gls{etc}) \cite{etc_2021} and \gls{iea} \cite{iea_2022} which all have production costs of green hydrogen below 2\kg in 2030. Vartiainen et al. \cite{vartiainen2022} find more favourable cost  of green hydrogen production of 0.77-1.99\kg in 2030. Compared to the optimised values in our study, they assume a higher number of full load hours of electrolysis and a larger capacity increase of solar and electrolysis. A \gls{jrc} report \cite{JRC_2022} shows hydrogen production costs above our results in 2030 at 1.66-3.86\kg due to slower capacity growth. In 2050, we find cost of green hydrogen production of 1.26-1.51\kg, which is in good agreement with results from \gls{etc} of 1.21\kg \cite{etc_2021}, Hydrogen council of 1.4\kg \cite{hydrogencouncil_2021} and BloombergNEF of 0.68-1.55\kg \cite{bloomberg_2020}. Vartiainen et al. \cite{vartiainen2022} find lower costs 0.33-0.99\kg due to greater capacity expansion. In contrast to our study, these studies assume an average electricity price and a fixed number of full load hours for electrolysis and thus cannot reflect the system advantage of electrolysis running at very low electricity prices. However, we only assume local learning for electrolysis and do not consider learning due to capacity expansion in other regions, while the other studies analyse global developments. \\
\\ Odenweller et al. \cite{Odenweller2022} show that fast scale-ups of electrolysis capacities until 2030 as we see in our \ce{1.5} scenarios may be infeasible. However, in this study we want to show the cost-optimal capacities that are necessary to achieve a given CO$_2$ budget. We do not consider hydrogen imports in this study. Imports would results in lower installed electrolysis and renewable capacities in Europe.  Seck et al. find hydrogen imports of 15\% \cite{seck2022}, \gls{irena} 17-50\% \cite{irena_g7} of total demand in 2050. The role of blue hydrogen differs between the studies as well. Studies by \gls{etc} and BloombergNEF \cite{etc_2021, bloomberg_2020} show that for most European countries, green hydrogen production is cheaper than blue hydrogen production, which is consistent with our results. Several studies \cite{seck2022, hydrogencouncil_2021, JRC_2022} find a combination of both blue and green hydrogen production cost optimal. For example, Seck et al. \cite{seck2022} find a share of 20-52\% of blue hydrogen in total production. In comparison to our study, Seck et al. \cite{seck2022} assume a seven times higher CO$_2$ storage potential and limit the installation pace of renewable capacities to historical levels, which makes blue hydrogen production more favourable. This is in agreement with our sensitivity analysis included in the Supplementary Material. \\
\\ Our CO$_2$-unconstrained scenario is less expensive than scenarios with a CO$_2$ budget. The study by Way et al. \cite{oxford2021} shows that a scenario that follows historical trends and does not allow the transformation of comparatively expensive sectors is more costly than a scenario with a \budget{1.5}. However, unlike the scenario in \cite{oxford2021}, in the CO$_2$-unconstrained scenario parts of the energy system can transform if it is cost-effective.

\section{Limitations of this study}
The presented results contain several limitations, first in the way the experience curves are modelled, and second in the general scenario assumptions. We have prioritised the limitations based on our subjective view of their impact on the results.
\begin{enumerate}
	\item A major impact on the results is the simplified assumption that hydrogen electrolysis costs are subject to local learning and renewable capacity costs are subject to global learning. If electrolysis are scaled up e.g. in the US and China, this accelerates technology learning and lowers the investment costs in Europe as well. 
	\item We assume no import of green hydrogen. This would lead to lower renewable and electrolysis capacities in Europe. However, globally, electrolysis would have to be built to satisfy the European demand, thus lowering investment costs.
	\item The results are further limited by the exogenous assumptions in the transport sector which specify the share of \gls{ice} for a certain period and the share of hydrogen demand for shipping. These fixed shares have a large impact on the results, especially on scenarios with a \budget{1.5}.  To stay within the given budget, hydrogen is required in 2020-2035 to produce synthetic fuels, while with a higher share of electric vehicles this would not be necessary. The impact of the transport transition pathways is analysed in the Supplementary Material. 
	\item To keep the computational effort reasonable, no network infrastructure is included. Networks may have potential bottlenecks that affect the results. We have included in the Supplementary Material a sensitivity analysis concerning the spatial resolution for the exogenous method (see Supplementary Material Figure \sref{fig:spatial_electrolysis}). A higher spatial resolution increases capacities of electrolysis, which result in lower investment costs of electrolysis for the endogenous and sequential method. The here presented results therefore provide an upper bound on investment costs for electrolysis and a lower bound on system costs with respect to the limitation of spatial resolution. 
	\item We only model technology learning for investment costs, but not for fixed operational and maintenance cost (\gls{fom}), efficiencies or lifetimes, which are assumed exogenously based on the year of construction and the respective technology. In particular, an improvement in the efficiency of electrolysis could lead to lower capacities or increased use of hydrogen.
	\item No spill-over effects between technologies are taken into account. These effects could be important for example for on- and offshore wind, but also for spill-over between different electrolysis types.
	\item We assume all plants are large (>100 MW) and disregard the effects of scaling the size of individual electrolysers.
	\item We do not limit the annual maximum expansion of electrolysis capacity. If this condition would be binding, it would lead to a greater use of grey or blue hydrogen and higher investment costs for electrolysis.
	\item We model a fixed weighted average cost of capital (\gls{wacc}) of 7\%, but do not model country-dependence or learning in financing that would decrease the \gls{wacc} as banks and investors become more comfortable with a new technology \cite{egli2018}. 
	\item Additional benefits of a faster reduction of CO$_2$ emissions, such as those related to new jobs or health are not considered within the optimisation. Taking these factors into account could significantly increase the costs of the \ce{2.0} compared to the \ce{1.5} scenarios.
	\item  We do not investigate the impact of different social discount rates. This is analysed in a previous publication \cite{victoria2021speed}.
	\item Although we investigate the robustness of the results by varying learning rates, we do not consider the change of a learning rate during a scenario, e.g. the learning rate of hydrogen electrolysis could be 16\% between 2020-2035 and decrease to 10\% thereafter. 
	\item We consider only one factor experience curves and not, for example, learning-by-research as an additional factor
\end{enumerate}
In this paper, we have classified all hydrogen produced by electrolysis as green, even though conventional power plants are still part of the generation mix, especially up to 2035. In all scenarios, the renewable generation is greater than the electricity demand of the electrolysis at every modelled hour. It is therefore possible, that all hydrogen produced via electrolysis is green, even though conventional power plants remain contributing to the generation. \\
\\Future work should investigate limitations with potentially large impacts such as imports of hydrogen, full
endogenisation of the transport sector and higher spatial resolution. 
%
\section{Conclusion}
In this study, we use a sector-coupled European model with endogenous cost reductions through experience curves for electrolysis and renewable generation technologies, assuming different CO$_2$ budgets. In all scenarios, CO$_2$ neutrality is required as an additional condition in the model by 2050. In a second part, we explore the trade-offs of three different methods modelling learning-by-doing.\\
\\ We find that in scenarios with ambitious climate targets, costs for green hydrogen electrolysis can be reduced to 1.26\kg by 2050. In our results, scenarios with a tight budget corresponding to \ce{1.5} warming require a rapid transformation of the energy system already in 2025 and high build out rates of both electrolysis and renewable generation. This indicates that, without negative emissions after 2050 or further efforts such as sufficiency measures, the scenarios with +1.5$^\text{o}$C warming are difficult to achieve. Hydrogen production shifts from grey to green hydrogen. Depending on climate targets, more than half of hydrogen demand is met by electrolysis in 2025 (\ce{1.5} scenario) or 2045 (\ce{2.0} scenario). The option of blue hydrogen is not used in any scenario. \\
\\ A rapid build-up of both electrolysis and renewable generation capacity is necessary by 2030 to stay within a \ce{1.5-2} target. While for the former, the European Union's targets for electrolysis are in line with our \ce{1.5-1.7} budget scenarios, for the latter, the proposed capacities for wind and solar in REPowerEU \cite{repowereu} are behind the capacities necessary to provide sufficient green electricity. \\
\\We show that ignoring the virtuous circle between capacity expansion and lower costs leads to delayed investments. A comparison of total costs of scenarios with different transformation speeds should consider cost reductions depending on the usage of a technology. A faster expansion of a technology than predicted in the exogenous cost assumptions thus leads to lower costs, a slower expansion to higher costs. In our results, total annual system costs are thereby up to 13\% higher and levelised cost of hydrogen increase by up to 67\% with the exogenous compared to the endogenous method. A middle ground approach where costs are updated sequentially offers the advantage of a lower computational effort compared to the endogenous method and  maintains the correlation between investment costs and capacities. \\
\\ Significant cost reductions in the production of green hydrogen to 1.26\kg by 2050 are possible. As further cost reductions and scale-up of electrolysis are expected in the coming years, endogenous cost modelling of electrolysis is important to compare total costs of different scenarios and to determine the right timing for investments. \newpage
\FloatBarrier
\section{Tables}\label{tables}
\begin{table}
	\begin{center}
		\begin{minipage}{280pt}
			\caption{Global capacity, European share of global capacity and learning rates (cost reduction for every doubling of cumulative capacity).}
			\label{tab:learning_rates}%
			\begin{tabular}{@{}llll@{}}
				\toprule
				Technology & Global capacity [GW] & European share [\%] & Learning rate [\%]\\
				\midrule
				solar PV & 707 \cite{irena_data2020} & 22 \cite{irena_data2020} & 24 \cite{cost_dea} \\ 
				onshore wind & 699 \cite{irena_data2020} & 26 \cite{irena_data2020} & 10 \cite{cost_dea} \\
				offshore wind & 34 \cite{irena_data2020} & 73 \cite{irena_data2020} & 10 \cite{cost_dea} \\
				H$_2$ electrolysis & 1 \cite{iea_electrolysis_2021} & local learning & 16 \cite{irena2020} \\
				\botrule
			\end{tabular}
		\end{minipage}
	\end{center}
\end{table}
\FloatBarrier
\section{Methods}\label{methods}
\subsection{Model description}\label{sec:model}
We use the open-source European sector coupled model PyPSA-Eur-Sec \cite{githubpypsa}, which minimises total system costs while optimising generation, storage and distribution capacity and dispatch. It covers energy and feedstock demand in the sectors electricity, heating, transport and industry. The model has already been presented in detail in various publications \cite{Synergies2018, victoria_early_2020, victoria2021speed, zeyen2021}, so in the following only the newly implemented features for this study are presented. \\
\\ The period from 2020 to 2050 is modelled, with perfect foresight and 7 investment periods at 5-year intervals. Three different CO$_2$ budgets of 25.7, 45 and 73.9 Gt CO$_2$ are assumed for the whole modelling horizon. These budgets correspond respectively to a warming of +1.5$^\text{o}$C, +1.7$^\text{o}$C, +2$^\text{o}$C, assuming a per capita share of the global CO$_2$ budgets (further explanations about the chosen budgets are given in an earlier publication \cite{victoria2021speed}). The CO$_2$ emission paths are not fixed, except that CO$_2$ neutrality (or negative emissions) is enforced for 2050, in line with the European Commission's target for net-zero greenhouse gas emissions. \\
\\Existing generation capacities (lignite, coal, gas, hydro, nuclear, solar, offshore and onshore wind) are taken from the \gls{irena} 2020 report \cite{irena2020} and the open-source package powerplantmatching \cite{powerplantmatching}. Costs, lifetime and efficiencies are assumed for the respective year of the Danish Energy Agency \gls{dea} \cite{cost_dea}. The investment costs of renewable generation capacity (solar, onshore and offshore wind) and hydrogen electrolysis are in the endogenous case not exogenously specified but are part of the optimisation in form of a piecewise-realised one-factor experience curve. This results in a mixed integer linear problem \gls{milp} which is further described below. The total system costs are not discounted by a social discount rate that reflects the value of future investments in order to investigate the impact of learning in isolation from other effects. A high discount rate leads to a significantly lower weighting of costs in 2050 and thus shifts investments into the future while endogenous learning leads to earlier investments.\\
\\ There are three different competing options for producing hydrogen in the model: (i) grey hydrogen (via steam methane reforming (\gls{SMR})), (ii) blue hydrogen (\gls{SMR} + carbon capture) and (iii) green hydrogen (via electrolysis). Hydrogen can be used for methanation, for heating (hydrogen boilers), electricity (fuel cells and retrofitted \gls{ocgt}), in the industry and in the transport sector. The demand pathway for hydrogen in parts of industry and transport is exogenously defined, while in all others sectors, hydrogen competes with other ways of supplying demand in these sectors. This means both demand for hydrogen (e.g. where it competes with heat pumps in the heating sector) and the supply side, for example if the hydrogen is produced via electrolysis or \gls{SMR}, are part of the optimisation. \gls{ocgt} and gas boilers for heating can be retrofitted to run flexibly with natural gas or hydrogen (see Figure \ref{fig:h2usage}). There are various electrolysis technologies. In this study, cost and efficiency assumptions of alkaline electrolysis cells (\gls{aec}) are used since they are currently the most common electrolysers available on the market. \\
 \\ In contrast to other studies, no maximum annual expansion rate of renewable generation capacity and no annual CO$_2$ emission paths are specified. Those constraints often reduce strongly the solution time of the optimisation problem, but also predetermine the transition paths. Since we consider multiple sectors and investment periods, the optimisation problem is aggregated spatially and temporally to make it computationally solvable. Spatially, energy transmission networks are reduced to a single node for Europe, while six different typical regions are used to represent the variability of  renewable generation. We optimise the power transfer capacity between transmission and distribution level. No existing grid infrastructure of distribution grids is assumed. Losses in distribution are neglected. Costs of the distribution grid of 500\kwe are applied. Electricity demands, heat pumps, resistive heaters, rooftop PV, home batteries and electric vehicles are connected to the distribution grid. All remaining technologies of the power sector (e.g. large scale storage, wind parks, conventional power plants, electrolysers) are connected to the high-voltage grid. In the Supplementary Material, we analyse the impact of spatial aggregation on our results by comparing scenarios with higher spatial resolution and the exogenous method with our results from the manuscript (see Figure \sref{fig:spatial_electrolysis}). The infrastructure cost contribute with 0.1-7.6\% to the total annualised costs only by a small margin. In terms of temporal aggregation, 10 typical days per investment period are considered. This allows, in contrast to most \gls{iam}s, the possibility to represent the temporal variation of the renewable generation and is comparable to other \gls{esm} like PRIMES using two or nine typical days \cite{primes} or Heuberger et al. \cite{heuberger2017} using 11 temporal days. The 10 typical days are obtained through k-medoids clustering using the Python package tsam \cite{tsam_github, tsam_paper}, so that they represent the average statistics of weather and demand, while also capturing more extreme events. To find the optimal solution we use the commercial solver Gurobi \cite{gurobi} using 12 threads.
\subsection{Endogenous learning}\label{sec:experience_curve}
\begin{table*}[h]
	\begin{center}
	\begin{minipage}{400pt}
	\begin{tabular}{m{4em} m{4cm} m{1cm} m{4cm}} 
		{\textbf{parameter}} & {\textbf{definition}} & {\textbf{variable}} & {\textbf{definition}} \\
		{} & {} & {} & {} \\
		\midrule
		$\overline{c_0}$ &  initial investment costs & $c$ &  investment costs \\
		$\overline{E_0}$ &  initial experience & $E$ &  experience, cumulative installed capacity \\
		$\alpha$ &  learning index &  $TC$ &  cumulative technology costs \\
		$LR$ &  learning rate &  $\delta$ &  continuous variables $\in [0,1]$, part of \gls{sos} \\
		$s$ &  technology (e.g. solar) &  $cap$ &  new build capacity per investment period \\
		$t$ &  investment period &  $inv$ &  investment cost per investment period \\
		$i$ &  interpolation point &  \\ 
		$N$ &  total number of interpolation points used for piece-wise linearisation &   &   \\
		$gf$ &  global factor, share of global capacity installed in Europe &   &   \\
		$(\overline{E_i}, \overline{TC_i})$ &  interpolation points of piece-wise linearisation &   &   \\
		$m_{s,j}$ &  slope of line segment $j$ for technology $s$ &   &   \\
		\bottomrule
	\end{tabular}
			\caption{Summary of parameters (\textit{left}) and variables (\textit{right}).}
	\label{tab:parameters}
		\end{minipage}
\end{center}
\end{table*}
Experience curves are an economic concept based on empirical evidence in which the specific investment costs $c$ decrease by a constant factor $\alpha$ with each doubling of experience $E$ 
\begin{equation}\label{eq:experience_curve}
	c(E)=\overline {c_0} \cdot \Big(\frac{E}{\overline {E_0}}\Big)^{-\alpha}  \quad \text{with $\alpha$ given by} \quad \alpha = \log_2 \Big(\frac{1}{1-LR}\Big).
\end{equation}
The constants $\overline {c_0}$ and $\overline {E_0}$ are fixed starting points, $LR$ is the so called learning rate. If for example the learning rate is 20\% ($LR=0.2$), the costs are reduced by 20\% for each doubling of cumulative experience. Typically learning rates range between 5\%-25\%. Smaller modular technologies (e.g. \gls{PV} or wind) tend to have higher learning rates than large-scale plants \cite{neij2008, wilson2020}. In this study the global cumulative capacity is used as a proxy for experience. $c$ represents the investment costs [EUR per MW], $c_0$ the initial investment costs [EUR per MW].\\
\\ Experience curves make \gls{esm} optimisation problems both non-linear and non-convex, which makes solving particularly challenging. There are two main approaches \cite{mattson} to integrate experience curves within \gls{esm}:
\begin{enumerate}
	\item direct non-linear implementation,
	\item piecewise linear approximation of cumulative costs.
\end{enumerate}
In this paper, we follow the latter and use special ordered sets of type 2 (\gls{sos}). Compared to the non-linear implementation this has several advantages: it can find a global minimum rather than getting stuck in a local one, it does not depend on the initial starting point of the solver, and it can be solved faster using commercial solver algorithms. \\
%
\begin{figure}[h]
	\centering
	\includegraphics[width=1.0\linewidth]{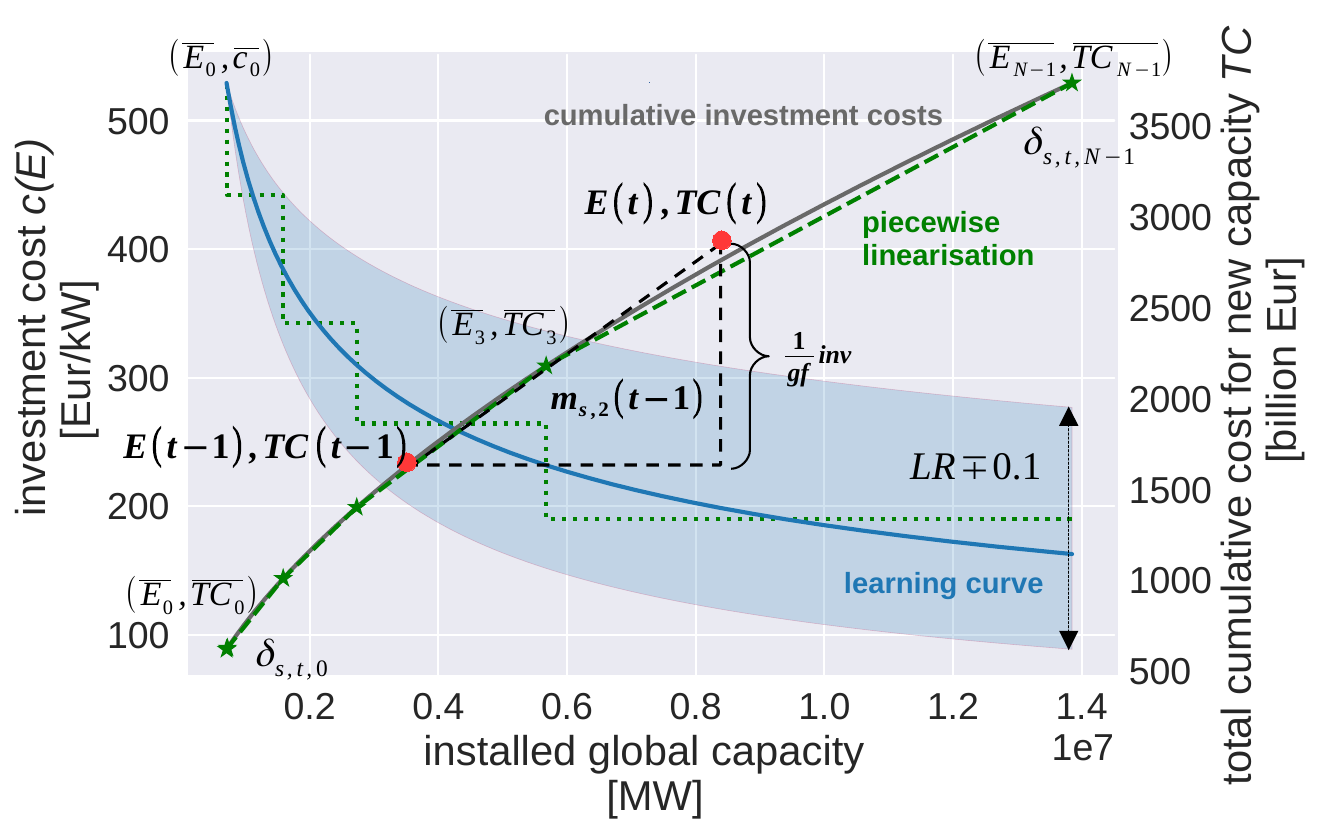}
	\caption{Experience curve and cumulative cost curve, as well as the implemented piecewise linearisation (\textit{hashed line}) with in this case four line segments.}
	\label{fig:linearisation}
\end{figure}
\\ \textbf{Total technology cost $TC$:} The cumulative investment costs for a technology can be obtained by integrating the experience curve (eq. \ref{eq:experience_curve}). For $\alpha \ne 1\rightarrow LR \ne 0.5$ this results in
\begin{equation}\label{eq_cum_cost}
	TC = \int_{E_0}^{E} c \,dE'= \frac{1}{1-\alpha} \Big(c(E) E - \overline{c_0 E_0}\Big).
\end{equation}
The cumulative costs $TC$ are stepwise linearised with a given number of line segments (see Figure \ref{fig:linearisation}). The greater the number of segments, the more precise the solution. However, this also increases the number of variables and thus the solution time.\\
\\Other models use the following number of segments: In ERIS \cite{eris2004} 6, global MARKAL 6-8, MARKAL-Europe 6-20 \cite{markal1999}. Heuberger et al. \cite{heuberger2017} 5. In this study we assume five line segments. This corresponds to six interpolation points $(\overline{TC_i}, \overline{E_i})$ (marked in Figure \ref{fig:linearisation} with a star) for $i \in$ [0, N-1] where $N$ is the number of interpolation points. \\
\\The definition of the line segments follows Barreto's approach \cite{barreto2001}: line segments at the beginning of the learning are shorter to capture the steep part of the experience curve more precisely. With each line segment the cumulative cost increase   doubles. \\
\\ \textbf{Special ordered set of type 2 \gls{sos}:}
We define a set of continuous variables $\delta_{s,t,i} \in [0,1]$ for each technology $s$ (e.g. solar \gls{PV}), investment period $t$ and interpolation point $i \in [0, N-1]$ of the linearisation. $N$ is the total number of interpolation points used for the piece-wise linearisation. The continuous variable indicates which line segment is active. Only two adjacent delta variables are non-zero at the same time $t$ for technology $s$ and their sum is equal one
\begin{equation}
	\sum_{i=0}^{N-1} \delta_{s,t,i} = 1.
\end{equation}
For example, if line segment $i$ is active then only $\delta_{s,t,i}$ and $\delta_{s,t,i+1}$ are non-zero. \\
\\ \textbf{Cumulative experience $E$:} The cumulative capacity $E_{s, t}$ of a technology $s$, time $t$ and interpolation points $i$ is defined as a summation of the product of the continuous variable $\delta$ and the x-position of the interpolation points $\overline{E_{s, i}}$
\begin{equation}
	E_{s, t} = \sum_{i=0}^{N-1} \delta_{s,t,i} \cdot \overline{E_{s, i}}.
\end{equation}
For example, if line segment $i$ is active, then (4) will interpolate between $\overline{E_{s,i}}$ and $\overline{E_{s,i+1}}$. \\
\\The new installed capacity $cap_{s, t}$ per investment period in Europe is the difference of the cumulative experience weighted by a global factor $gf$
\begin{equation}
	cap_{s, t} = gf \cdot (E_{s, t} - E_{s, t-1}).
\end{equation}
The global factor is one for local learning (assumed for hydrogen electrolysis in this study). For global learning (as considered for PV, onshore and offshore wind in the following results) it represents today's share of European compared to global capacities. \\
\\ \textbf{Linear combination:} If no time-delay for the learning effects would be considered, one could express the cumulative cost $TC$ similar to the cumulative experience $E$ with the help of the \gls{sos} variables $\delta$ and the y-position of the interpolation points $\overline{TC_{s,i}}$ as
\begin{equation}
	TC_{s,t} = \sum_{i=0}^{N-1} \delta_{s,t,i} \cdot \overline{TC_{s, i}}.
\end{equation}
\textbf{Temporal-delayed learning effect}: In this study we consider a temporal-delayed learning effect which means that the investment cost decrease in an investment period $t$ depends on the cumulative installed capacities at the previous investment period $t-1$. This represents learning effects more realistically, as investment costs do not decrease immediately in the same reference year as the emerging technology is employed, but are subject to a time lag. One should be aware that this results in an overestimating of the cumulative cost curve shown exemplary in Figure \ref{fig:linearisation}. The cumulative costs are defined with the delayed learning as
\begin{equation}
	TC_{s, t} = TC_{s, t-1} + m_{s,t-1, i} \cdot cap_{s,t}.
\end{equation}
Here $m_{s,t-1, i}$ is the slope of the line segment at the previous investment period $t-1$. As the total costs are the integral of the experience curve, the slope $m_{s, t, i}$ is equivalent to the specific investment costs. \\
\\ The overall investment costs per investment period $inv_{s, t}$ are defined as
\begin{equation}
	inv_{s, t} = gf \cdot (TC_{s, t}-TC_{s, t-1}).
\end{equation}
\section{Data availability}
The specific model runs and scenario data for this study are archived at Zenodo \url{https://zenodo.org/record/6645232#.Yt6L6tJBwkI}
 under a CC-BY-4.0 license.
\section{Code availability}
The PyPSA-Eur-Sec model is available under MIT license  via Github \url{https://github.com/PyPSA/pypsa-eur-sec}. Model documentation \url{https://pypsa-eur-sec.readthedocs.io/en/latest/}. All the technology assumptions are available via Github \url{https://github.com/PyPSA/technology-data}, version v0.3.0 is used in this study. The source code and input data for this study are openly available at Zenodo \url{https://zenodo.org/record/6645232#.Yt6L6tJBwkI} under a CC-BY-4.0 license.
\bibliography{sn-bibliography}
\clearpage
\newpage\null\thispagestyle{empty}
\section{Author contributions}
\textbf{Elisabeth Zeyen}: Methodology, Software, Validation,
Formal analysis, Investigation, Data Curation, Writing -
Original Draft, Visualisation, Conceptualisation \\
\textbf{Marta Victoria}: Writing- Review \& Editing, Supervision, Conceptualisation, Methodology \\
\textbf{Tom Brown}: Conceptualisation, Methodology, Resources,
Writing- Review \& Editing, Supervision, Funding acquisition, Project administration
\section{Competing interest declaration}
The authors declare no competing interests.
\section{Acknowledgement}
The authors thank Johannes Hampp, Falko Ueckerdt, Markus Millinger, Lina Reichenberg, Fredrik Hedenus and Niclas Mattsson for helpful discussions and suggestions. We thank the anonymous reviewers for the detailed feedback which helped to improve the paper.
\section{Additional information}
\backmatter
\textbf{Extended data} is available for this paper at \url{https://zenodo.org/record/6645232#.Yt6L6tJBwkI}. \\
\textbf{Supplementary information} is available. \\
\textbf{Correspondence and requests for materials} should be addressed to Elisabeth Zeyen.

%
%
%
%


\printglossary[type=\acronymtype]
\clearpage
\newpage\null\thispagestyle{empty}
\end{document}


\title[Supplementary Material -  Endogenous learning for green hydrogen in a sector-coupled energy model for Europe]{\center{Supplementary Material}}
\subtitle{Endogenous learning for green hydrogen in a sector-coupled energy model for Europe}


\author*[1,2]{\fnm{Elisabeth} \sur{Zeyen}}\email{e.zeyen@tu-berlin.de}

\author[3,4]{\fnm{Marta} \sur{Victoria}}\email{t.brown@tu-berlin.de}

\author[1,2]{\fnm{Tom} \sur{Brown}}\email{mvp@mpe.au.dk}

\affil*[1]{\orgdiv{Department of Digital Transformation in Energy Systems, Faculty of Process Engineering}, \orgname{TU Berlin}, \orgaddress{\street{Einsteinufer 25 (TA 8)}, \city{Berlin}, \postcode{10587}, \state{Berlin}, \country{Germany}}}

\affil[2]{\orgdiv{Institute for Automation and Applied Informatics (IAI)}, \orgname{Karlsruhe Institute of Technology (KIT)}, \orgaddress{\street{Forschungszentrum 449}, \city{Eggenstein-Leopoldshafen}, \postcode{76344}, \state{Baden-Württemberg}, \country{Germany}}}

\affil[3]{\orgdiv{Department of Mechanical and Production Engineering}, \orgname{Aarhus University}, \orgaddress{\street{Inge Lehmanns Gade 10}, \city{Aarhus}, \postcode{8000}, \country{Denmark}}}

\affil[4]{\orgdiv{Novo Nordisk Foundation CO2 Research Center}, \orgaddress{\street{Gustav Wieds Vej 10}, \city{Aarhus}, \postcode{8000},  \country{Denmark}}}

\abstract{}

\maketitle
\thispagestyle{empty}
\newpage
\tableofcontents
\pagenumbering{gobble}
\newpage
\pagenumbering{arabic}
\setcounter{page}{1}
\section{Method}
\subsection{Perfect foresight}
In this study, a time span of 2020-2050 with 7 investment periods of 5 years each ([2020,2025,...,2045,2050]) is considered and the total system costs are minimised. The modelling of these multiple investment periods is implemented in one single optimisation problem (perfect foresight) and allows a change in the installed capacities of the assets in each investment step. The investment costs of each asset are annualised and distributed over the years in which the asset is active (from build year until the end of the lifetime). This allows the costs in the individual investment periods to be weighted (e.g. with a social discount rate) and the costs of savings do not have to be taken into account at the end of the modelling horizon (which is the case if the total investment is considered instead of the net present value (\gls{npv}) in each time step). \\
\\ The annualised investment costs $c_s$ [$\frac{\text{\officialeuro}}{MW a}$] of an asset $s$ are defined as 
\begin{equation}
	c_s = k_s \cdot \frac{i_s}{1-\frac{1}{(1+i_s)^{L_s}}}.
\end{equation}
$k_s$ represent the specific investment cost [$\frac{\text{\officialeuro}}{MW}$], $i_s$ the discount rate, and $L_s$ the lifetime of the asset. \\
\\ The annual costs of one single year $a$ consists of fixed and operational costs. Fixed annualised investment costs of the power capacities $G_{s}$, energy capacities $E_{s}$ and inter-connector capacities $F_{l}$ are considered for all active assets (build year $b_s$ < $a$ < end of lifetime $b_s+L_s$). Operational costs $o_t$ are defined for power dispatch through generation and storage $g_t$, as wells as power flow through inter-connectors $f_t$ at every time step $t$. Inter-connectors represent the conversion from one energy carrier to another (e.g. an electrolysis converting electricity to hydrogen). The total annual costs $f_a$ is defined as
\begin{align}\label{eq:objective}
	f_a = &\sum_{s|b_s \leq a \leq b_s+L_s} c_{s}\cdot G_{s} + \sum_{s|b_s \leq a \leq b_s+L_s}  c_{s}\cdot E_{s} + \sum_{l|b_l \leq a \leq b_l+L_l}  c_{l}\cdot F_{l} &\text{(fixed costs)}\\ + &\sum_{s|b_s \leq a \leq b_s+L_s,t} o_{s} \cdot g_{s,t} + \sum_{l |b_l \leq a \leq b_l+L_l, t} o_{l,t} \cdot f_{l,t}. &\text{(operational costs)} \nonumber
\end{align}
The total system costs for the whole investment period are minimised resulting in the objective function
\begin{equation}
	\min \sum_a w_a f_a.
\end{equation}
$w_a$  represents a weighting of the annualised costs depending on the duration of the investment period and the social discount rate $r$. The annual cost of one year a then discounted by
\begin{equation}
	d_a = \frac{1}{(1+r)^{a_0 - a}}
\end{equation}
For example for the year $a$=2022, starting year $a_0$=2020 and a social discount rate of $r$=0.02
\begin{equation}
	d_{2022}= \frac{1}{(1+0.02)^{(2022-2020)}} = 0.96.
\end{equation} 
The weighting $w_a$ is the sum of all the years within the investment period, so e.g. for the period 2020-2025
\begin{equation}
	w_{2020} = \sum_{a=2020}^{2025} d_a = 4.8.
\end{equation} 
In this study a social discount rate of $r=0$ is assumed which results in an equal weighting of all investment periods of $w_{a}$ = 5.
\subsection{Demand}
\begin{figure*}[!ht]
	\centering
	\subfloat[\centering industry demand]{{\includegraphics[width=0.45\textwidth]{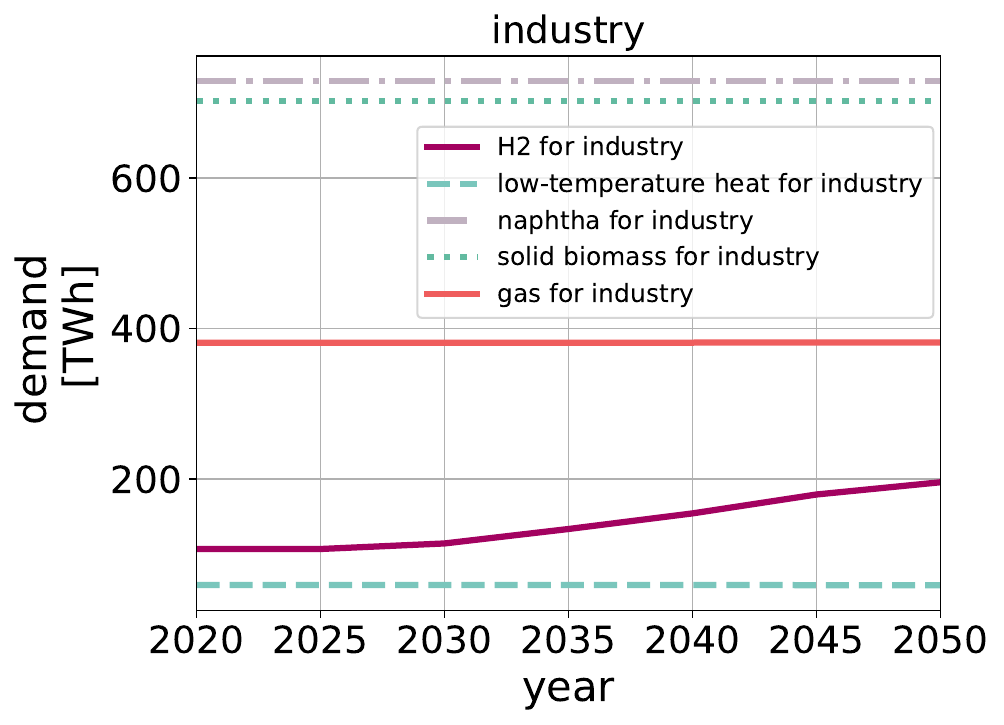} }}%
	\qquad
	\subfloat[\centering CO$_2$ emissions]{{\includegraphics[width=0.45\textwidth]{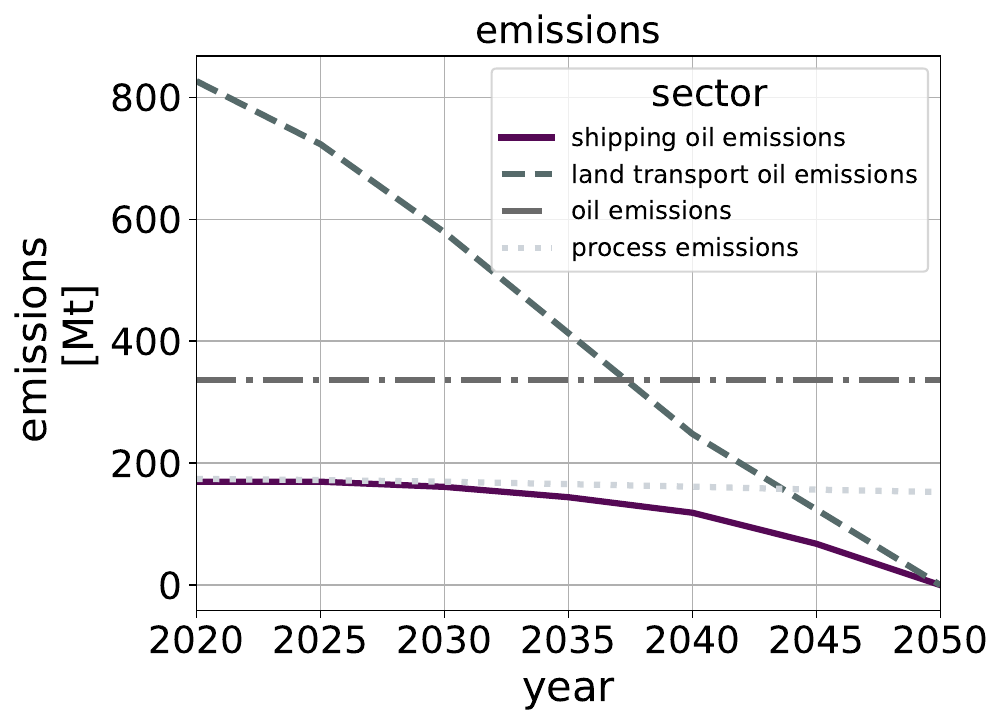} }}%
	\qquad
	\subfloat[\centering transport demand]{{\includegraphics[width=0.45\textwidth]{demand_land_transport .pdf} }}%
	\qquad
	\subfloat[\centering transport demand for shipping and aviation]{{\includegraphics[width=0.45\textwidth]{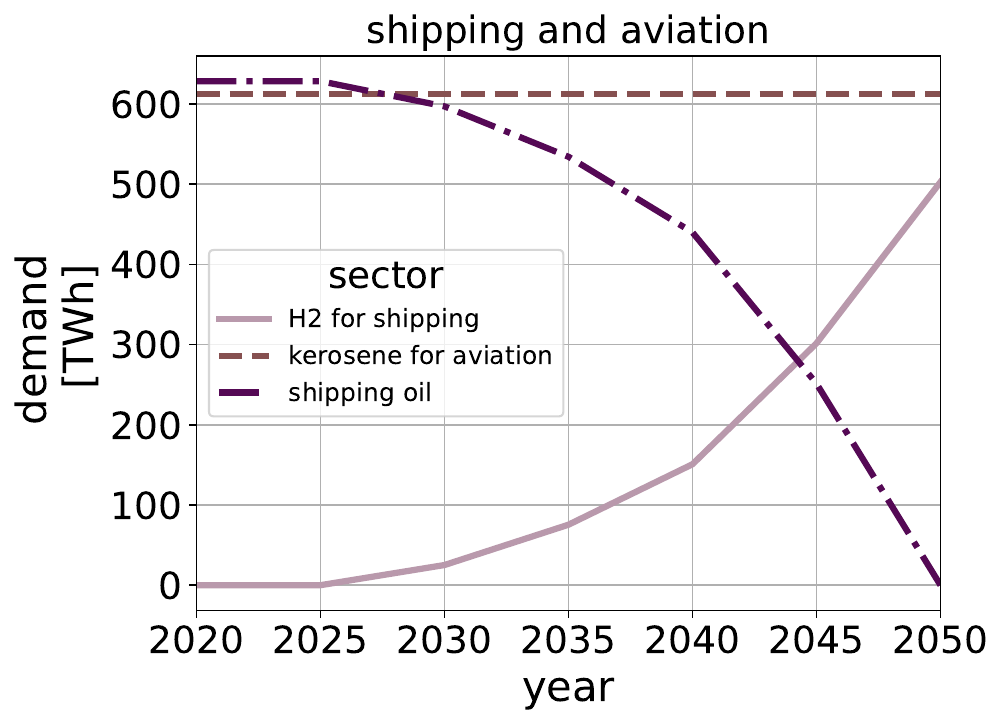}}}%
	\qquad
	\subfloat[\centering heat demand]{{\includegraphics[width=0.45\textwidth]{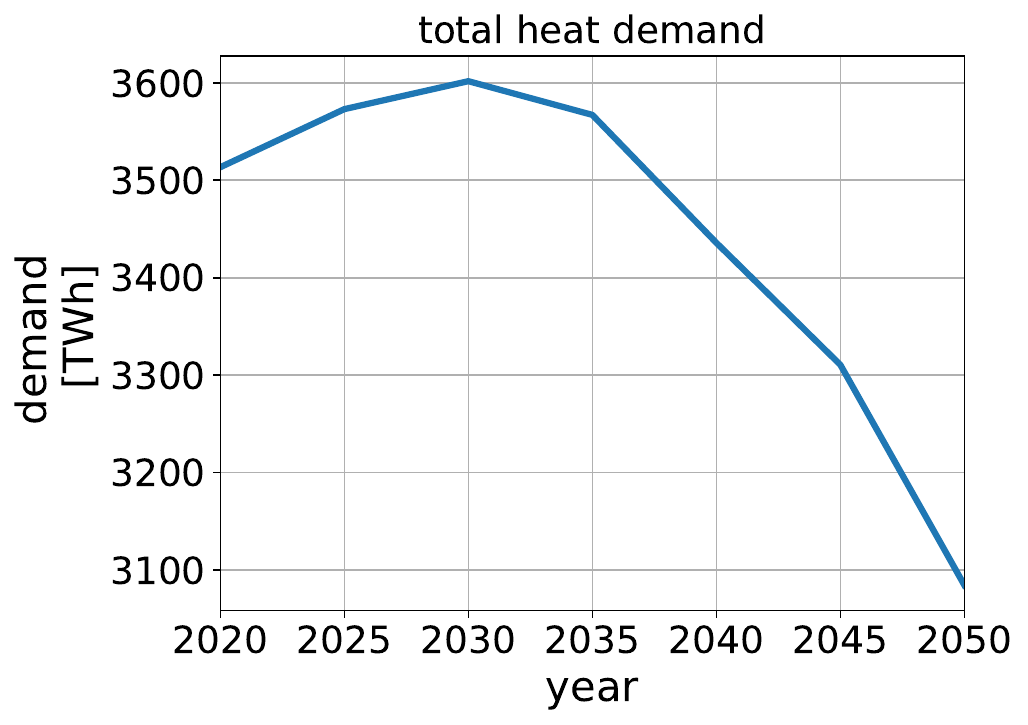} }}%
	\qquad
	\subfloat[\centering heat demand by heating system]{{\includegraphics[width=0.45\textwidth]{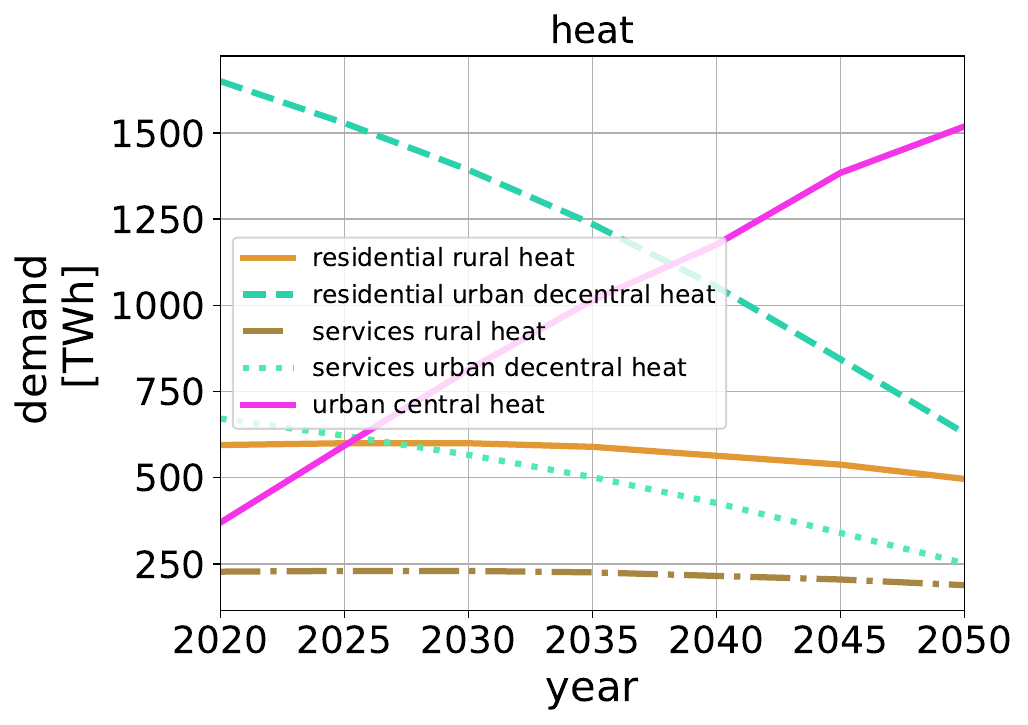} }}%
	\qquad
	\caption{Exogenous demand assumptions during the planing horizon for industry, transport and heating sector. As well as exogenous CO$_2$ emissions of industry processes and transport.}%
	\label{fig:demands}
\end{figure*}


\FloatBarrier
\section{Further sensitivity analysis}
\FloatBarrier
\subsection{Limit build out rates of renewable generation capacity}\label{sec:sensi_build_out}
\begin{figure}
	\centering
	\includegraphics[width=0.7\linewidth]{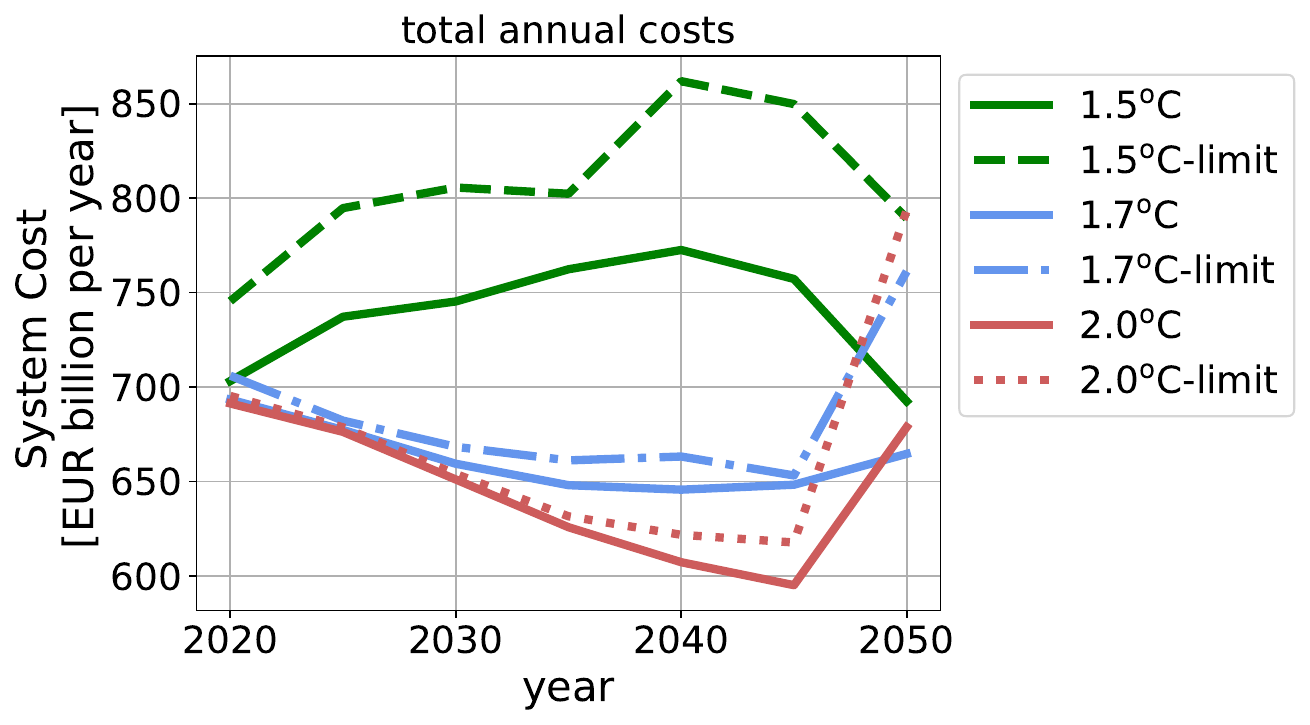}
	\caption{Comparing total annual system costs for scenarios with and without renewable build out rates.}
	\label{fig:totalcostsperyearlimit}
\end{figure}

\begin{figure}
	\centering
    \includegraphics[width=1.\textwidth]{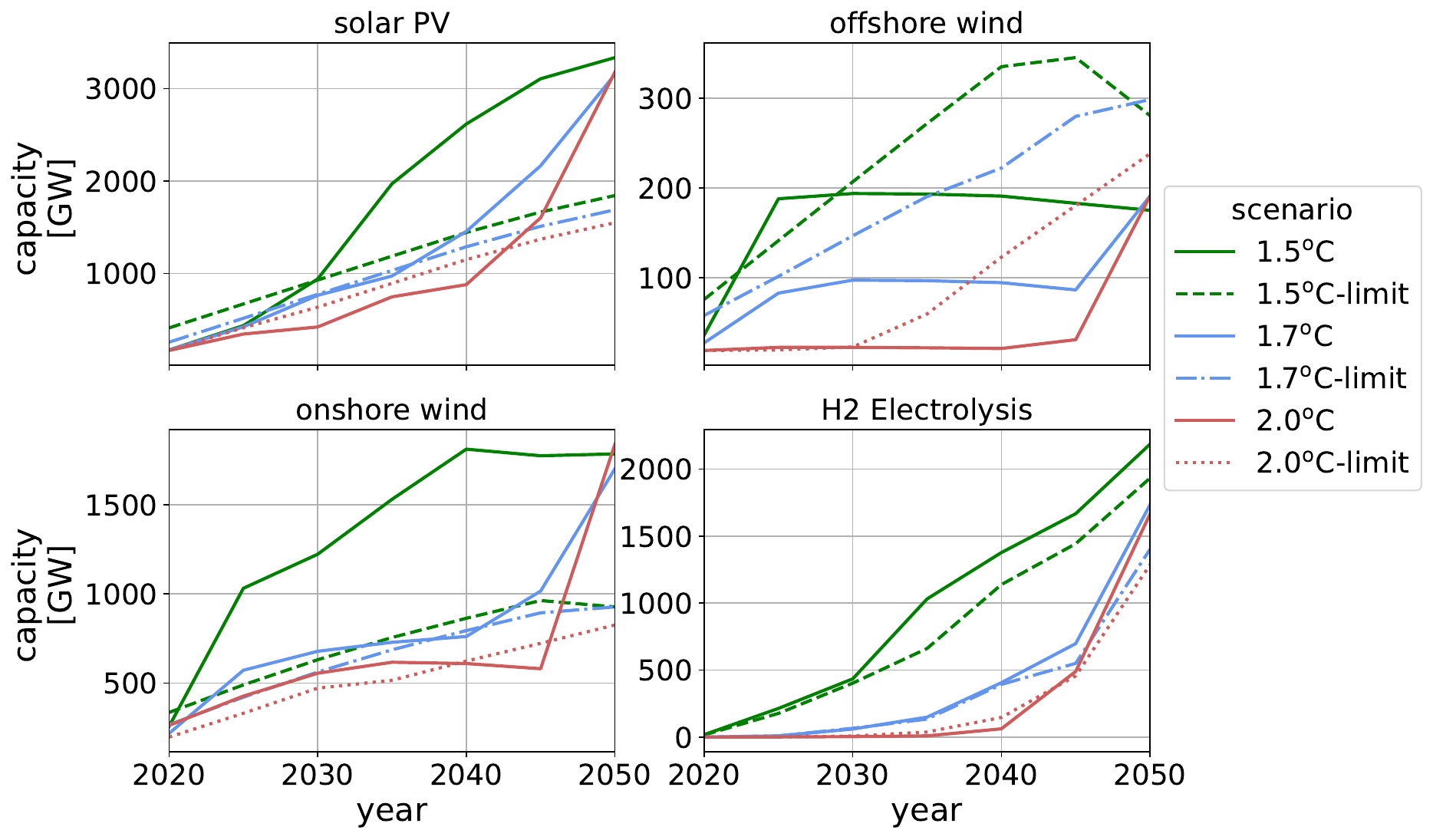} 
	\caption{Installed capacities of renewable generation and electrolysis for different CO$_2$ budgets assuming global learning for renewable generation and limited growth rates.}
	\label{fig:capacities_renewables_limit}
\end{figure}

\begin{figure*}
	\centering
	\subfloat[\centering Electricity]{{\includegraphics[width=0.7\textwidth]{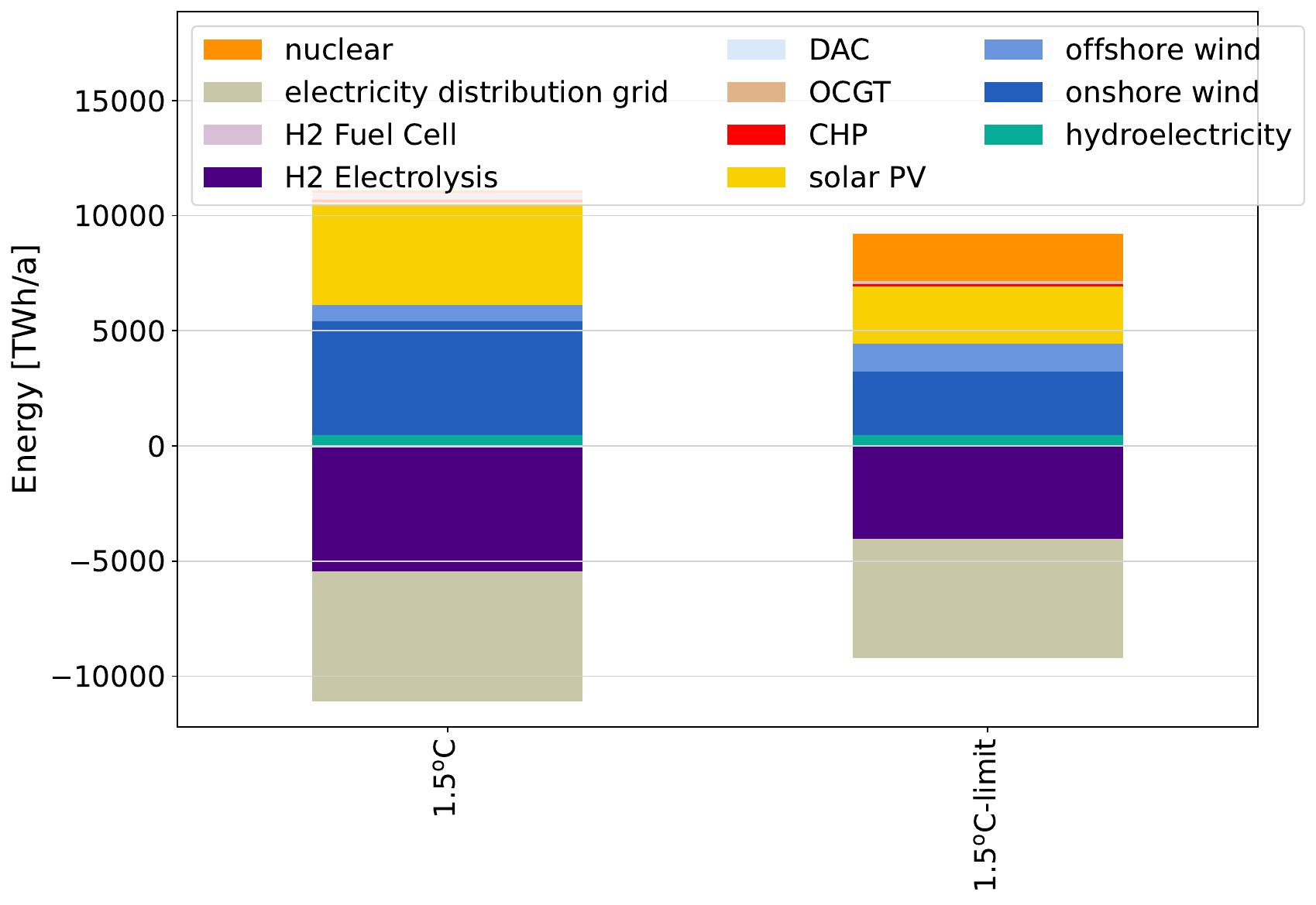} }}%
	\qquad
	\subfloat[\centering hydrogen]{{\includegraphics[width=0.7\textwidth]{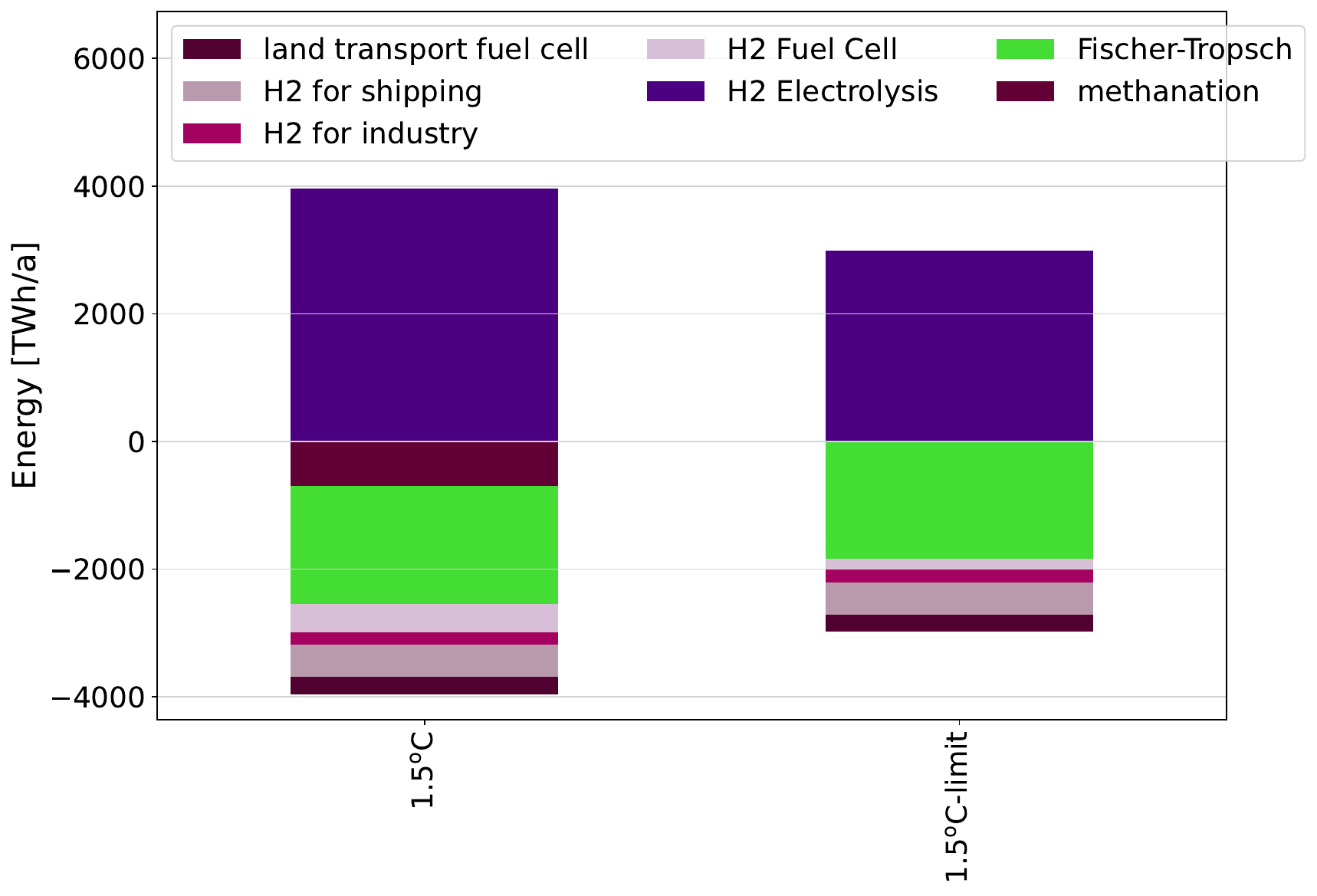} }}
	\caption{Comparing electricity and hydrogen supply and demand for scenarios with a \budget{1.5} with and without growth limit of renewables.}
	\label{fig:balances_limit}%
\end{figure*}

The high expansion rates of renewable energies are challenging as new plants often have long planning phases, are delayed or the construction is stopped by lawsuits or by environmental protection reasons. In the following, we therefore examine scenarios in which the expansion rates of onshore and offshore wind as well as solar \gls{PV} are limited respectively to 31 GW, 13 GW and 52 GW annual new capacity additions in Europe. These expansion rates are derived from the maximum historical expansion rates in Germany (4.9 GW in 2017, 2.1 GW in 2015, 8.2 GW in 2012 \cite{limitres}), which are scaled up for Europe in proportion to the population. The expansion of renewables is not constrained in the main results, as this often predetermines the transition paths and larger expansion rates than the historical ones are possible per se. \\
\\  The installed capacities are significantly lower for solar PV and onshore wind with limited expansion rates (see figure \ref{fig:capacities_renewables_limit}). Offshore installations are increasing to compensate for the low capacities of solar PV and onshore wind. The total costs are about 4-9\% higher than in the non-limited scenarios, depending on the budget (see figure \ref{fig:totalcostsperyearlimit}). Less hydrogen is produced and used since the electricity prices are higher. Nuclear power is used since not all of the electricity demand can be covered by renewables due to the growth limit (see figure \ref{fig:balances_limit}). This causes higher system costs.
\FloatBarrier
\subsection{Costs and transition speed of the transport sector}\label{sec:sensi_transport}
\begin{figure*}[!ht]
	\centering
	\subfloat[\centering annualised system cost without costs for vehicles and infrastructure]{{\includegraphics[width=0.45\textwidth]{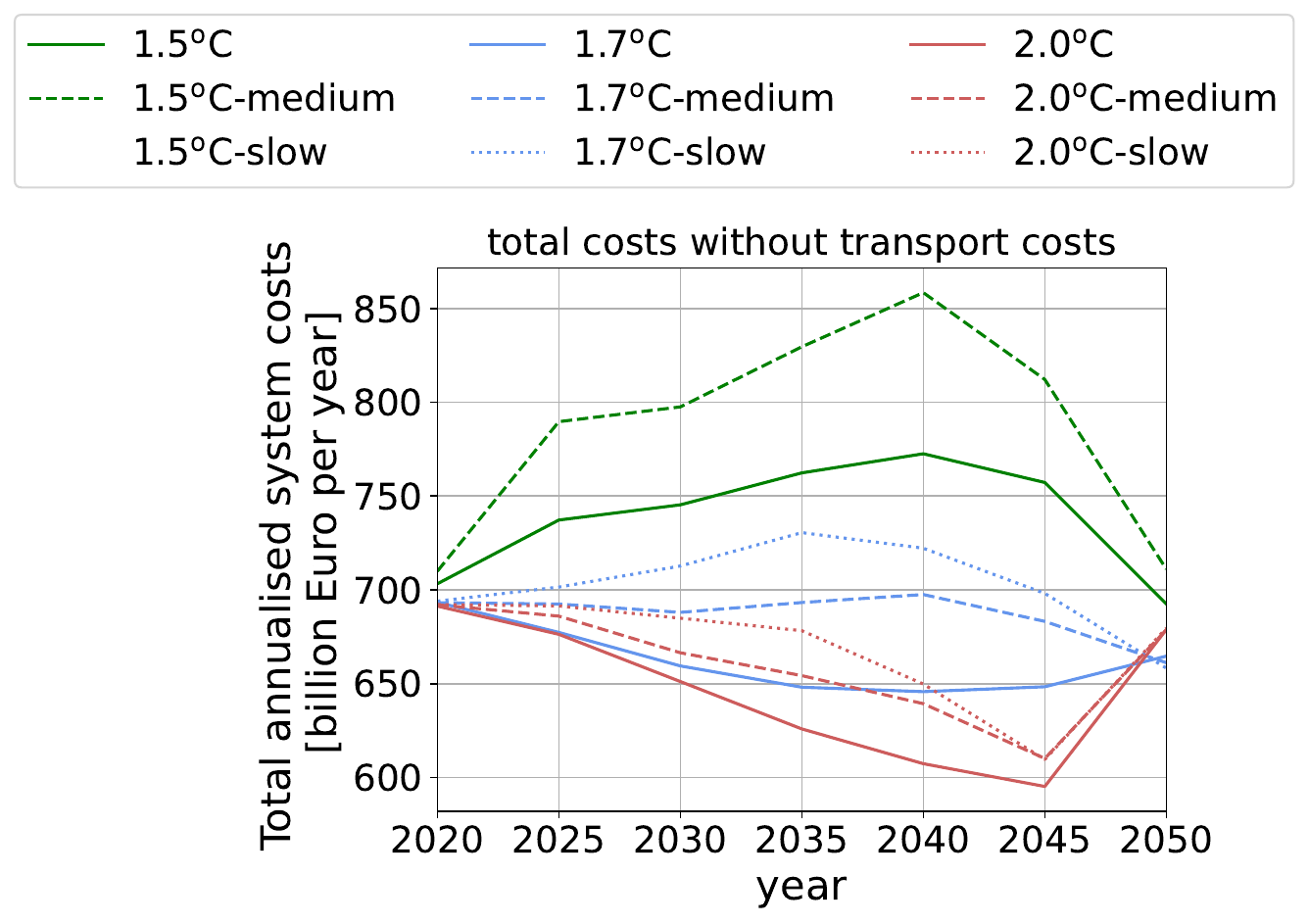} }}%
	\qquad
	\subfloat[\centering annualised system cost including costs for vehicles and infrastructure cost]{{\includegraphics[width=0.45\textwidth]{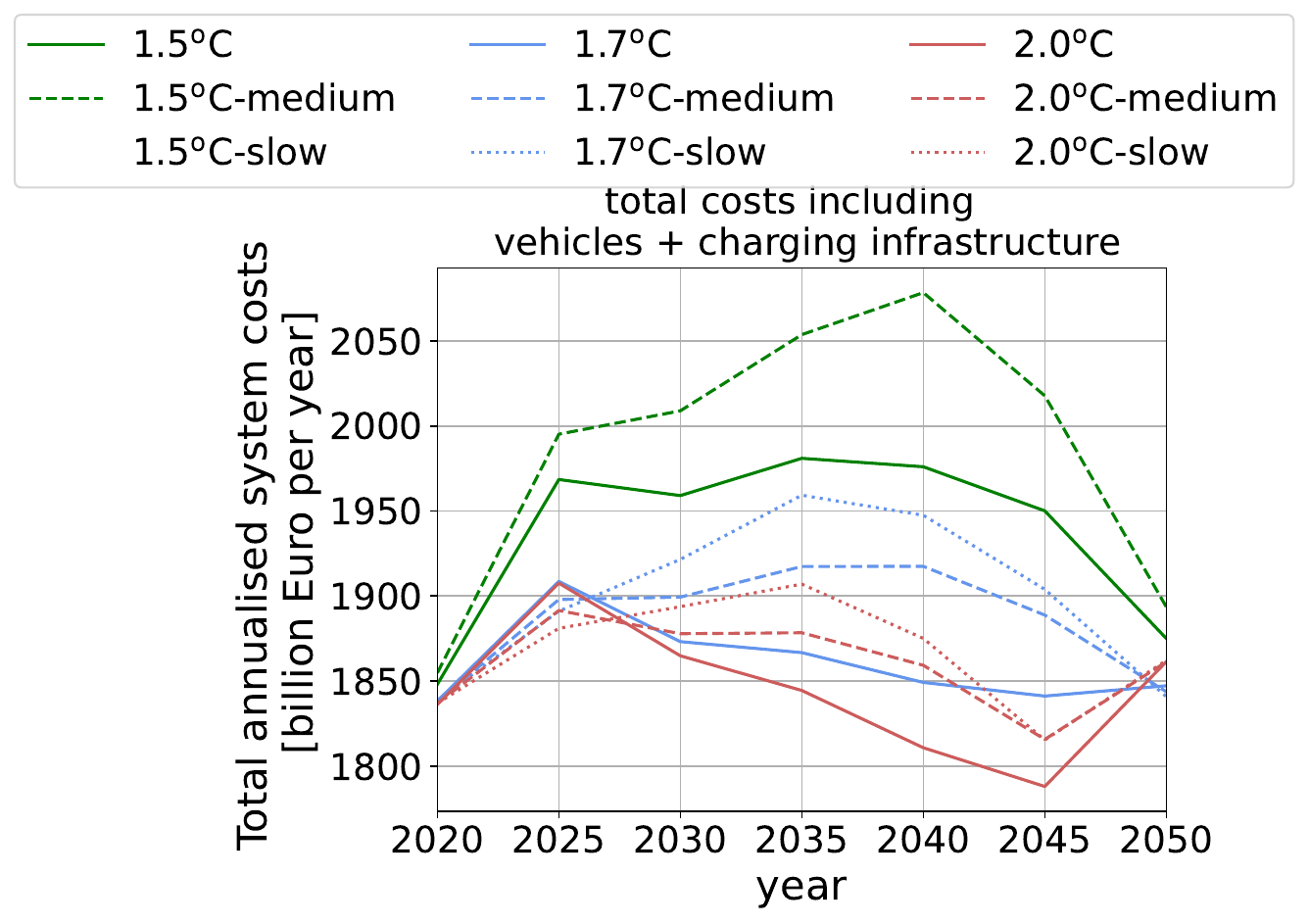} }}%
	\caption{Total annualised costs for three different transition speeds of the transport sector and without and including costs for charging infrastructure and cars. The slow transport scenario with a \budget{1.5} is not feasible and therefore not displayed in the graphics.}
	\label{fig:sensi_transport}
\end{figure*}

The proportions of internal combustion engine, electric and fuel cell cars are exogenously specified in this study. In this sensitivity analysis, on the one hand, the additional costs of cars and charging infrastructure are estimated, and on the other hand, the effects of a slower transformation of land transport are illuminated. \\
\begin{figure}
	\centering
	{{	\includegraphics[width=1\textwidth]{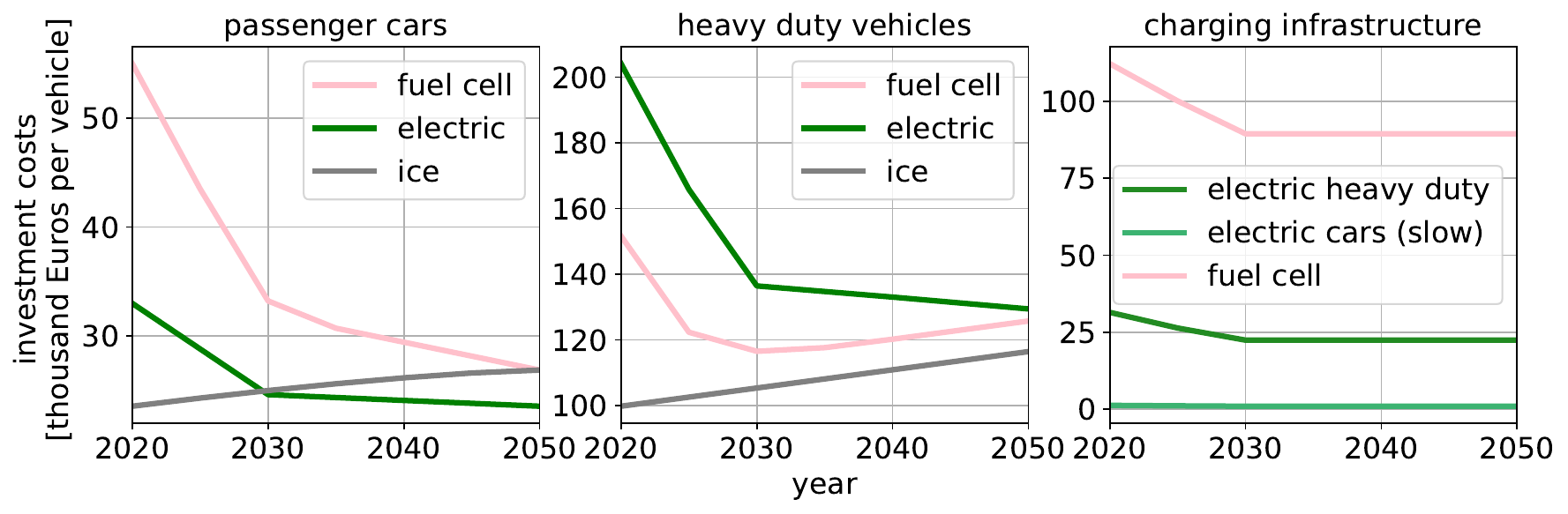}} }%
	\caption{Investment costs for passenger and heavy duty vehicles depending on engine type (electric vehicles (\textit{electric}), internal combustion engine (\textit{ice}) and fuel cell (\textit{fuel cell}) cars), as well as investment costs for charging infrastructure. No infrastructure costs for current internal combustion engines are assumed.}
	\label{fig:investmentcars}
\end{figure}
\\ Vehicles are divided into (i) light duty vehicles (passenger cars, freight transport, 2-wheelers) and (ii) heavy duty vehicles (buses, motor coaches, trolleys, heavy duty freight). Total number of vehicles and demand shares are taken from the JRC-IDEES 2015 \cite{jrcidees} and considered fixed throughout the transition. The cost and lifetime assumptions for passenger cars, heavy duty vehicles and charging infrastructure are taken from the Fraunhofer ISE study \cite{fraunhofer2020} (see Figure \ref{fig:investmentcars}). The investment costs are annualised assuming a discount rate of 7\%. For the passenger cars one charger per car is assumed, for the heavy duty vehicles fast charging with 20 vehicles per charger is assumed. This is a rather conservative assumption, e.g. compared to the Bloomberg report \cite{bloomberg_ev} which assumes a rise from today's 5-20 vehicles per charger to 30-40 vehicles per charger by 2050. No costs are assumed for the existing infrastructure for internal combustion vehicles. The resulting annualised costs for cars and charger infrastructure for the different transport scenarios are added after the optimisation to the annualised system costs. \\
\\ In our base scenarios we consider a fast change in the transport sector. This assumes that by 2040, all vehicles are completely electrified or hydrogen-powered. In order to shed light on the impacts of a slower transformation of the land transport sector, two further transformation paths are highlighted in this sensitivity analysis with a medium and slow transformation of the sector. For all three budgets the fast transition of the land transport sector is cost-optimal. The assumption of a fast change in the transport sector, which is stated in the main results, is therefore justified for all three budgets. If costs for vehicles and charging infrastructure are included, total costs are between 1-2\% higher with a slower transformation from the \ce{2.0} to the \budget{1.5} (see Figure \ref{fig:sensi_transport}). The costs for cars and charging infrastructure make up a large part of the total costs with a share of 59-67\%. The proportionally small cost difference of 1-2\% of total costs can be explained by the large share of transport sector costs. However, a faster transition of the transport sector leads to savings of 538-1722 billion euros over the entire planning horizon. With a faster transition of the land transport sectors, electrolysis capacities are built up earlier. In 2050, investment costs and installed capacities of electrolysis are comparable for different speeds of the transport sector.\\
\begin{figure}[!ht]
	\centering
	\subfloat[\centering fast]{{\includegraphics[width=0.28\textwidth]{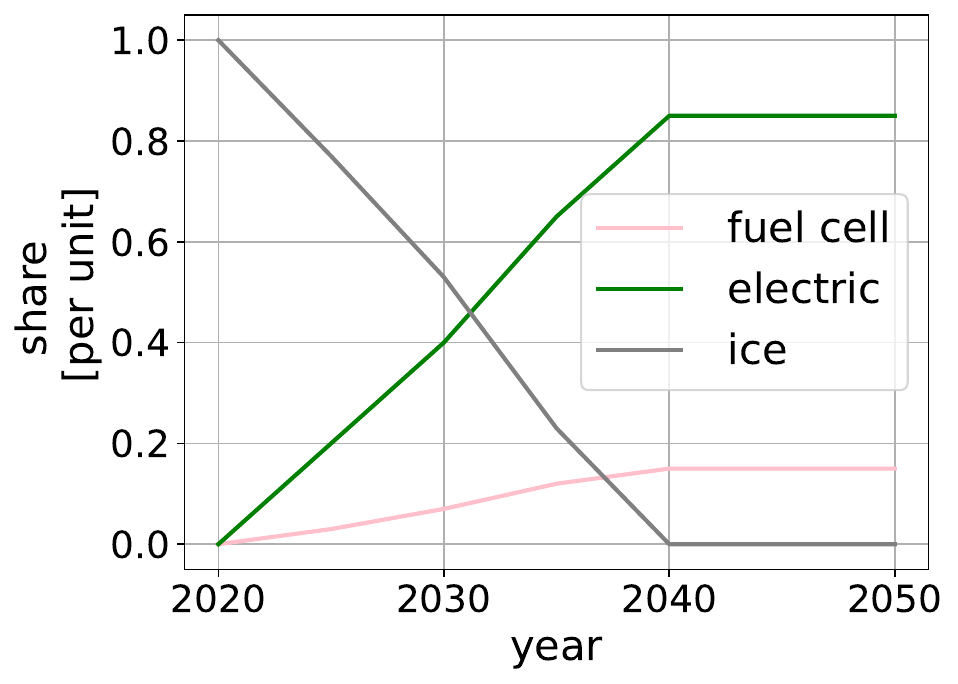} }}%
	\qquad
	\subfloat[\centering medium]{{\includegraphics[width=0.28\textwidth]{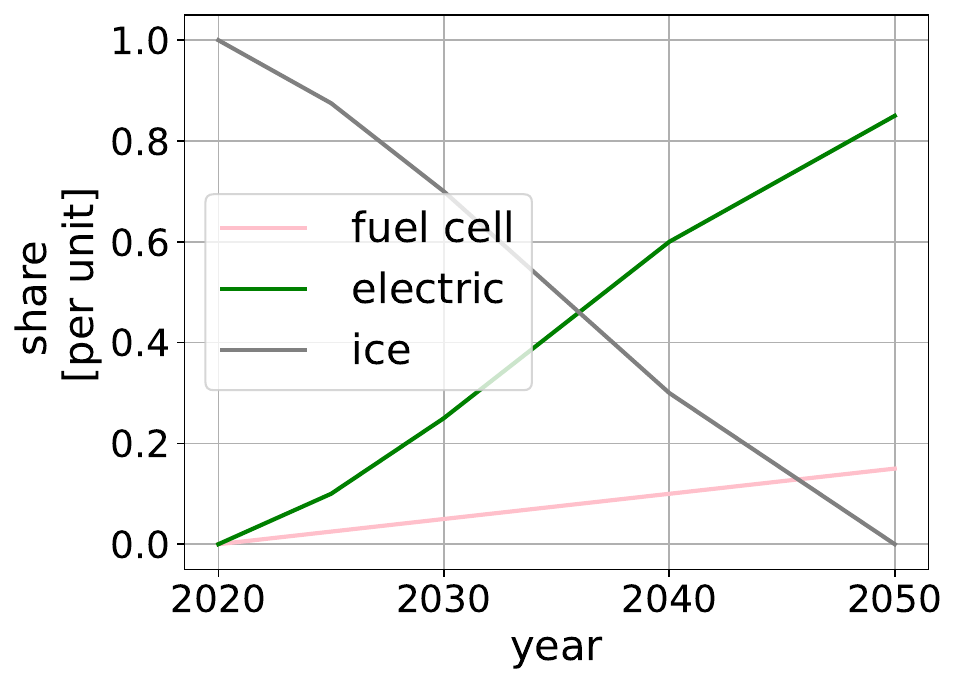} }}%
	\qquad
	\subfloat[\centering slow]{{\includegraphics[width=0.28\textwidth]{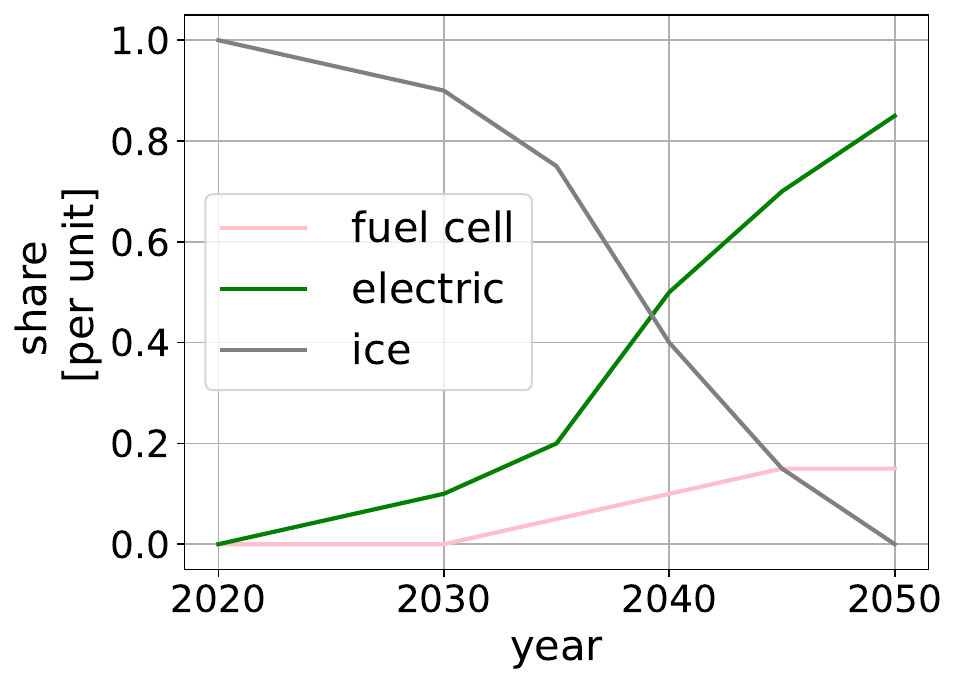} }}%
	\caption{Shares of electric, fuel cell and internal combustion engine (\textit{ice}) for the different speeds of transformation of the transport sector. In the main result section, the fast transition of the transport sector is assumed.}
	\label{fig:sensi_transport_demand}
\end{figure}
\\The investment costs for electrolysis in 2050 are similar between the scenarios. Our results for the investment costs of hydrogen electrolysis should therefore be robust for 2050, even in the case of a slower transformation of the land transport sector. In 2030, investment costs of electrolysis is even lower in the scenario with a slower transformation of the transport sector, as the demand for hydrogen is higher. The higher demand for hydrogen is driven by the larger demand for fuel for internal combustion engine vehicles, which has to be partly met by synthetic fuels in order to stay within climate targets.

\FloatBarrier
\subsection{Endogenous learning batteries and blue hydrogen production}\label{sec:sensi_batteries}
The choice on which technologies endogenous learning is applied impacts the results. Two further scenarios are therefore considered as a sensitivity analysis. First, to better represent the competition between blue and green hydrogen production, a scenario with endogenous learning is assumed for \gls{SMR} with carbon capture (LR$_{\text{SMR}}$=10\%, capture rate 90\%) and no learning on electrolysis and renewable. Second, endogenous learning is assumed for batteries (LR$_{\text{battery, energy}}$=18\%, LR$_{\text{battery, power}}$=19\%) in combination with endogenous learning of solar \gls{PV}. In both scenarios, endogenous learning does not lead to increased use of either technology (\gls{SMR} + CC or batteries). \\
\\ To investigate the conditions under which blue hydrogen production is used, we perform a sensitivity analysis on three key parameters for blue hydrogen production (i) the investment cost, (ii) the CO$_2$ sequestration potential and the (iii) the capture rate. The following results are performed with the exogenous method and a CO$_2$ budget of \ce{1.7}. 
\paragraph{Investment costs} We vary the investment costs between 0 to 572 Eur/kW$_{\text{H}2}$ (our base cost assumption). For CO$_2$ storage potential and carbon capture rate we assume our base assumptions of 200 Mt$_{\text{CO}_2}$ per year and 90\% respectively. The option of blue hydrogen is used if the investment costs are below 286 EUR/kW$_{\text{H}2}$. However, most of the hydrogen is produced via green hydrogen, even if blue hydrogen production is free of charge (see Figure \ref{fig:balances-blueh2-investment-costs-smrcc-h2}). There are three reasons for this. First, electrolysis offers the system advantage in the way that electricity from renewable sources can be used, which would otherwise be curtailed. Second, blue hydrogen production does not capture all CO$_2$. Therefore, additional \gls{dac} or negative emissions from biomass with carbon and capture are needed to reach net zero emissions in 2050. Third, the limited CO$_2$ storage potential, which is always a binding condition in our scenarios. 
\paragraph{CO$_2$ sequestration potential} We vary the CO$_2$ storage potential per year between 200 (base assumption) and 2000 Mt$_{\text{CO}_2}$/a. It is unclear how large the CO$_2$ storage potential is in Europe. Estimates vary between 108-7147 GtCO2 \cite{co2potential}. We have assumed a sequestration potential of 200 Mt/a as a base value to account for uncertainties about CO$_2$ infrastructure and storage potential.  Base assumptions for investment costs and carbon capture rate are assumed. As the storage potential increases, the amount of synthetic fuels produced and thus the amount of hydrogen production is reduced. Hydrogen is still mainly produced via electrolysis in 2050 (see Figure \ref{fig:balances-blueh2-co2-sequestration-h2}). Above a storage potential of 800 Mt$_{\text{CO}_2}$/a, about a small amount of 18 TWh$_{\text{H}_2}$ (1\% of total production) grey hydrogen is still produced in 2050. From a storage potential of 2000 Mt$_{\text{CO}_2}$/a, blue hydrogen is produced. In this case, the share of blue hydrogen in total production is 19\%. It should be noted that with an annual sequestration of 2000 Mt$_{\text{CO}_2}$ the CO$_2$ storage potential is exploited after 54 years based on a conservative assumption of a total storage potential of 108 Gt$_{\text{CO}_2}$.
\begin{figure}
	\centering
	\includegraphics[width=0.7\linewidth]{"balances-blueH2-investment costs SMR+CC-H2"}
	\caption{Varying the investment costs of blue hydrogen production from 0 to 572 \kw$_{\text{H}_2}$ (our base cost assumptions). CO$_2$ sequestration potential and capture rate are set at base assumptions (200 Mt$_{\text{CO}_2}$ per year, and 90\% respectively).}
	\label{fig:balances-blueh2-investment-costs-smrcc-h2}
\end{figure}
\begin{figure}
	\centering
	\includegraphics[width=0.7\linewidth]{"balances-blueH2-CO2 sequestration-H2"}
	\caption{Varying CO$_2$ sequestration potential per year from  200-2000 Mt$_{\text{CO}_2}$ per year. \gls{SMR} investment costs and capture rate are set at base assumptions (572 \kw$_{\text{H}_2}$ and 90\% respectively).}
	\label{fig:balances-blueh2-co2-sequestration-h2}
\end{figure}
\paragraph{Carbon capture rate} There is a wide range of assumed possible carbon capture rates of blue hydrogen production ranging from 53-99\%, with higher rates being achieved primarily with autothermal reforming. In the following, we compare our base assumption of 90\% capture rate with an assumed capture rate of 100\%. In our scenarios, varying only the carbon capture rate has no impact on the results (see Figure \ref{fig:balances-blueh2-capture-rate-h2}).
\begin{figure}
	\centering
	\includegraphics[width=0.7\linewidth]{"balances-blueH2-capture rate-H2"}
	\caption{Comparing carbon capture rate blue hydrogen production from 100\% to our base assumptions of 90\%.  CO$_2$ sequestration potential and \gls{SMR} investment costs are set at base assumptions (200 Mt$_{\text{CO}_2}$/a, and 572 \kw$_{\text{H}_2}$ respectively).}
	\label{fig:balances-blueh2-capture-rate-h2}
\end{figure}
\paragraph{Combined impact of two parameters on the results}
In the following, we make the most optimistic assumptions for two parameters and vary the third. First, we consider scenarios in which we vary the CO$_2$ sequestration potential with no investment costs for the production of blue hydrogen and a carbon capture rate of 100\%. With a storage potential of 200 Mt$_{\text{CO}_2}$/a, blue hydrogen is used under these assumptions in the years 2040-2045 with a maximum share of hydrogen production of 17\% (see Figure \ref{fig:balances-blueh2-noinvestment-cost--100-cc-h2}). With increasing storage potential, the share of blue hydrogen production also increases to a share of 80\% in 2050 at a storage potential of 2000 Mt$_{\text{CO}_2}$/a. A larger part of the hydrogen is converted back into electricity in fuel cells (1343 TWh$_{\text{H}_2}$ in 2050) in this scenario. The CO$_2$ storage potential is fully utilised. With a sequestration potential greater than 2000 Mt$_{\text{CO}_2}$/a, hydrogen production would be completely switched to blue hydrogen. \\
\\ In a second step, we consider a high sequestration potential (2000 Mt$_{\text{CO}_2}$/a) and a high carbon capture rate (100\%) and vary the assumptions on the investment costs of blue hydrogen production. With our base cost assumptions of 572\kw$_{\text{H}_2}$ 23\% of the hydrogen is produced as blue hydrogen in 2050 (see Figure \ref{fig:balances-blueh2-high-store--100-cc-h2}). The share is increasing with decreasing investment costs. With investment costs of 286\kw$_{\text{H}_2}$ the share of blue hydrogen production is 49\%.
\begin{figure}
	\centering
	\includegraphics[width=0.9\linewidth]{"balances-blueH2-no investment cost + 100p CC-H2"}
	\caption{Varying the CO$_2$ sequestration potential from 200-2000  Mt$_{\text{CO}_2}$/a with 100\% carbon capture rate and no investment costs for blue hydrogen production. This is compared to our base scenario (right panel) with investment costs of blue hydrogen production of 572 \kw$_{\text{H}_2}$, CO$_2$ sequestration potential of 200 Mt$_{\text{CO}_2}$/a, 90\% carbon capture rate.}
	\label{fig:balances-blueh2-noinvestment-cost--100-cc-h2}
\end{figure}
\begin{figure}
	\centering
	\includegraphics[width=0.9\linewidth]{"balances-blueH2-high store + 100p CC-H2"}
	\caption{Varying the investment costs of blue hydrogen production assuming high CO$_2$ sequestration potential of 2000 Mt$_{\text{CO}_2}$/a and 100\% carbon capture rate. This is compared to our base scenario (right panel) with investment costs of blue hydrogen production of 572 \kw$_{\text{H}_2}$, CO$_2$ sequestration potential of 200 Mt$_{\text{CO}_2}$/a, 90\% carbon capture rate of blue hydrogen production.}
	\label{fig:balances-blueh2-high-store--100-cc-h2}
\end{figure}

\FloatBarrier
\subsection{Investment costs of electrolysis}\label{sec:sensi_c0}
\begin{figure}
	\centering
	\includegraphics[width=0.7\linewidth]{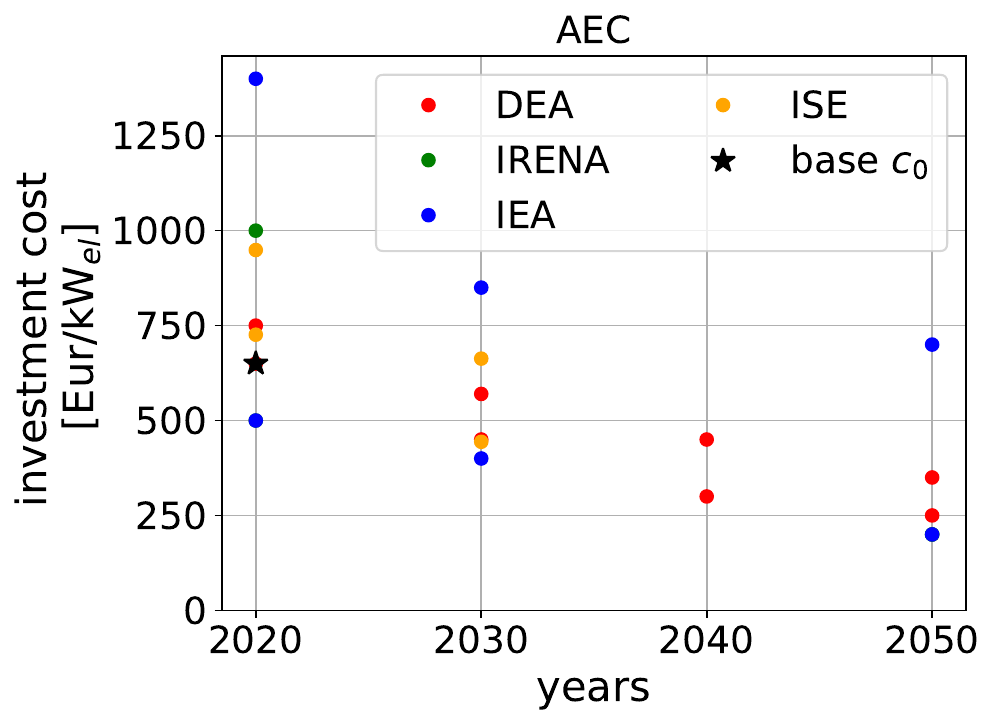}
	\caption{Investment costs of alkaline electrolysis (\gls{aec}) from Danish Energy Agency (\gls{dea}) \cite{cost_dea}, International Renewable Energy Agency (\gls{irena}) \cite{irena2020}, International Energy Agency (\gls{iea}) \cite{IEA2019}, and Fraunhofer Institute for Solar Energy Systems (\gls{ise}) \cite{ise_fraunhofer_cost}. If the source contains small and large system sizes both values are shown. Our base initial cost assumptions $c_0=$ 650\kwe \ are from \gls{dea} for a 100 MW plant in the year 2020.}
	\label{fig:c0aec}
\end{figure}

\begin{figure}
	\centering
	\begin{subfigure}{0.45\textwidth}
		\includegraphics[width=1\linewidth]{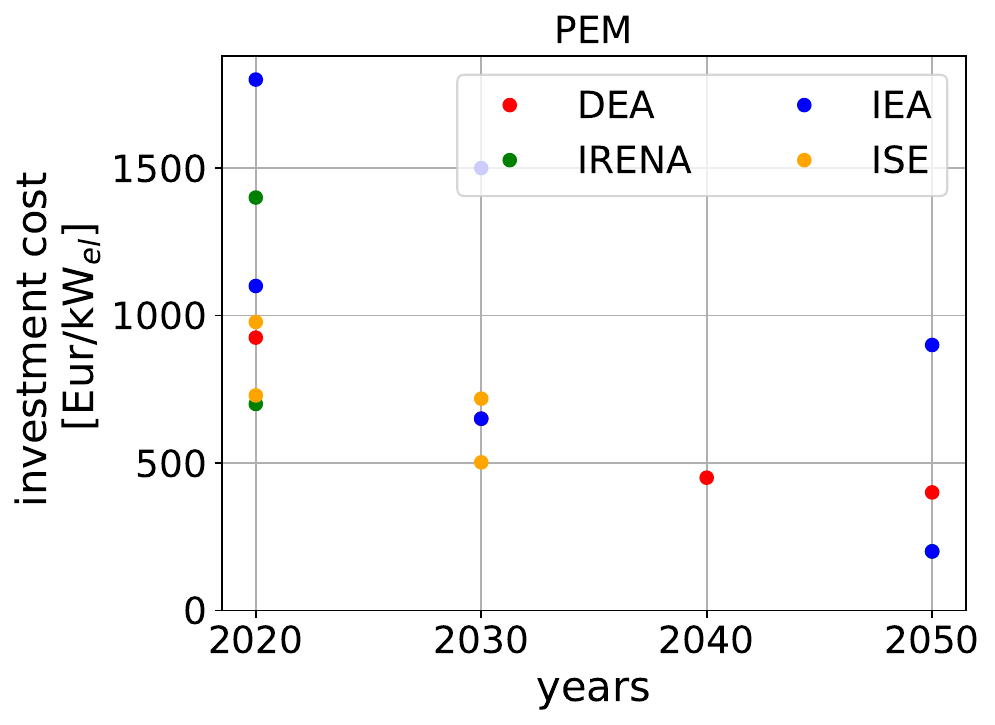}
		\subcaption{\gls{pem}}
		\label{fig:c0pem}
	\end{subfigure}
	\begin{subfigure}{0.45\textwidth}
		\includegraphics[width=1.\linewidth]{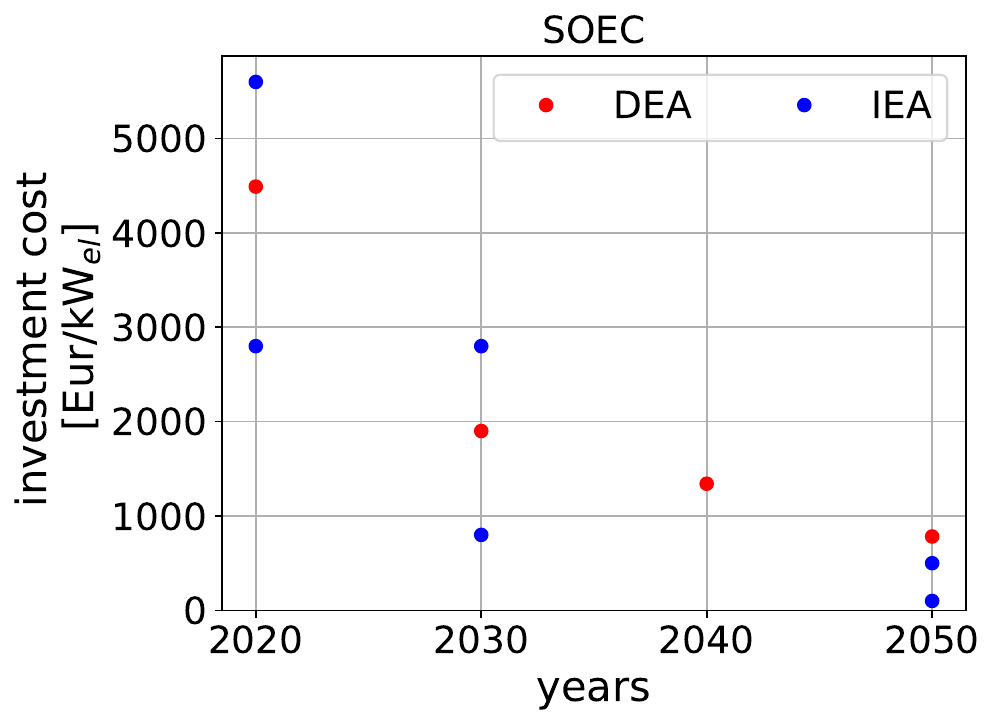}
		\subcaption{\gls{soec}}
		\label{fig:c0soec}
	\end{subfigure}	
	\caption{Investment costs of \gls{pem} and \gls{soec} from Danish Energy Agency (\gls{dea}) \cite{cost_dea}, International Renewable Energy Agency (\gls{irena}) \cite{irena2020}, International Energy Agency (\gls{iea}) \cite{IEA2019}, and Fraunhofer Institute for Solar Energy Systems (\gls{ise}) \cite{ise_fraunhofer_cost}.}
	\label{fig:c0_other}
\end{figure}

\begin{figure}
	\centering
	\includegraphics[width=0.7\linewidth]{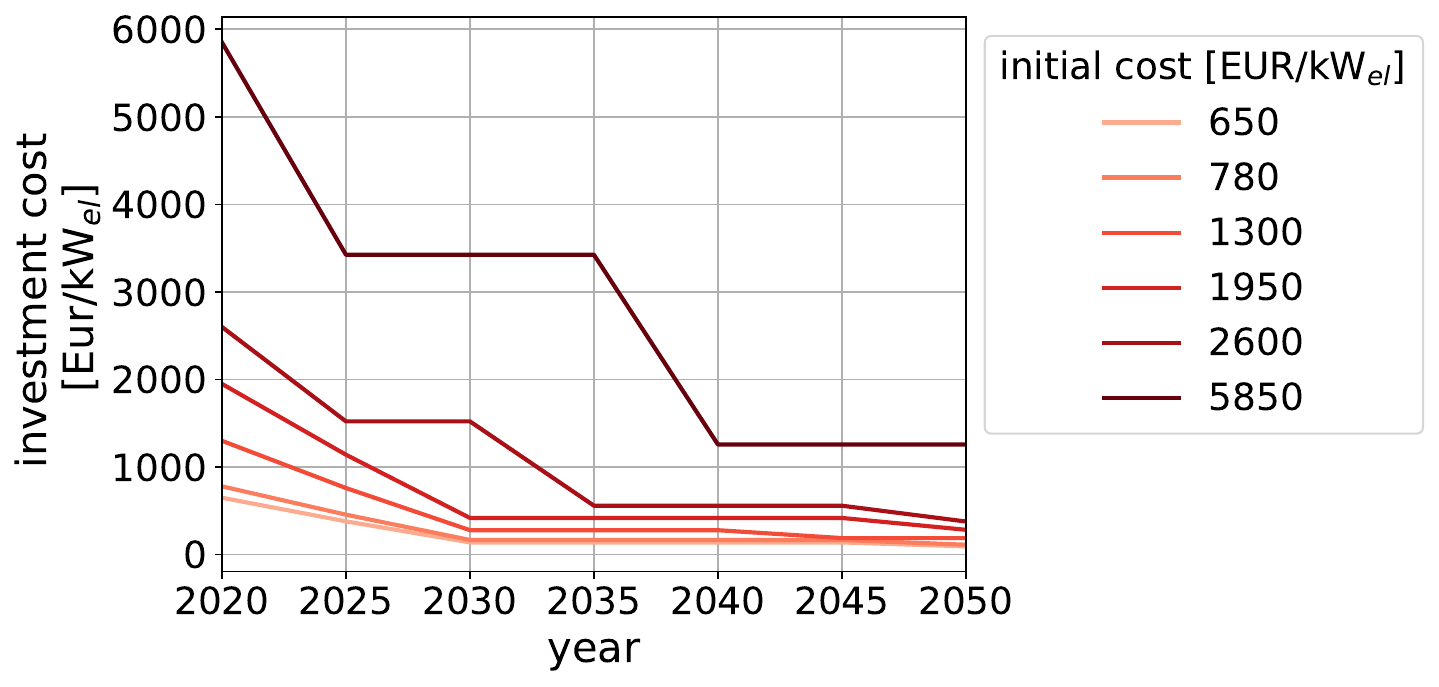}
	\caption{Investment costs for the endogenous method with base learning rate of 16\% and varying initial cost assumptions for scenarios with a \ce{1.7} budget.}
	\label{fig:c0sensi1p7}
\end{figure}

\begin{figure}
	\centering
	\includegraphics[width=1.\linewidth]{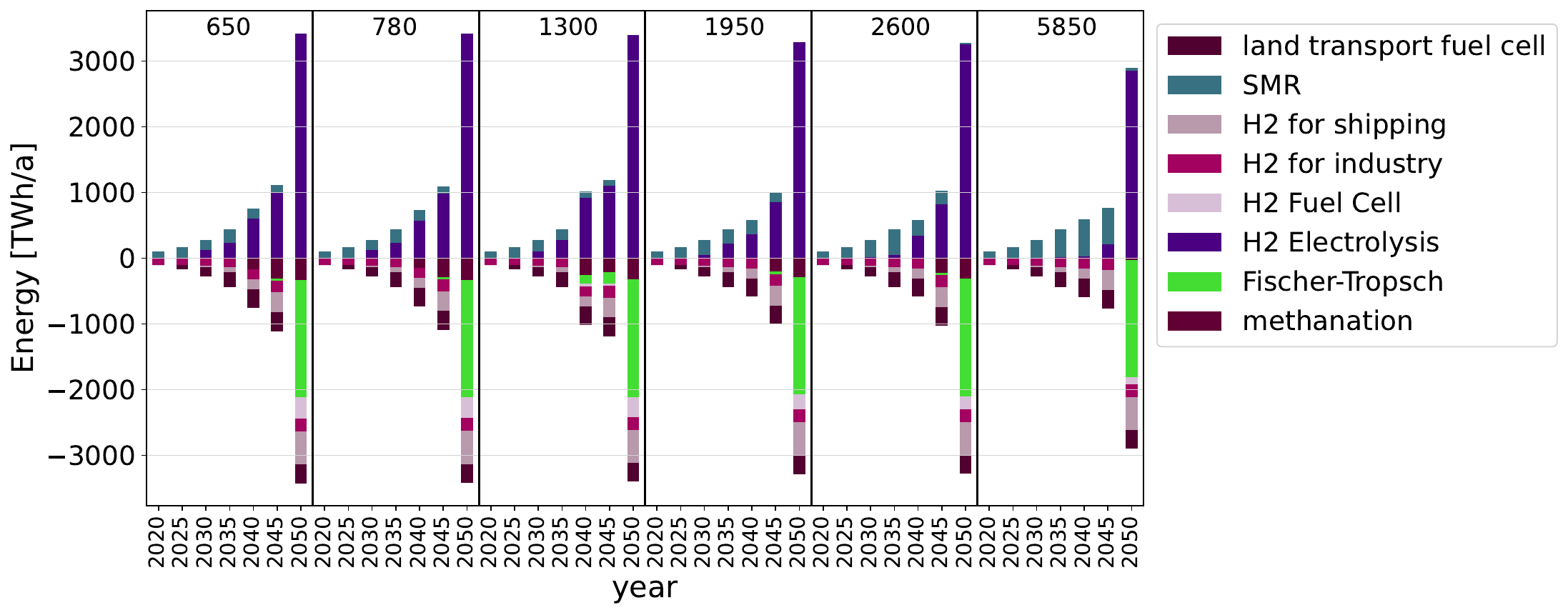}
	\caption{Hydrogen balance for different initial investment costs (given in \kwe in the plot) of electrolysis.}
	\label{fig:eb-c0sensi1p7}
\end{figure}

The technology assumptions of electrolysis are taken from the Danish Energy Agency (\gls{dea}) \cite{cost_dea} for a 100 MW alkaline electrolysis (\gls{aec}) plant. These costs, also called engineering, procurement and construction (\gls{epc}) price, consist of the equipment and installation costs. They include the expenses for stack, power electronics, gas conditioning, balancing of the plant and labour. The assumptions are for an electrolysis which produces hydrogen with a pressure of 35 bar and an additional waste heat steam of 50$^\text{o}$C from electricity (400 V$_{\text{AC}}$) and purified water. The annual operational and maintenance costs are estimated in \cite{cost_dea} based on current projects to be 2\% of the investment costs. The cost of replacing the stack is not included in the fixed operational and maintenance cost (\gls{fom}) since it is assumed that the stack does not need to be replaced within the technical lifetime (lifetime of \gls{aec} stack is assumed to be more than 100 000 hours, electrolysis with 4000 full load hours and a lifetime between 25-35 years). The investment cost do not include costs for water purification, transformer costs or  connection fees to the transmission system operators. \\
\\We are taken most of our technology assumptions of the \gls{dea}'s technology catalogue since firstly it contains a large number of technologies and thus avoids picking particularly optimistic or particularly pessimistic assumptions for individual technologies, and secondly, this data is constantly updated. There are currently a few electrolysis units with capacities above 100 MW that are already operational and several projects are expected to be connected to the grid in the next two years \cite{electrolyser_china, electrolyser_dk}. \\
\\ In the following scenarios, we vary the assumptions about the initial investment costs firstly to reflect uncertainties regarding these (see Figure \ref{fig:c0aec} for cost assumptions from different sources) and secondly to represent other types of electrolysis such as polymer electrolyte membrane (\gls{pem}) and solid oxide electrolyser cells (\gls{soec}) with current higher investment costs (see Figure \ref{fig:c0_other}). We assume the same learning rate of 18\% for all following scenarios. The learning rates of \gls{pem} and \gls{soec} are subject to higher uncertainty, as these technologies are not yet as mature as \gls{aec}. The learning rates for \gls{pem} vary between 13-30\% \cite{cost_dea, reksten2022, schmidt2017, irena2020} and for \gls{soec} between 0-44\% \cite{wei2017, boehm2019, tinoco2012}. Higher learning rates than those assumed here result in a steeper learning curve and would thus reduce the costs for \gls{pem} and \gls{soec} more quickly. This effect is not analysed in the following sensitivity analysis. The effects of unit-scaling also differ between the electrolysis types. \gls{soec} investment costs depend more on the system size compared to \gls{aec} or \gls{pem} \cite{boehm2020}. This is due to the fact that a smaller share of the total costs depends on the stack, which has a low cost impact on the system size due to its modularity. We use the endogenous method to represent the learning an scenarios corresponding to a \budget{1.7}. Our base initial investment cost $c_0$ of 650 Eur/kW$_{\text{el}}$ for \gls{aec} in 2020 according to \gls{dea} are increased by factors of 1.2 to 9 times. \\
\\Electrolysis is installed even with very high investment costs of 5850\kwe and the investment costs decrease in all scenarios over the modelling horizon (see Figure \ref{fig:c0sensi1p7}). Initial costs of 780-1300\kwee, corresponding for example to uncertainties in cost assumptions of \gls{aec} or low to medium cost estimates for \gls{pem}, lead to investment costs of electrolysis in the range of 114-189\kwe in 2050 and comparable volumes of hydrogen demand and supply (see Figure \ref{fig:eb-c0sensi1p7}). Cost assumptions in the range of 1950-2600\kwee, corresponding to high cost assumptions of \gls{pem} or low cost estimates for \gls{soec}, lead to investment costs of 284-379\kwe in 2050 and increased use of grey hydrogen in the period 2020-2045. High investment costs of 5850\kwe (corresponding to the highest cost assumptions in the datasets we discussed for \gls{soec}), decrease hydrogen production by 15\% in 2050 and 73\% of the hydrogen is produced via grey hydrogen in 2045. Investment cost of electrolysis are 1257\kwe in 2050.
\FloatBarrier
\subsection{Spatial resolution}\label{sec:sensi_spatial}
\begin{figure}
	\centering
	\begin{subfigure}{0.45\textwidth}
		\includegraphics[width=1\linewidth]{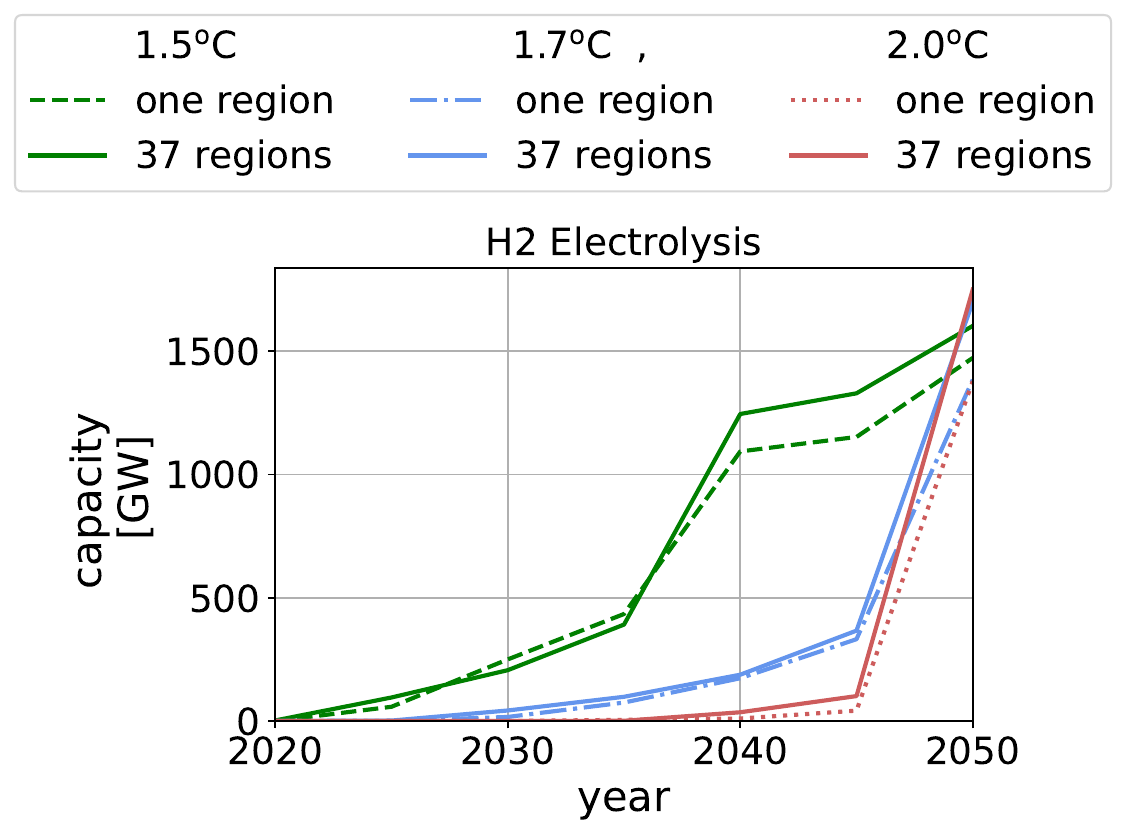}
		\subcaption{Electrolysis}
		\label{fig:spatial_electrolysis}
	\end{subfigure}
	\begin{subfigure}{0.45\textwidth}
		\includegraphics[width=1.\linewidth]{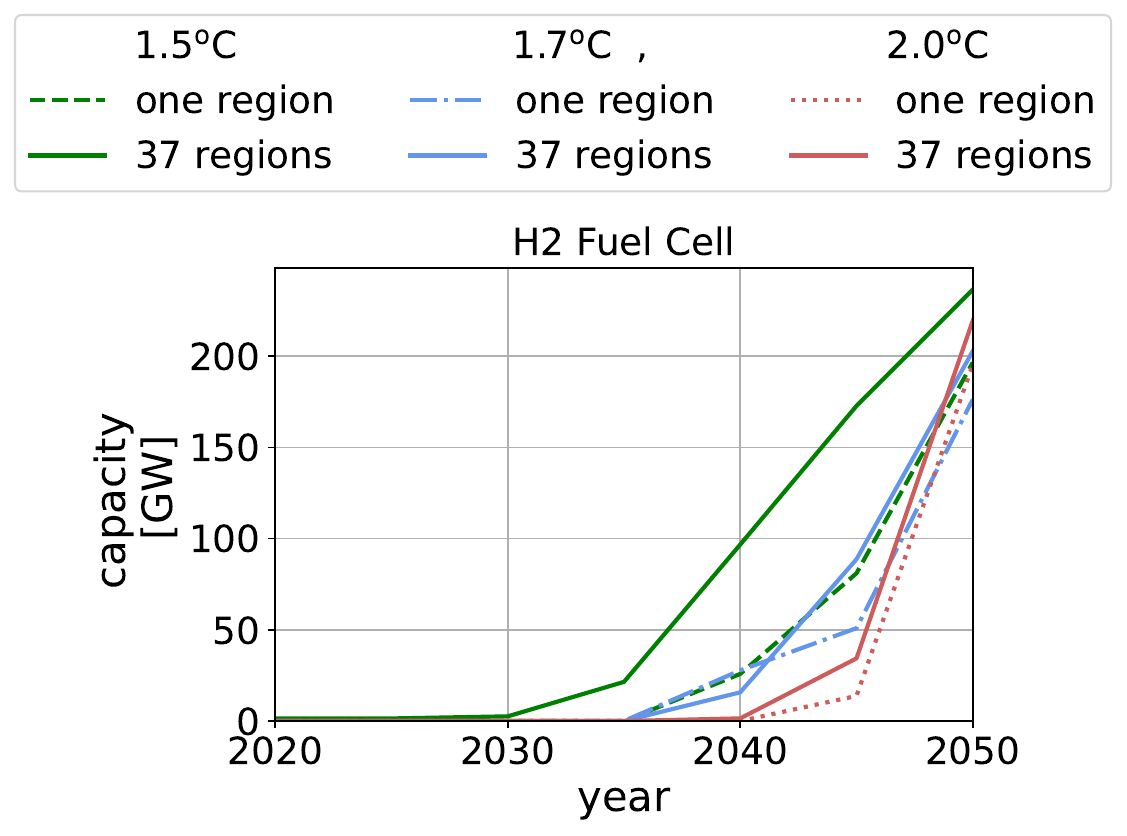}
		\subcaption{Fuel cell}
		\label{fig:spatail_fuel_cell}
	\end{subfigure}	
	\caption{Installed capacities of electrolysis and fuel cells for the exogenous method, base learning rate assumptions, comparing scenarios without any modelled grid infrastructure with scenarios with a spatial resolution of 37 regions.}
	\label{fig:spatial_h2}
\end{figure}

\begin{figure}
	\includegraphics[width=1.\linewidth]{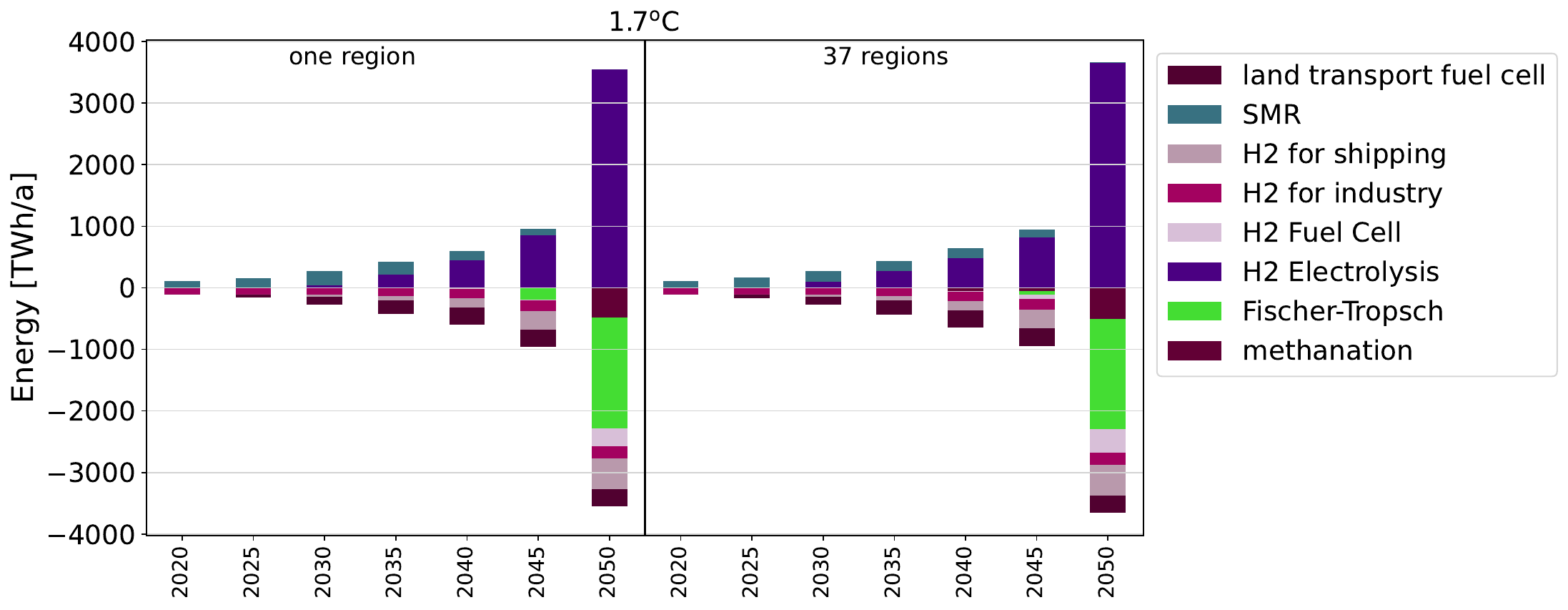}
	\caption{Energy balance hydrogen for a \ce{1.7} scenario, comparing a scenario without any modelled grid infrastructure with a scenario with a spatial resolution of 37 regions.}
	\label{fig:eb_spatial_H2_1p7}
\end{figure}

\begin{figure}
	\centering
	\includegraphics[width=1\linewidth]{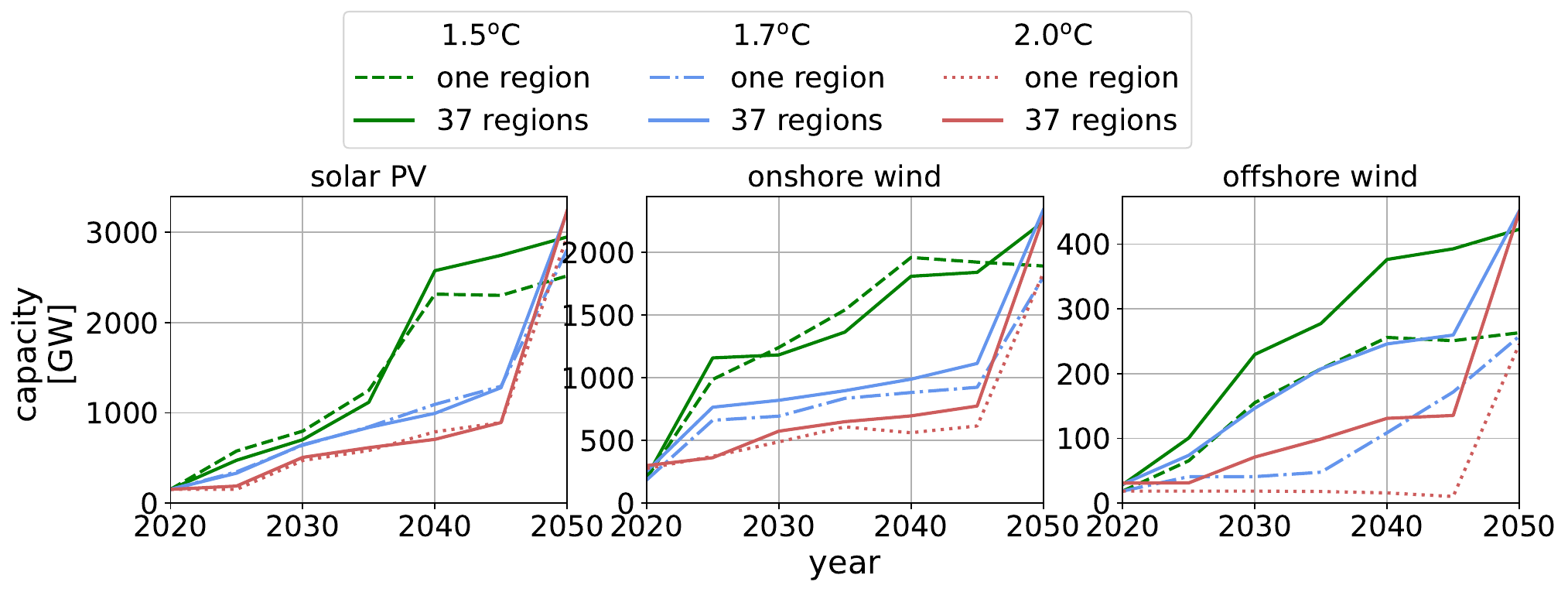}
	\caption{Installed capacities of renewable generation for the exogenous method, base learning rate assumptions, comparing scenarios without any modelled grid infrastructure with scenarios with a spatial resolution of 37 regions.}
	\label{fig:spatial_RES}
\end{figure}

\begin{figure}
	\centering
	\includegraphics[width=1\linewidth]{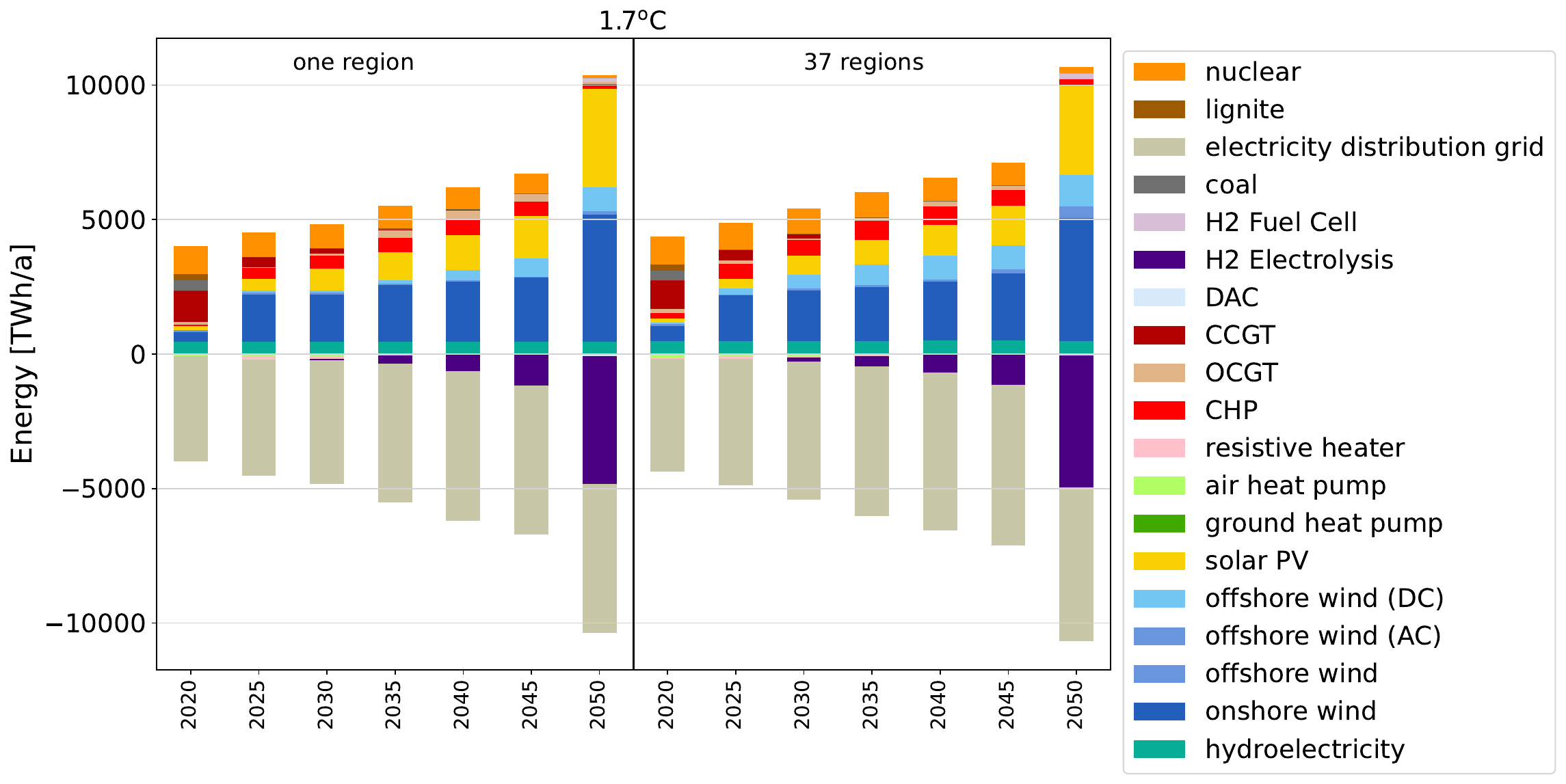}
	\caption{Energy balance electricity (transmission level) of the \ce{1.7} scenario, comparing a scenario without any modelled grid infrastructure with a scenario with a spatial resolution of 37 regions.}
	\label{fig:eb_spatial_AC_1p7}
\end{figure}

\begin{figure}
	\centering
		\includegraphics[width=0.7\linewidth]{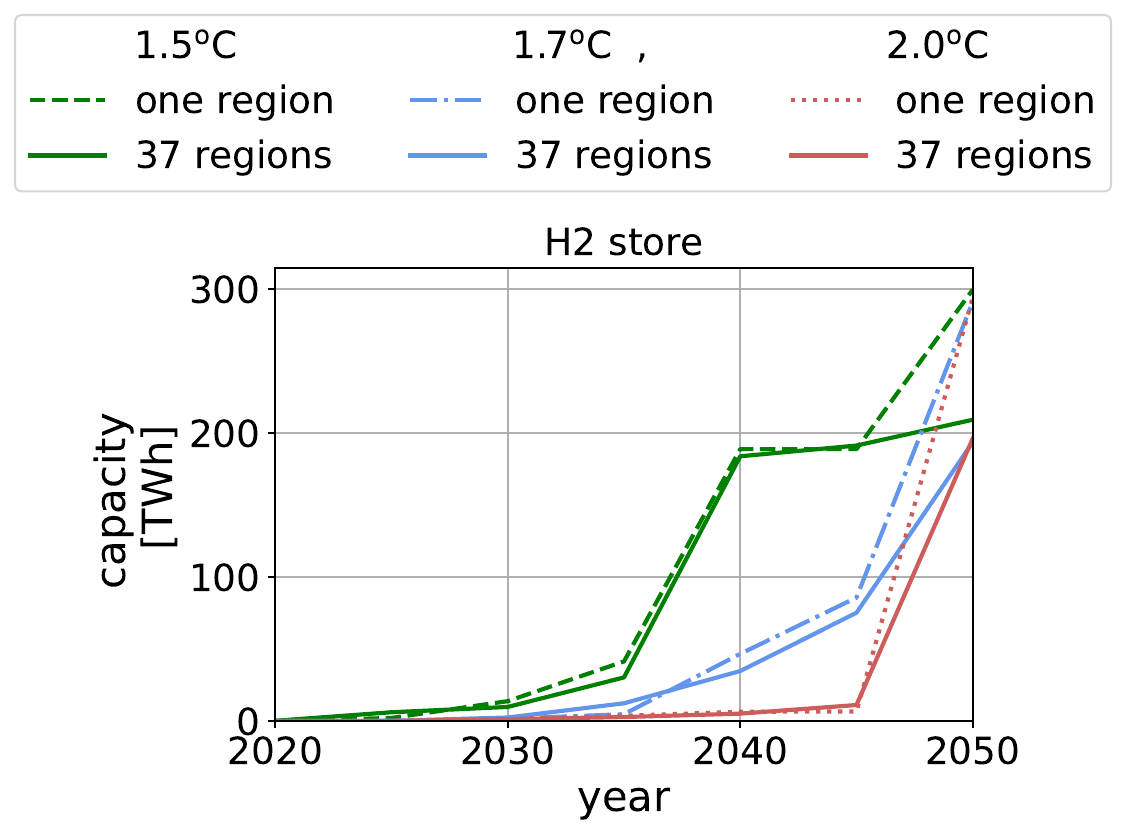}
	\caption{Installed capacities of hydrogen storage for the exogenous method, base learning rate assumptions, comparing scenarios without any modelled grid infrastructure with scenarios with a spatial resolution of 37 regions.}
	\label{fig:spatial_store}
\end{figure}

\begin{figure}
	\centering
	\includegraphics[width=1\linewidth]{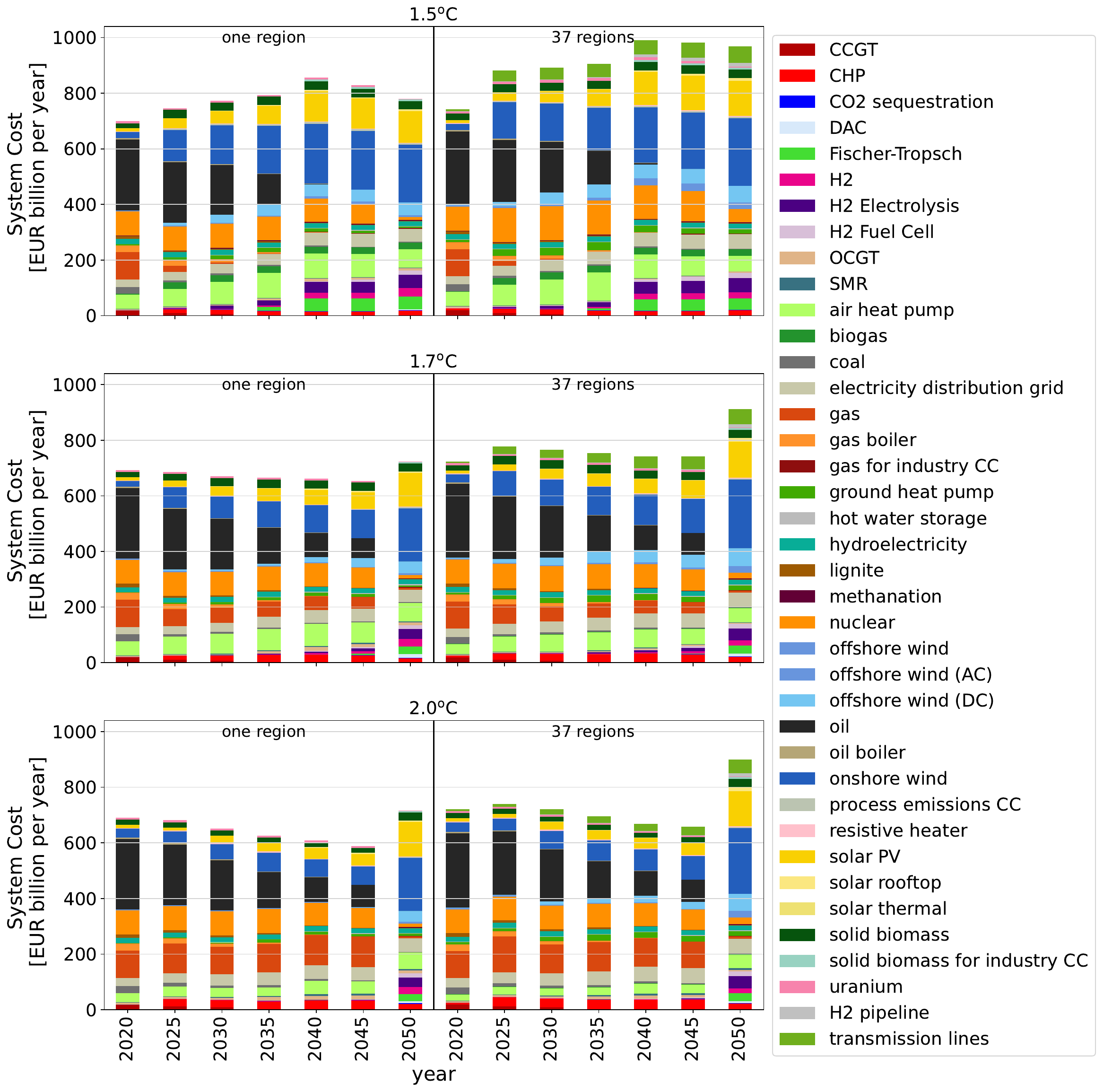}
	\caption{Annualised system costs for the three different budgets, comparing a scenario without any modelled grid infrastructure with a scenario with a spatial resolution of 37 regions.}
	\label{fig:costsspatial}
\end{figure}

In order to reduce the computational complexity with endogenous learning as a mixed integer problem (\gls{mip}), no grid infrastructure is represented in the main results and the renewable generation is modelled in six typical regions. This simplification can lead, on the one hand, to potential grid bottlenecks not being represented, and on the other hand to an underestimation of renewable generation by clustering different capacity factors of various regions \cite{Frysztacki2021}. In order to examine the impact of these assumptions on our results, scenarios with a higher spatial resolution of 37 regions are compared with the presented main scenarios of one region for the exogenous method. Current capacities for the electricity grid are assumed, which can be further expanded. The trade-offs between electricity and hydrogen grid are examined in more detail and with a higher spatial resolution in another publication by Neumann et al. \cite{neumann2022}.\\
\\ With a higher spatial resolution, the capacities of electrolysis  increase by 9-27\% in 2050 from the tight \ce{1.5} to the \ce{2.0} budget and the production of hydrogen rises by up to 2\% in 2050 compared to the scenarios without grid infrastructure (see Figures \ref{fig:spatial_electrolysis}, \ref{fig:eb_spatial_H2_1p7}). The larger volume of hydrogen is used for re-electrification in fuel cells (see Figure \ref{fig:spatail_fuel_cell}). The endogenous or sequential method in which investment costs depend or are adjusted based on the installed capacity would result in lower investment costs of electrolysis at higher spatial resolution. \\
\\  Overall the electricity produced by renewables generation in 2050 reduces by 2-3\% from the \ce{2.0} to the \ce{1.5} scenario with a higher spatial resolution since grid bottlenecks limit the distribution of renewable generation. The combination of different capacity factors results in lower generation of solar PV and onshore wind generation in the scenario without grid infrastructure and a higher generation of offshore wind. Larger renewable generation capacities are built with a higher spatial resolution. Solar PV, onshore and offshore wind capacities increase with a higher spatial resolution by 1-12\%, 19-30\% and 61-83\%  respectively in 2050 (see Figures \ref{fig:spatial_RES}, \ref{fig:eb_spatial_AC_1p7}). More electricity is fed into the grid by nuclear power plants in 2050 (278-558 TWh with high spatial resolution compared to 115-120 TWh with low spatial resolution). This change in generation mix leads to lower hydrogen storage capacities with a higher spatial resolution (see Figure \ref{fig:spatial_store}). \\
\\ Total system cost over the whole modelling horizon increase by 12-16\% from the \ce{2.0} to the \ce{1.5} scenarios with a higher spatial resolution (see Figure \ref{fig:costsspatial}). The higher costs are caused primarily by higher costs for nuclear, renewable energies and costs for the electricity distribution and transmission network. Costs for the electricity and hydrogen grid contribute with a share of 0.1-7.6\% to the annualised total costs. The largest costs are in 2050 with a share of 7.6\% of the total costs for the \budget{1.5}. In the \ce{1.5} scenario cost of about 12 billion Euros per year account for the hydrogen network, 61 billion Euros per year for the electricity transmission grid.

\FloatBarrier
\section{Further analysis of main results}\label{sec:graphics_appendix}
\subsection{Investment cost and installed capacities of renewable generation}
\subsubsection{Composition of the investment costs}\label{sec:cost_res_composition}
In the following, we provide an overview of the composition of the investment costs for wind and solar. The exogenous cost assumptions are from the Danish Energy Agency technology data \gls{dea} \cite{cost_dea} which also provides a detailed description of each technology. The technology assumptions depend on the build year of the respective asset. External grid connection costs are added to these investment costs in the model that do not undergo any learning. For offshore wind grid connection costs depend on the location and connection type (AC or DC), for solar and onshore wind additional grid connection costs of 133\kw \ are added. \\
\\The investment costs of onshore wind consist of the costs for equipment (turbine, foundation, cables), installation and development, cost of land, internal grid connection,  decommissioning cost of existing turbines and other costs (for example compensation of neighbours living close to the wind park). The investment cost of offshore wind include cost for equipment (turbine, foundation, cables, grid connection), installation, project development and other costs (e.g. insurances, sea right fees, contingencies). Operating and Maintenance cost of wind farms include insurance, service agreement, repairs not covered by service agreement, land rent and administration. Solar \gls{PV} investment costs include costs for the equipment (\gls{PV} module, inverter, transformer, grid connection, balance of the plant), the installation and other costs (e.g. costs for permits, surveys, studies, planning, legal expenditures). Operational and maintenance cost of solar PV include insurance, land rent, cleaning of the modules, asset management and grass cutting.
\subsubsection{Investment cost and capacities of solar and wind in our main results}
\begin{figure*}[!ht]
	\centering
\includegraphics[width=1\linewidth]{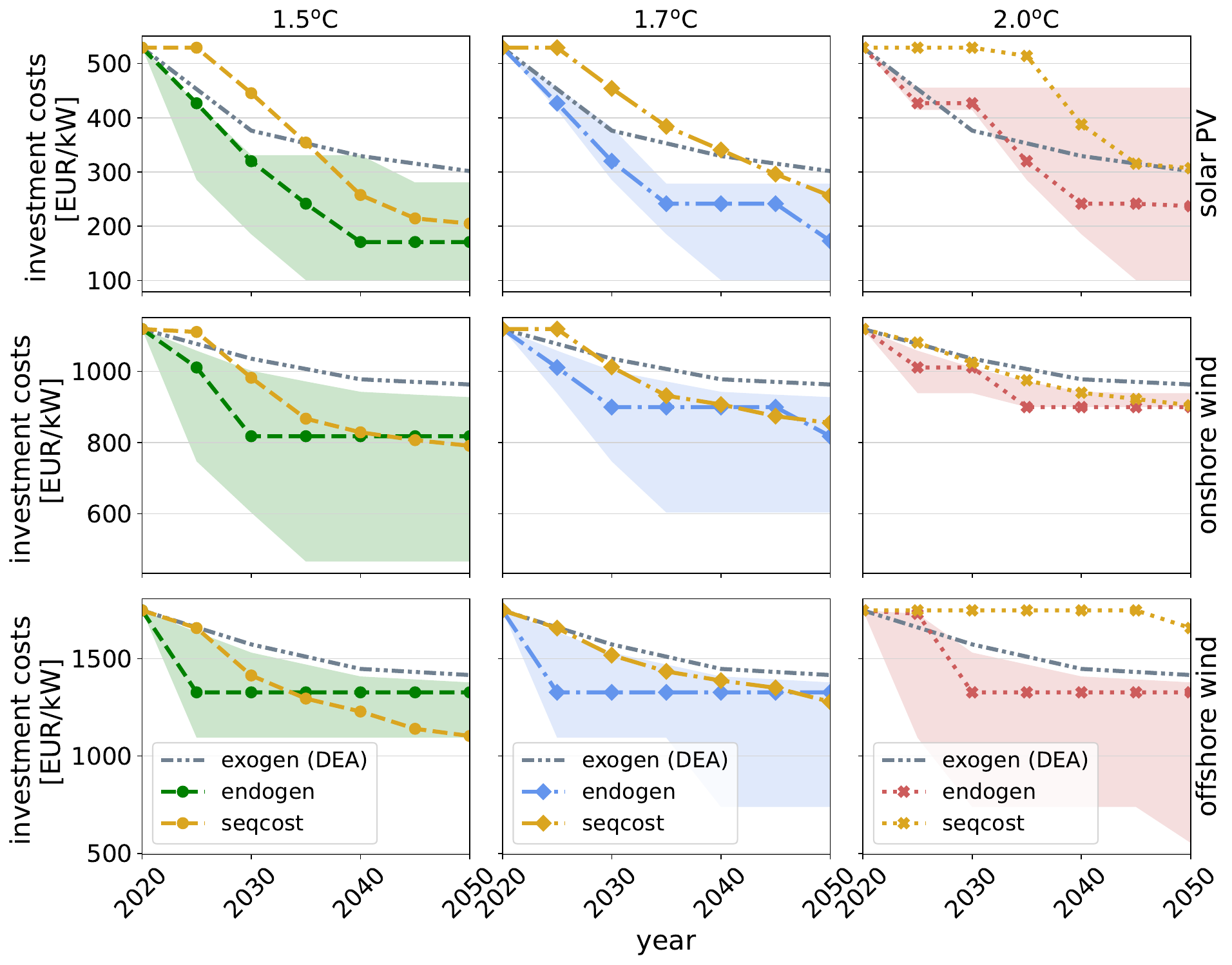}
	\caption{Investment costs for renewable generation (\textit{rows})  for different carbon budgets (\textit{columns}) and the three different methods (i) exogenous (\textit{exogen}), (ii) sequential cost (\textit{seqcost}) and (iii) endogenous (\textit{endogen}). Contour area indicates scenarios with $\pm 10\%$ variation of the learning rate for all technologies with endogenous learning. These investment costs are without grid connection costs which are added. For them no learning is assumed.}%
	\label{fig:investment_cost_renewables}
\end{figure*}

\begin{figure*}[!ht]
	\centering
	\includegraphics[width=1\linewidth]{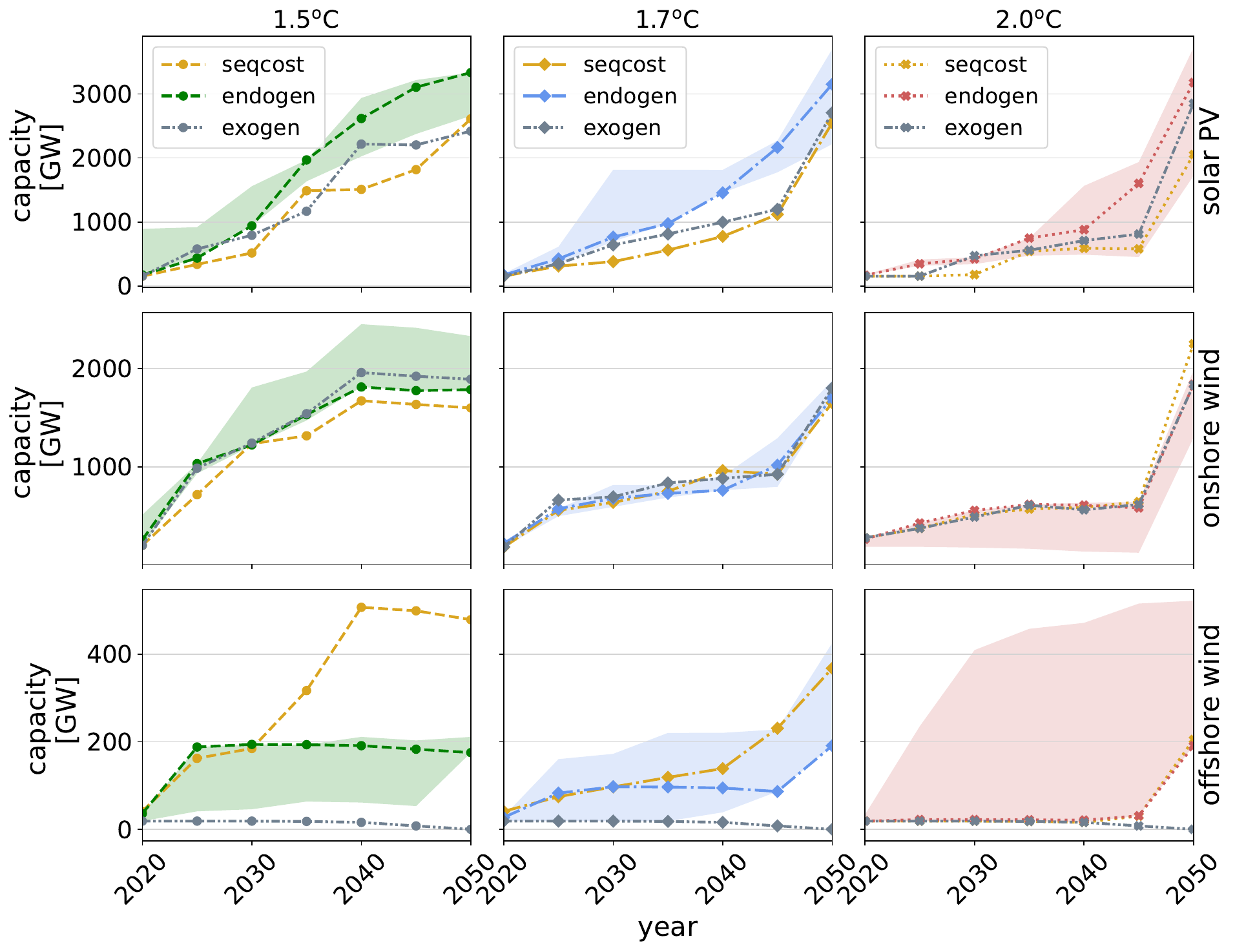}
	\caption{Installed capacities of renewable generation (\textit{rows}) for different CO$_2$ budgets (\textit{columns}) and the three different methods (i) exogenous (\textit{exogen}), (ii) sequential cost (\textit{seqcost}) and (iii) endogenous (\textit{endogen}). Contour area indicates scenarios with $\pm 10\%$ variation of the learning rate for all technologies with endogenous learning.  Trade off between offshore and onshore wind depending on the learning rate which results in a large contour area when varying learning rate. Values are also shown in Tables \ref{tab:capacities_1p5}, \ref{tab:capacities_1p7}, \ref{tab:capacities_2p0}.}
	\label{fig:capacities_renewables}
\end{figure*}

The investment costs for renewable generation depend on the installed capacities in the sequential and endogenous method. For both methods, the resulting investment costs for all three carriers (solar PV, on- and offsore wind) are below the exogenous cost assumptions of \gls{dea} if a stricter budget of \ce{1.5} to \ce{1.7} is assumed. This is caused since larger capacities are built than assumed for the exogenous cost projections. The actual course of cost reduction differs between the endogenous and sequential method. With the endogenous method, investment costs decrease in earlier investment periods compared to the sequential method (see Figure \ref{fig:investment_cost_renewables}) since capacities are scaled up faster (see Figure \ref{fig:capacities_renewables}). This is because the sequential method does not have the foresight of how far costs can decrease, but only updates the assumptions on investment costs after optimisation, depending on the installed capacities. In contrast, the endogenous method sees the potential cost reduction and decrease in overall costs through a rapid expansion of renewable capacities. In addition, the technology mix differs between the  methods. While the endogenous method expands solar PV more due to the anticipation of higher cost reductions from a higher learning rate, the sequential method expands more offshore wind. The exogenous method generally leads to lower renewable capacities compared to the other two methods, as the investment costs are higher.

\FloatBarrier
\clearpage
\subsection{Annualised total system cost}
\begin{figure}[H]
	\centering
	\includegraphics[width=0.9\linewidth]{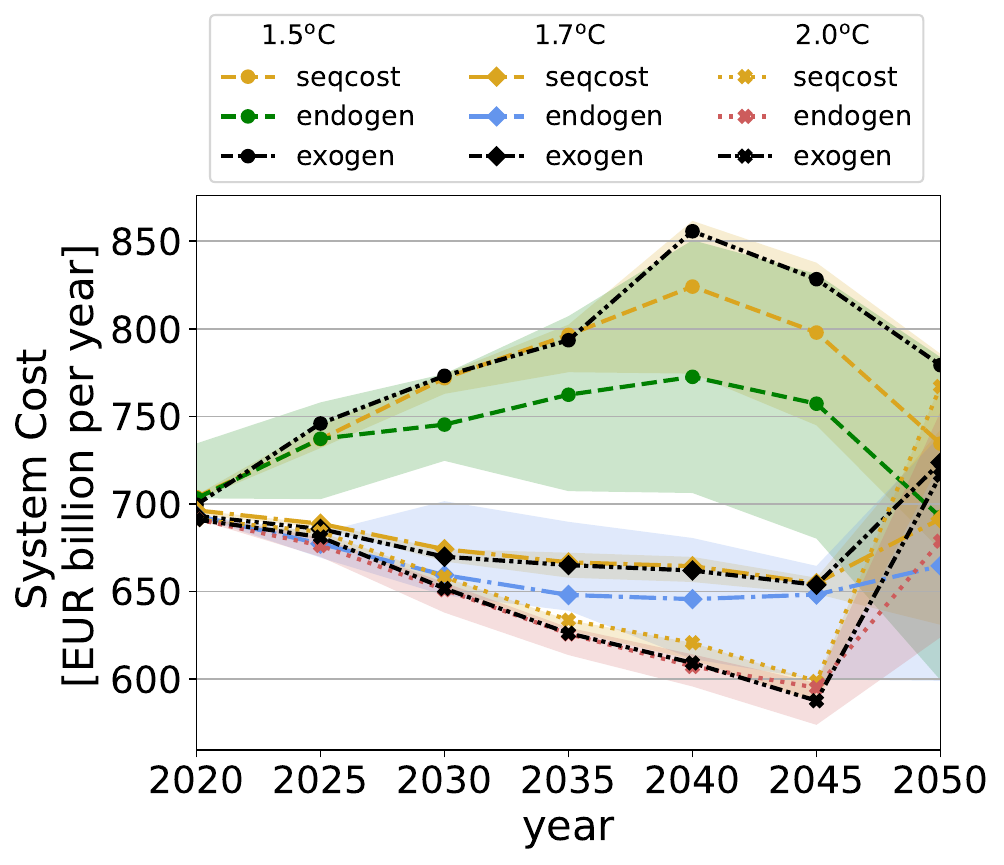}
	\caption{Annualised total system costs without estimated cost of climate change damage. Contour area shows scenarios in which the learning rate is varied by $\pm$10\%. If endogenous learning is considered, total system costs in 2050 are of comparable size for the different budgets.}
	\label{fig:costs}
\end{figure}

\begin{figure}[h]
	\centering
	\includegraphics[width=0.45\linewidth]{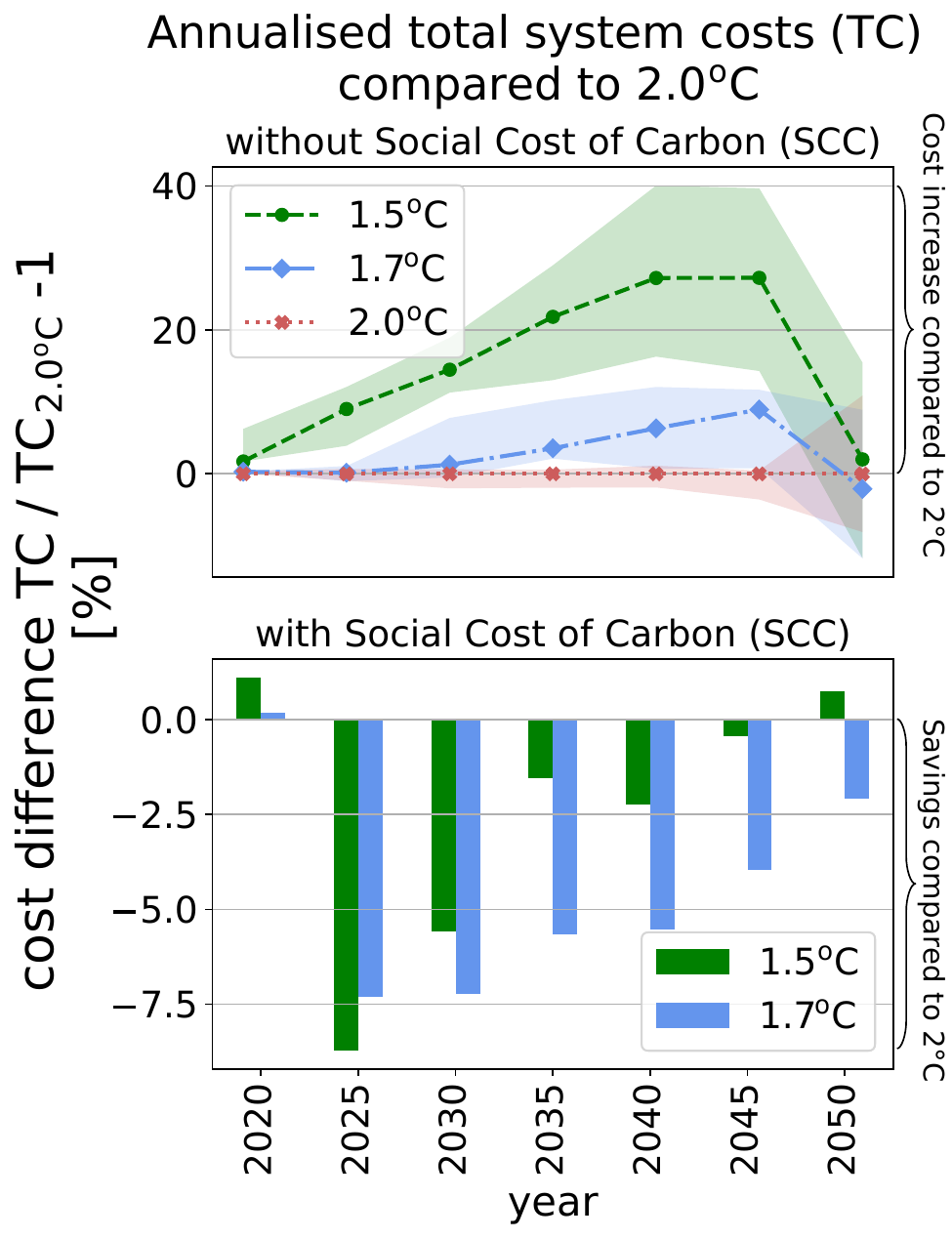}
	\caption{Total annualised system costs compared to the endogenous \budget{2} scenario with and without estimated costs of climate change damage with social cost of carbon (\gls{scc}) of 120\ \officialeuro \ per tonne CO$_2$.}
	\label{fig:total_cost_scc}
\end{figure}

\begin{figure}
	\centering
	\begin{subfigure}{0.45\textwidth}
		\includegraphics[width=1\linewidth]{figures/results/total_cost_method_bar_v2.pdf}
	\caption{base learning rate}
	\label{fig:cost_bar_methods1p5}
	\end{subfigure}
	\begin{subfigure}{0.45\textwidth}
	 \includegraphics[width=1\linewidth]{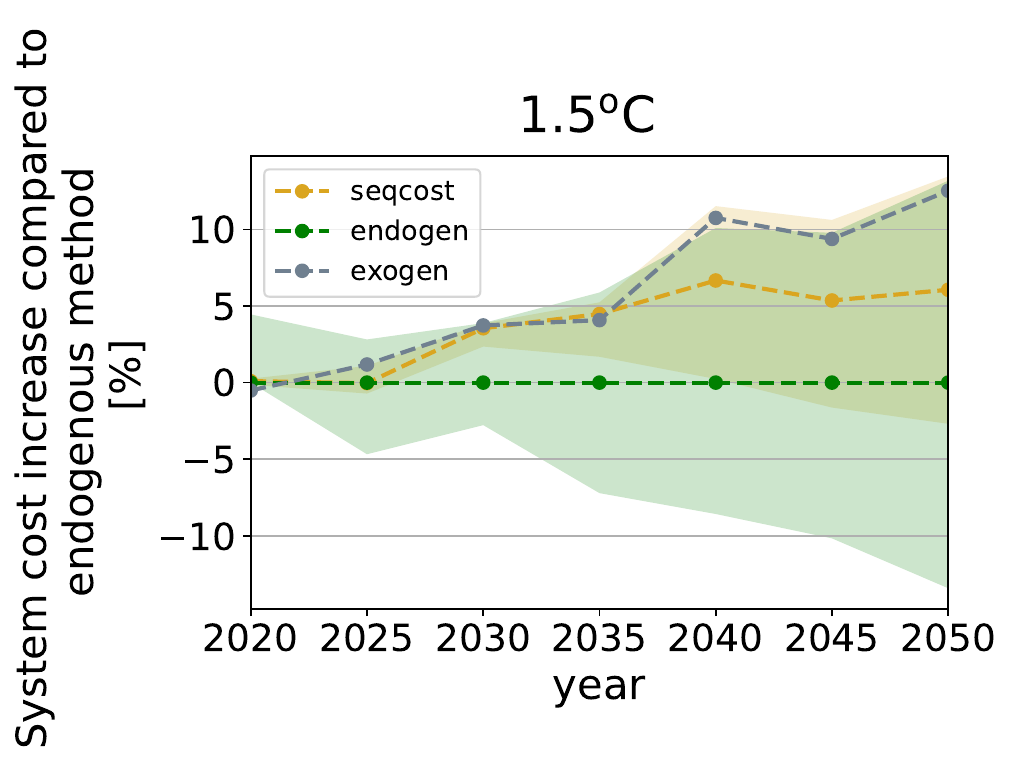} %
	\caption{with varied learning rate}%
	\label{fig:cap_cost_h2_var}
	\end{subfigure}
\caption{Difference in total system costs for \ce{1.5} scenarios compared to the endogenous method with base learning rates. Exogenous (\textit{exogen}) and sequential (\textit{seqcost}) method result in  higher total annualised system costs compared to the endogenous scenarios. The contour area shows scenarios which vary the base learning rate by $\pm$10\%.}
\label{fig:total_cost_methods_1p5}
\end{figure}

\begin{figure}
	\centering
	\begin{subfigure}{0.45\textwidth}
		\includegraphics[width=1\linewidth]{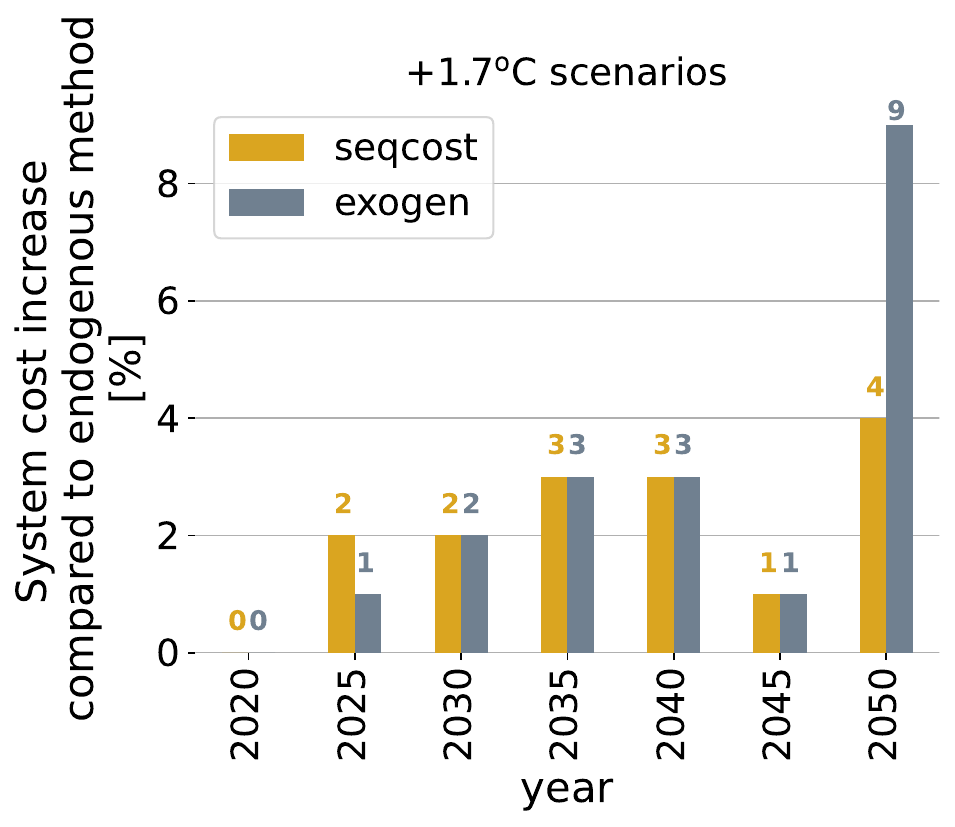}
		\subcaption{\ce{1.7}}
		\label{fig:cost_bar_methods1p7}
	\end{subfigure}
	\begin{subfigure}{0.45\textwidth}
		\includegraphics[width=1.\linewidth]{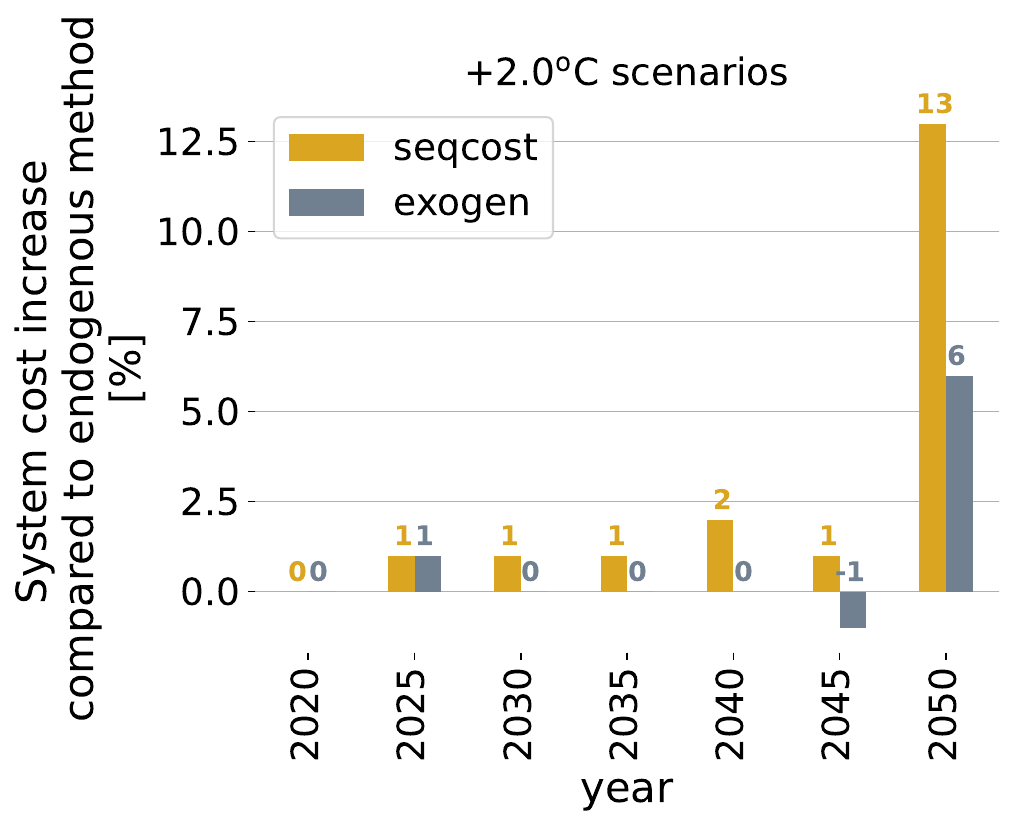}
		\subcaption{\ce{2.0}}
		\label{fig:cost_bar_methods2p0}
	\end{subfigure}	
	\caption{Difference in total system costs for \ce{1.7} and \ce{2.0} scenarios compared to the endogenous method with base learning rates. Exogenous (\textit{exogen}) and sequential (\textit{seqcost}) method result in  higher total annualised system costs compared to the endogenous scenarios.}
	\label{fig:cost_bar_methods}
\end{figure}

\begin{figure}
	\centering
	\begin{subfigure}{0.45\textwidth}
		\includegraphics[width=1\linewidth]{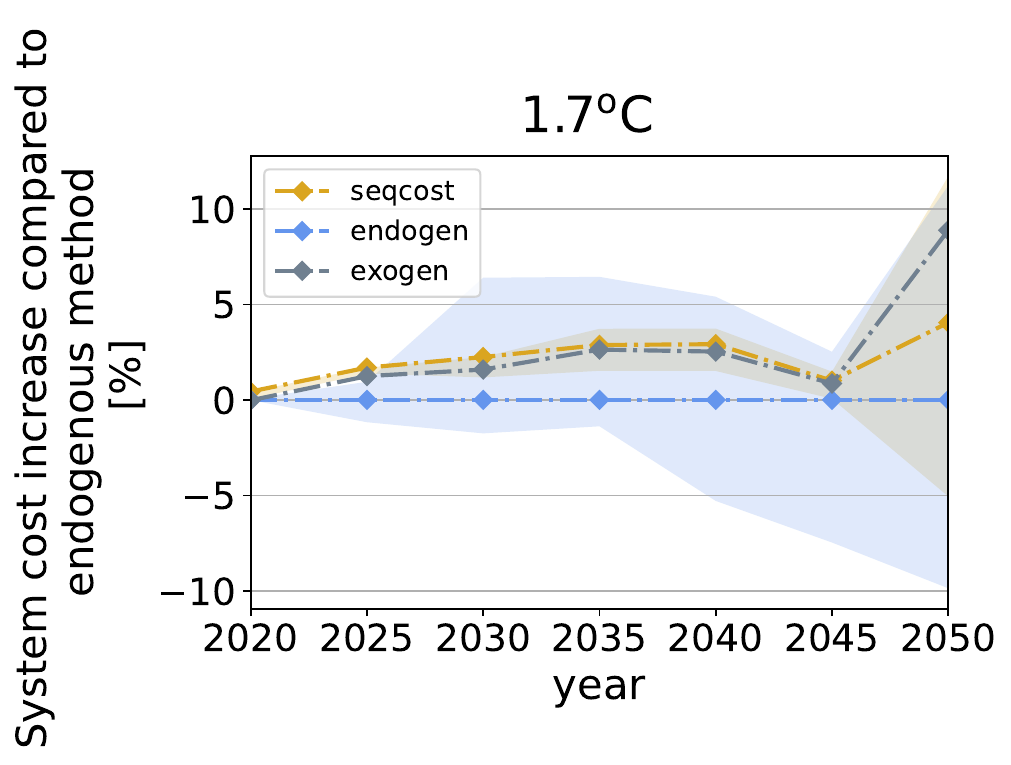}
		\subcaption{\ce{1.7}}
	\end{subfigure}
	\begin{subfigure}{0.45\textwidth}
		\includegraphics[width=1.\linewidth]{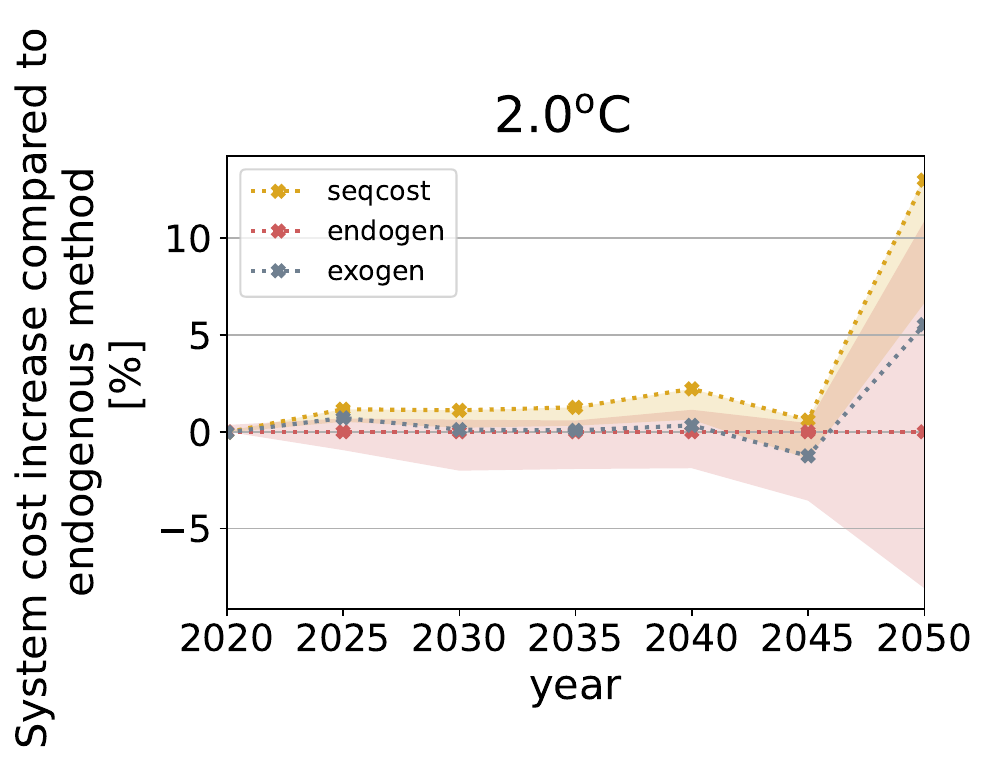}
		\subcaption{\ce{2.0}}
	\end{subfigure}	
	\caption{Difference in total system costs for \ce{1.7} and \ce{2.0} scenarios compared to the endogenous method. Contour area shows scenarios which vary the base learning rate by $\pm$10\%. Exogenous (\textit{exogen}) and sequential (\textit{seqcost}) method result in  higher total annualised system costs compared to the endogenous scenarios.}
	\label{fig:cost_bar_methods_withvar}
\end{figure}

Total annualised system costs without estimated costs of climate change damage for the three different budgets and three different methods are shown in Figure \ref{fig:costs}. The \budget{1.5} leads to higher annualised costs compared to the \ce{1.7} and \ce{2.} scenarios if estimated costs of climate change damage are not included. However, in 2050 the total annualised costs only vary between 665-692 billion Euros per year between the budgets in scenarios with the endogenous method (see also upper plot in Figure \ref{fig:total_cost_scc}). The endogenous method leads to lower total costs, especially with the \budget{1.5}, since cost reductions due to faster decarbonisation are better represented compared to the sequential and exogenous methods (see Figure \ref{fig:total_cost_methods_1p5}, Figure \ref{fig:cost_bar_methods} and Figure \ref{fig:cost_bar_methods_withvar}).
\FloatBarrier
\subsubsection{Annualised total system cost difference between the three budgets}
\FloatBarrier
\subsubsection{Annualised total system cost difference between the methods}\label{seq:cost_bar_diff}
\FloatBarrier
\subsection{Energy balances}\label{sec:balances}
\begin{figure*}[!ht]
	\includegraphics[width=1\textwidth]{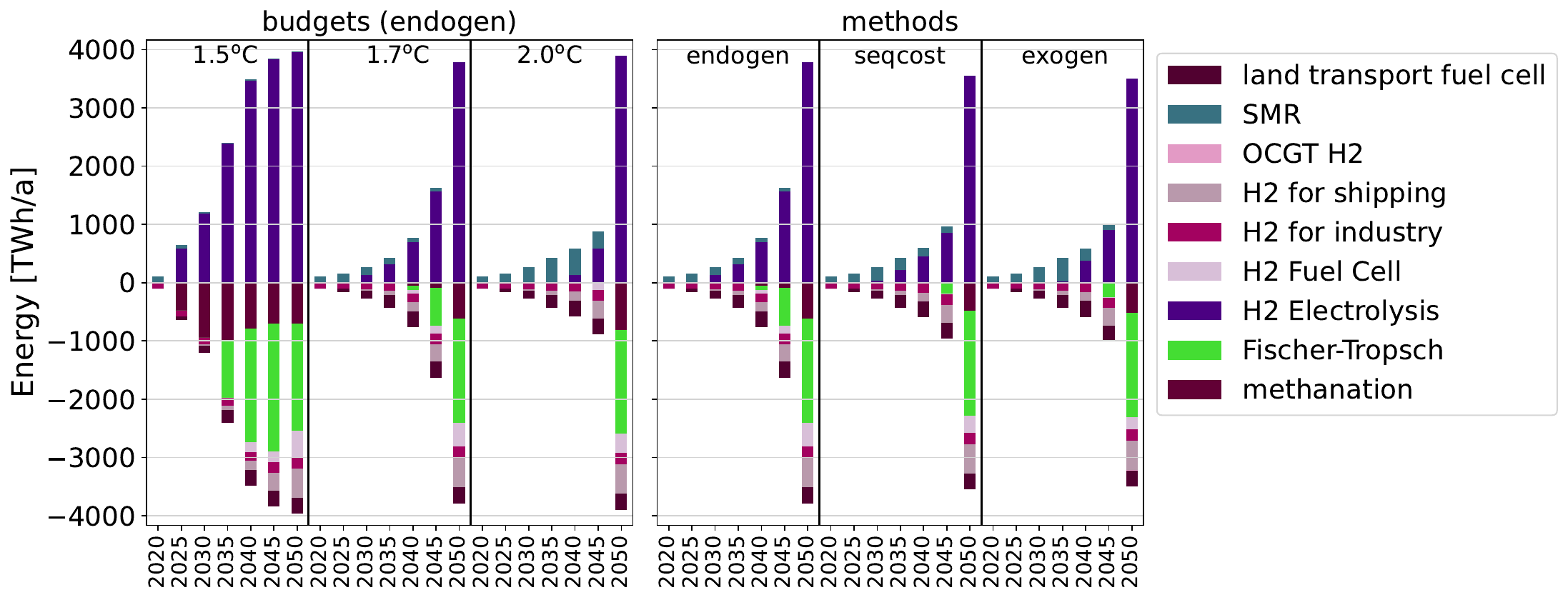}
	\caption{Energy balance for hydrogen. Supply site is positive, usage is negative. Left plot shows the three different budgets with the endogenous method, right plot the three different methods endogenous (\textit{endogen}), sequential cost (\textit{seqcost}) and exogenous (\textit{exogen}). Values are also displayed in Tables \ref{tab:energy_balance_H2_endogen}, \ref{tab:energy_balance_H2_method}.}
	\label{fig:eb_H2}
\end{figure*}

In the following, we show the energy balances for all considered energy carriers (e.g. hydrogen). We compare the three different budgets with the endogenous method and base learning rates, as well as the three different methods to model technology-learning (endogenous, sequential cost and exogenous) for the \ce{1.7} scenarios. 
\paragraph{Hydrogen}
The stricter the CO$_2$ budget, the more hydrogen is used, especially in the period between 2025-2045 (see Figure \ref{fig:eb_H2}). In scenarios with a \budget{1.5}, most of the hydrogen is already produced as green hydrogen in 2025. With a \budget{2.0}, production switches from grey to green hydrogen later in 2040. The option of producing blue hydrogen is not used in any of the scenarios. In all scenarios, hydrogen is used in 2050 for the production of synthetic fuels and feedstocks, for methanation and in the transport sector. A small part is converted back into electricity in fuel cells. However, the production of synthetic fuels and the methanation of hydrogen starts earlier in scenarios with a \ce{1.5} budget, from 2030, with a \ce{1.7} budget from 2045 and with a \budget{2.0} from 2050. \\
\\ With the endogenous method, the production changes from grey to green hydrogen at an earlier point in time compared to the sequential and exogenous method. Thus, with the endogenous method, the majority of hydrogen is already produced green in 2035, while with the sequential and exogenous method, the majority of hydrogen is still produced via \gls{SMR}. The volume produced also differs between the methods. With the endogenous method, 1600 TWh$_{\text{H}_2}$ of hydrogen are already produced in 2045, while with the sequential and exogenous method only 1000 TWh$_{\text{H}_2}$ are produced. The larger volume of hydrogen in scenarios with the endogenous method is mainly used for the production of synthetic fuels.

\begin{figure*}[!ht]
	\includegraphics[width=1\textwidth]{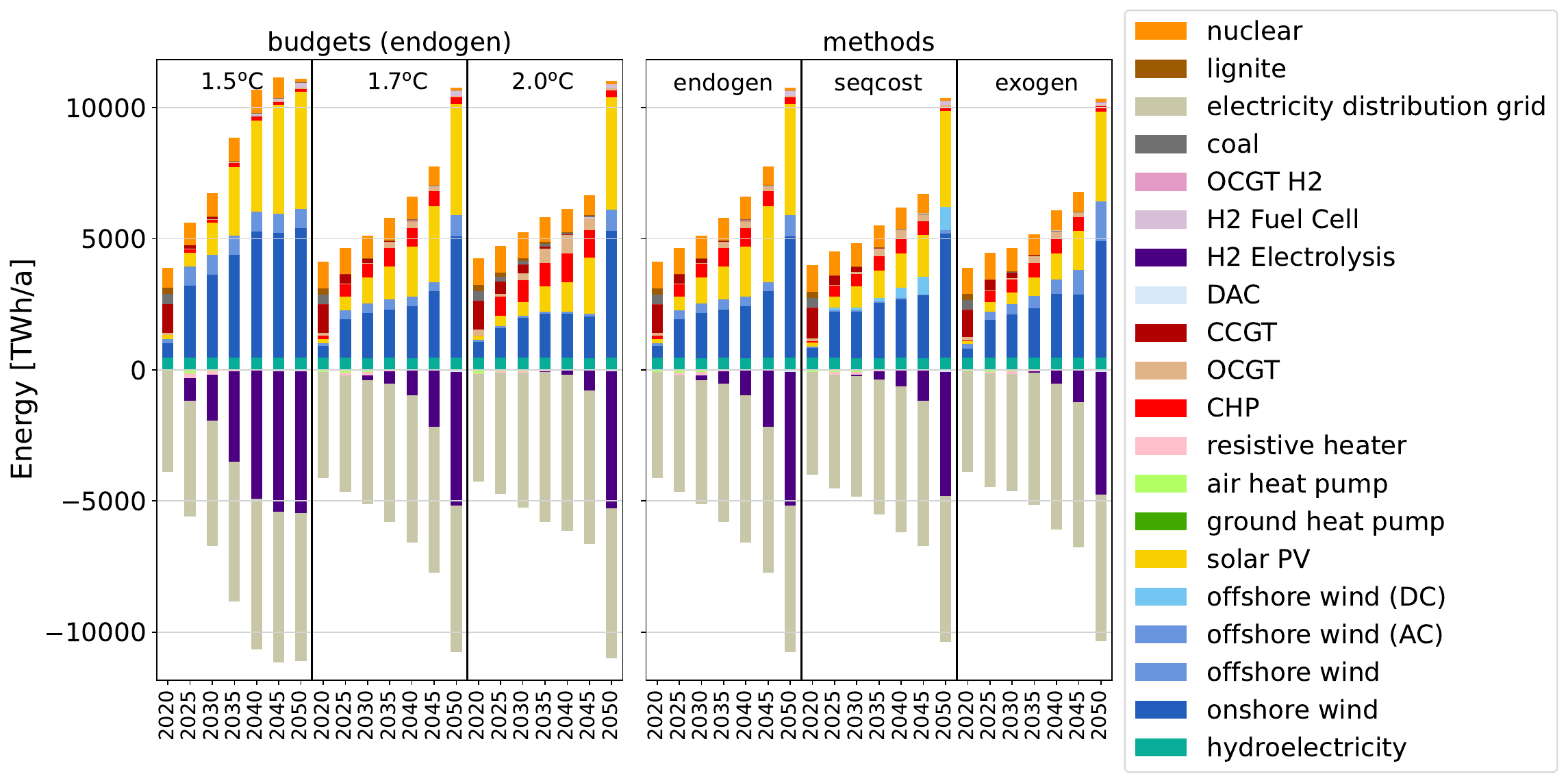}
	\caption{Energy balance for electricity (transmission level). Supply site is positive, usage is negative. Left plot shows the three different budgets with the endogenous method, right plot the three different methods endogenous (\textit{endogen}), sequential cost (\textit{seqcost}) and exogenous (\textit{exogen}). Values are also displayed in Tables \ref{tab:energy_balance_AC_endogen}, \ref{tab:energy_balance_AC_method}.}
	\label{fig:eb_AC}
\end{figure*}

\begin{figure*}[!ht]
	\includegraphics[width=1\textwidth]{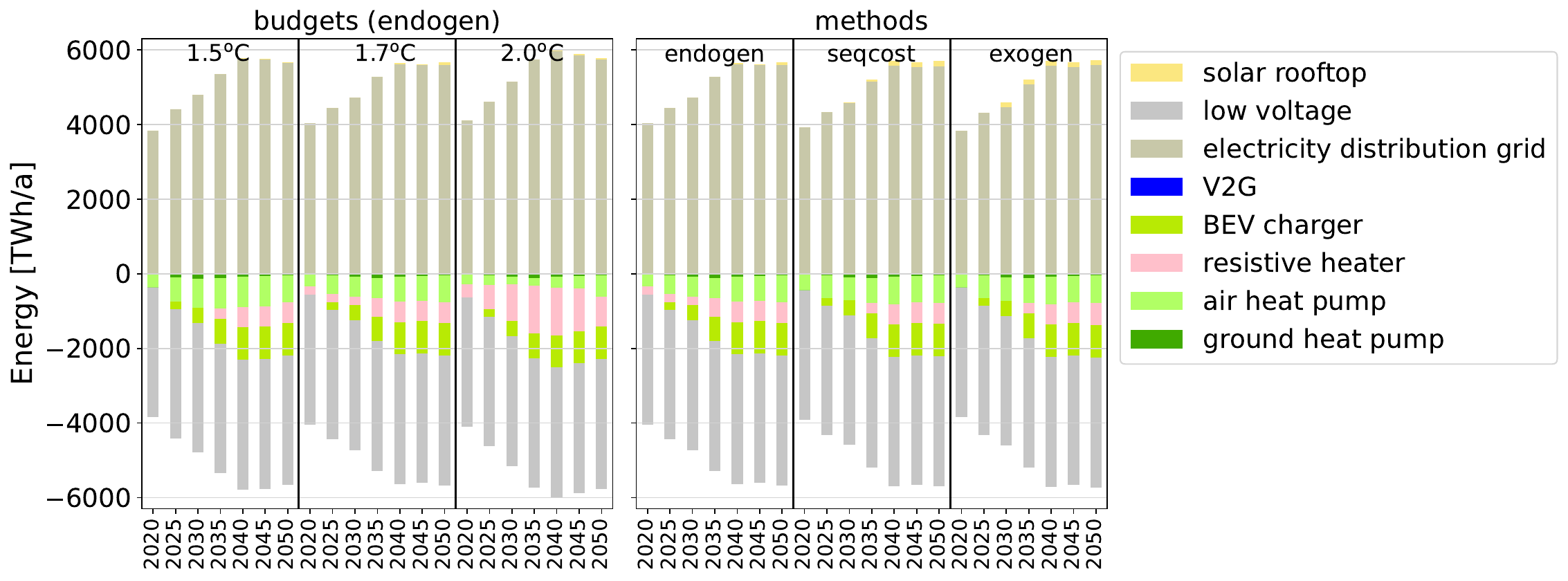}
	\caption{Energy balance for electricity (distribution level). Supply site is positive, usage is negative. Left plot shows the three different budgets with the endogenous method, right plot the three different methods endogenous (\textit{endogen}), sequential cost (\textit{seqcost}) and exogenous (\textit{exogen}). Values are also displayed in Tables \ref{tab:energy_balance_low_voltage_endogen}, \ref{tab:energy_balance_low_voltage_method}.}
	\label{fig:eb_low_voltage}
\end{figure*}

\begin{figure*}[!ht]
	\includegraphics[width=1\textwidth]{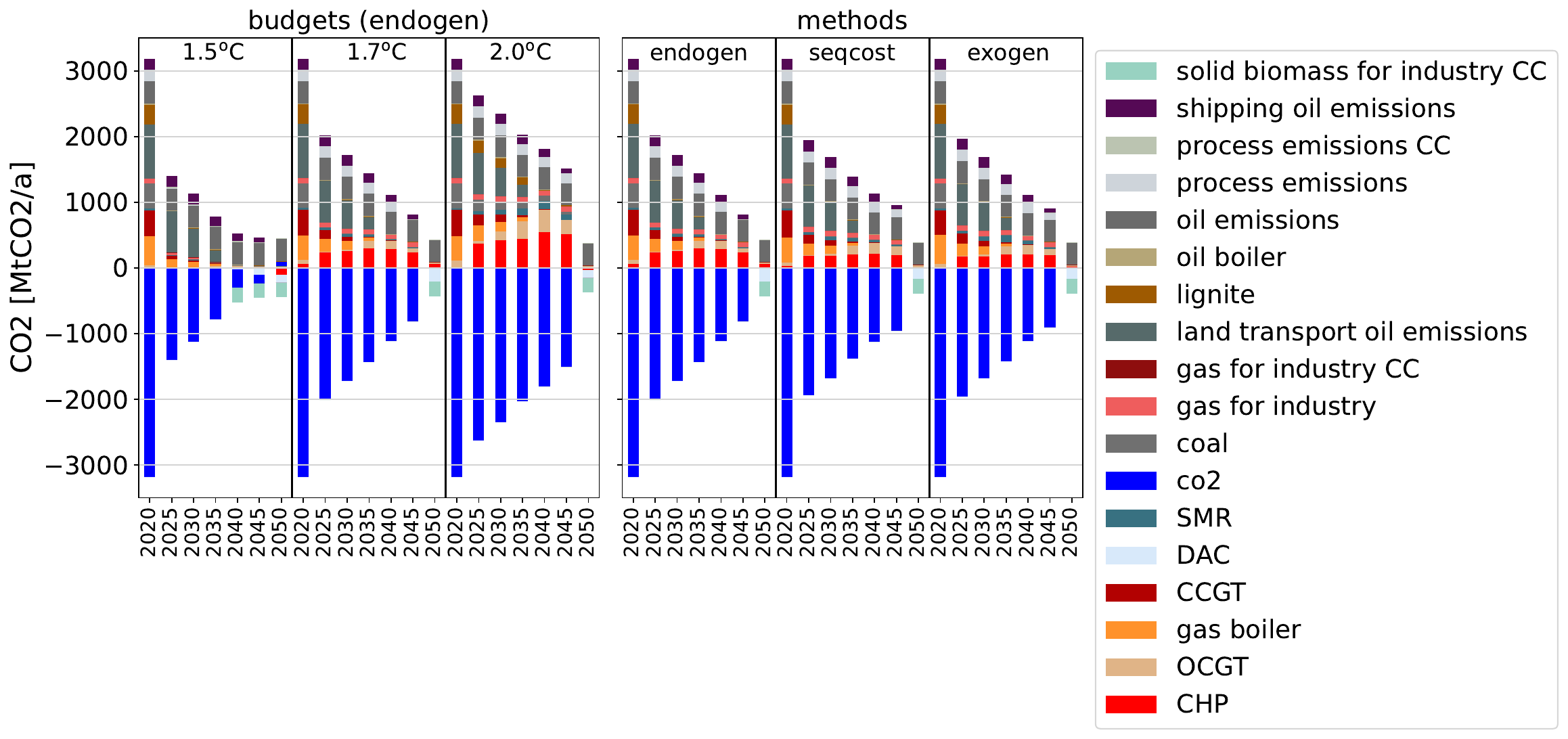}
	\caption{Energy balance for carbon dioxide. Supply site is positive, usage is negative. Left plot shows the three different budgets with the endogenous method, right plot the three different methods endogenous (\textit{endogen}), sequential cost (\textit{seqcost}) and exogenous (\textit{exogen}). Values are also displayed in Tables \ref{tab:energy_balance_co2_endogen}, \ref{tab:energy_balance_co2_method}.}
	\label{fig:eb_co2}
\end{figure*}

\begin{figure*}[!ht]
	\includegraphics[width=1\textwidth]{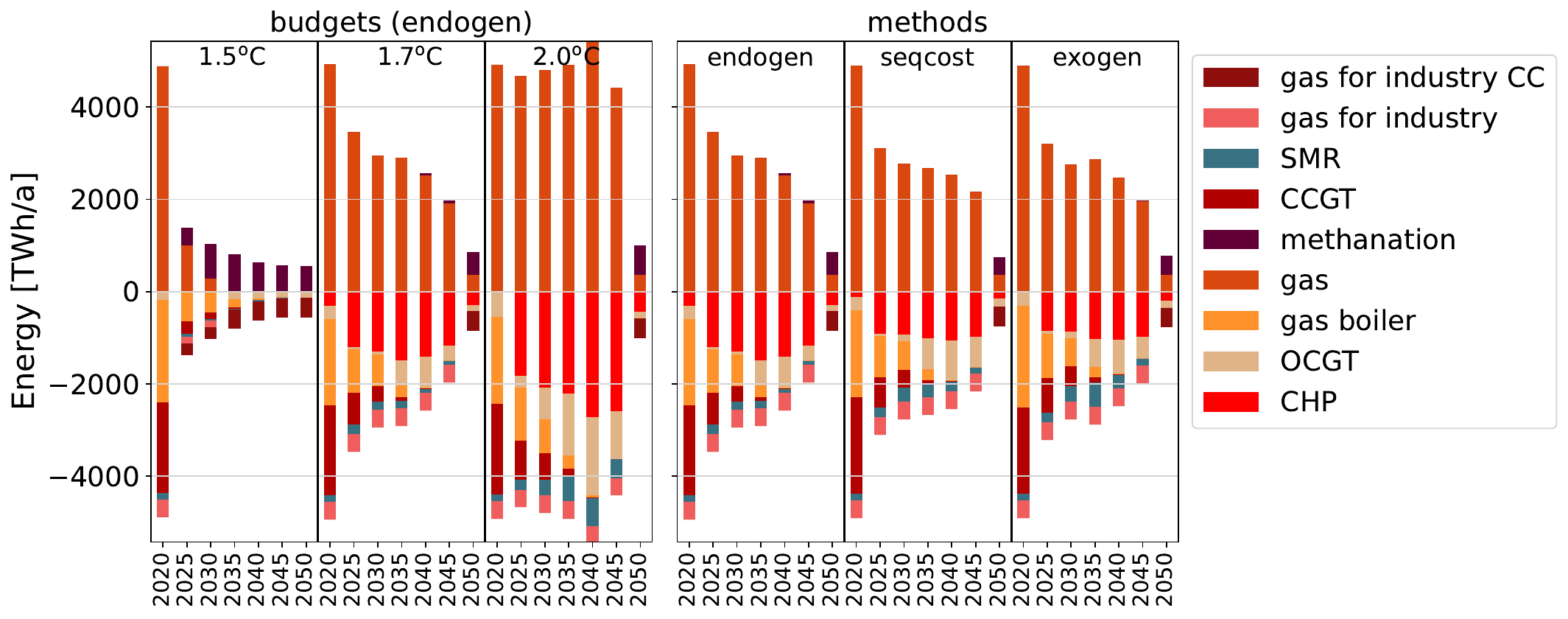}
	\caption{Energy balance for methane. Supply site is positive, usage is negative. Left plot shows the three different budgets with the endogenous method, right plot the three different methods endogenous (\textit{endogen}), sequential cost (\textit{seqcost}) and exogenous (\textit{exogen}). Values are also displayed in Tables \ref{tab:energy_balance_gas_endogen}, \ref{tab:energy_balance_gas_method}.}
	\label{fig:eb_gas}
\end{figure*}

\begin{figure*}[!ht]
	\includegraphics[width=1\textwidth]{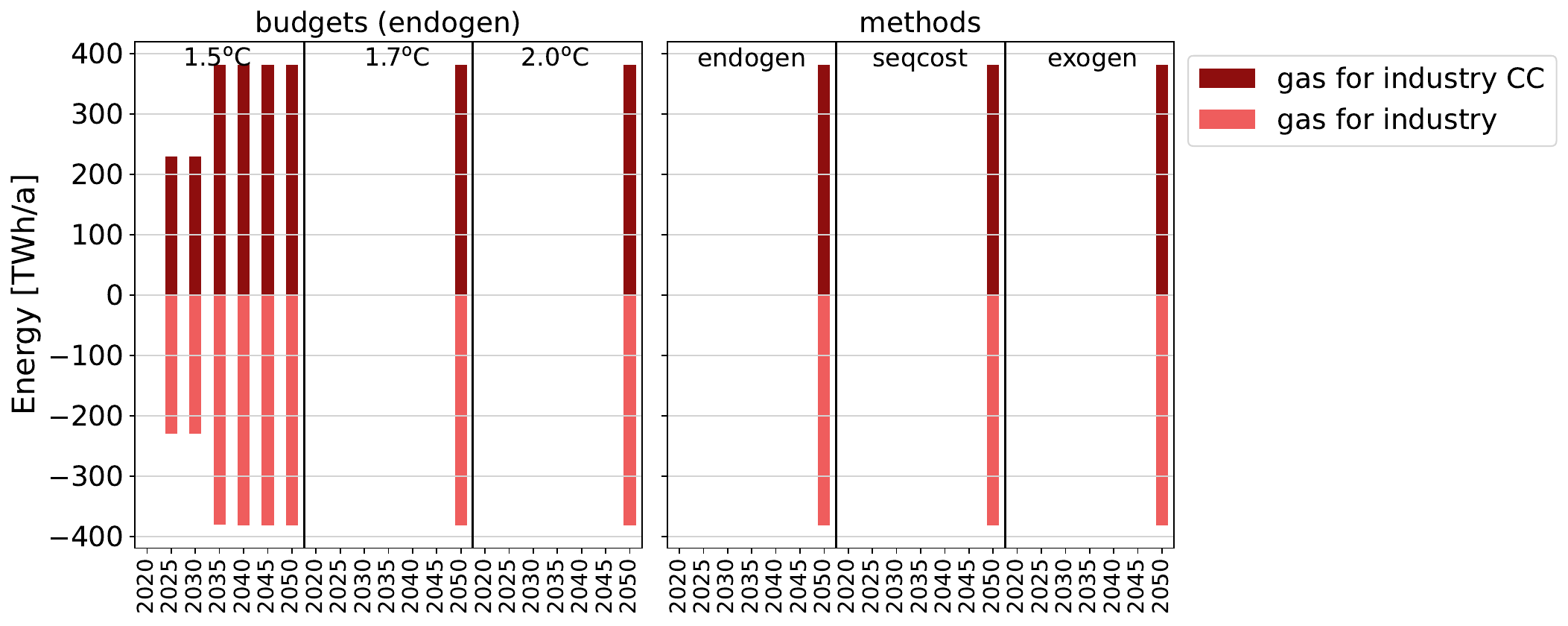}
	\caption{Energy balance for methane for industry. Supply site is positive, usage is negative. Left plot shows the three different budgets with the endogenous method, right plot the three different methods endogenous (\textit{endogen}), sequential cost (\textit{seqcost}) and exogenous (\textit{exogen}). Values are also displayed in Tables \ref{tab:energy_balance_gas_for_industry_endogen}, \ref{tab:energy_balance_gas_for_industry_method}.}
	\label{fig:eb_gas_for_industry}
\end{figure*}

\begin{figure*}[!ht]
	\includegraphics[width=1\textwidth]{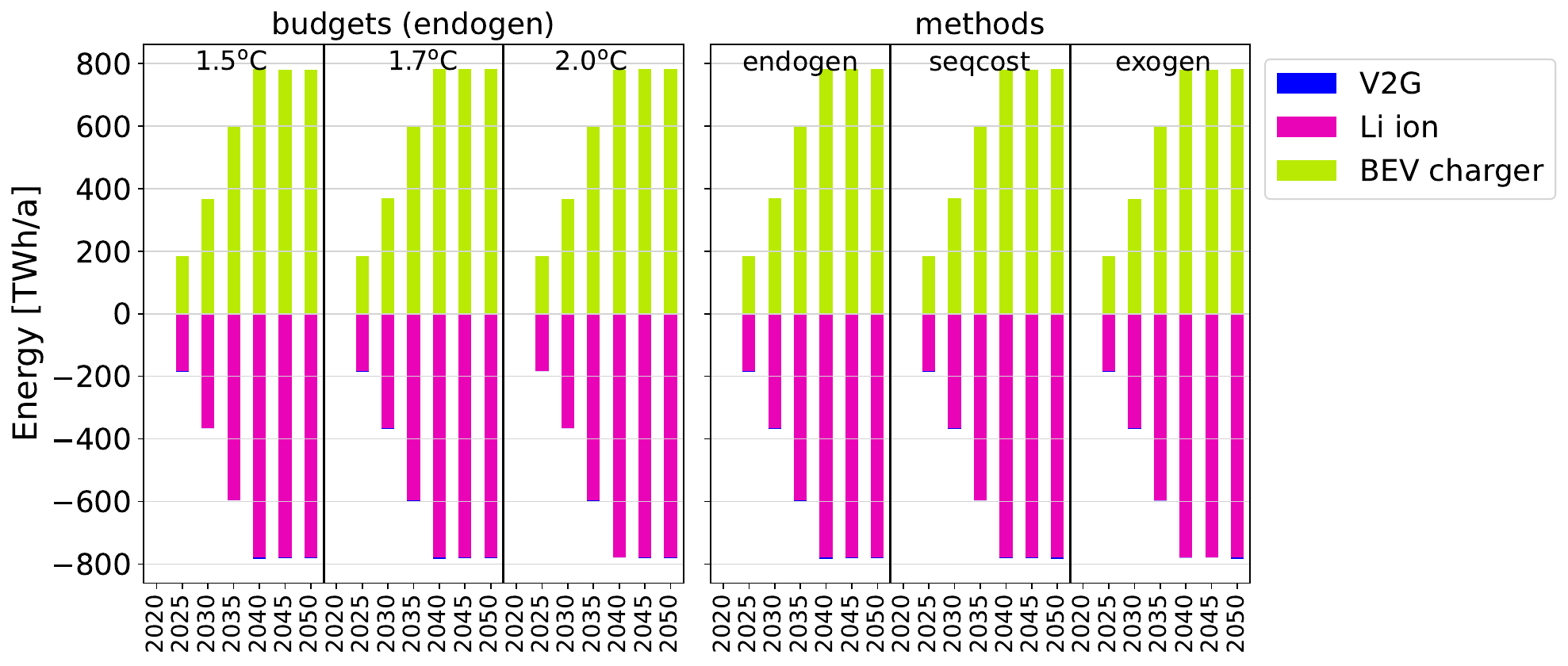}
	\caption{Energy balance for electric vehicles. Supply site is positive, usage is negative. Left plot shows the three different budgets with the endogenous method, right plot the three different methods endogenous (\textit{endogen}), sequential cost (\textit{seqcost}) and exogenous (\textit{exogen}). Values are also displayed in Tables \ref{tab:energy_balance_Li_ion_endogen}, \ref{tab:energy_balance_Li_ion_method}. The share of electric vehicles is exogenously fixed for all scenarios.}
	\label{fig:eb_Li_ion}
\end{figure*}

\begin{figure*}[!ht]
	\includegraphics[width=1\textwidth]{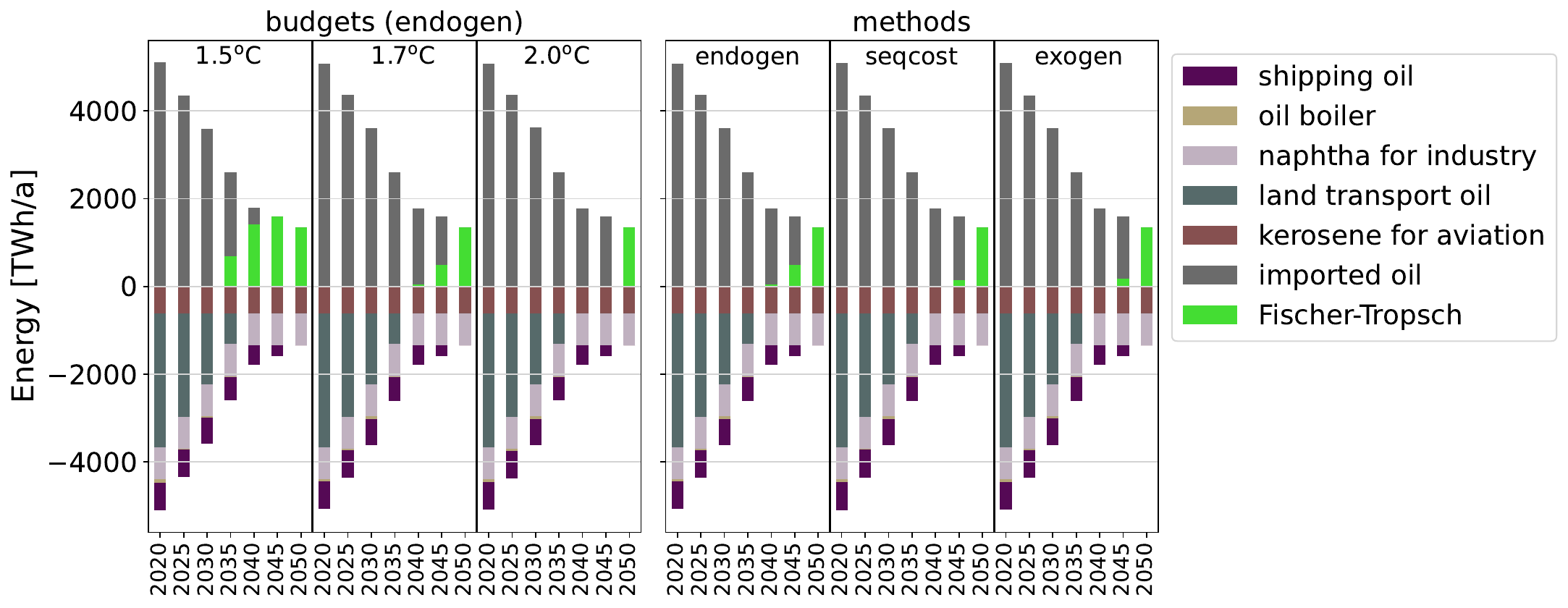}
	\caption{Energy balance for oil. Supply site is positive, usage is negative. Left plot shows the three different budgets with the endogenous method, right plot the three different methods endogenous (\textit{endogen}), sequential cost (\textit{seqcost}) and exogenous (\textit{exogen}). Values are also displayed in Tables \ref{tab:energy_balance_oil_endogen}, \ref{tab:energy_balance_oil_method}.}
	\label{fig:eb_oil}
\end{figure*}

\begin{figure*}[!ht]
	\includegraphics[width=1\textwidth]{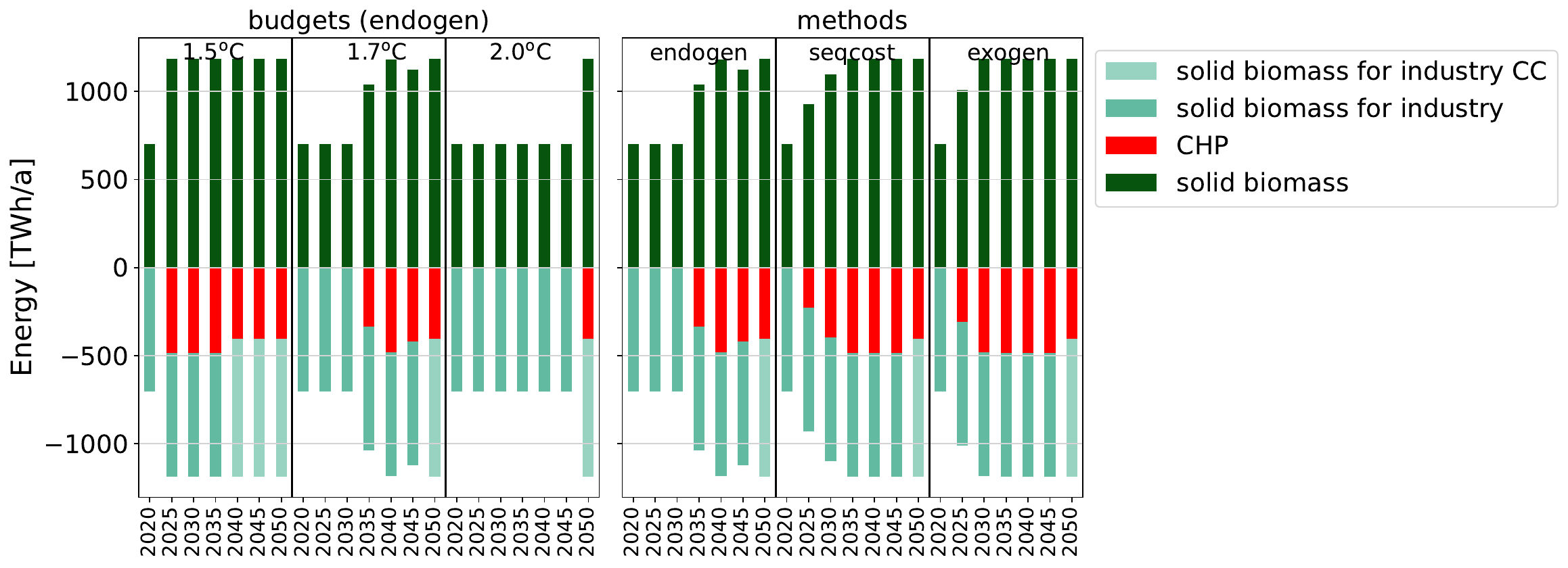}
	\caption{Energy balance for solid biomass. Supply site is positive, usage is negative. Left plot shows the three different budgets with the endogenous method, right plot the three different methods endogenous (\textit{endogen}), sequential cost (\textit{seqcost}) and exogenous (\textit{exogen}). Values are also displayed in Tables \ref{tab:energy_balance_solid_biomass_endogen}, \ref{tab:energy_balance_solid_biomass_method}.}
	\label{fig:eb_solid_biomass}
\end{figure*}

\begin{figure*}[!ht]
	\includegraphics[width=1\textwidth]{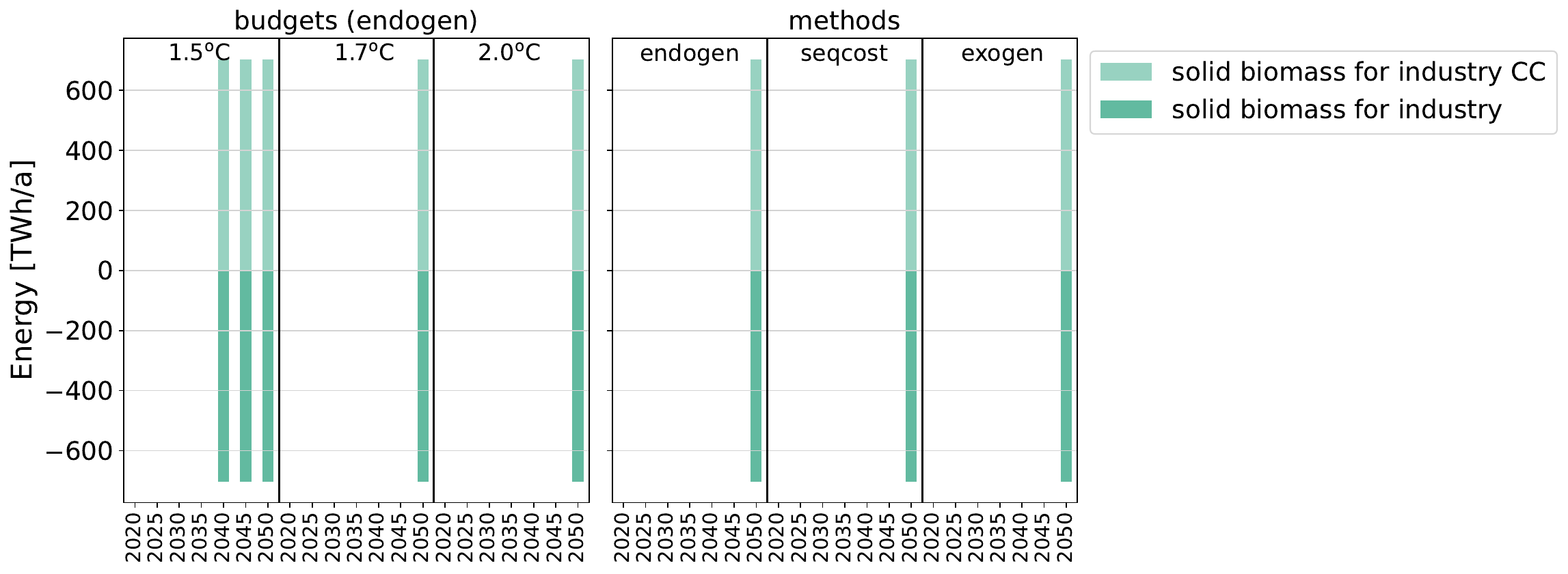}
	\caption{Energy balance for solid biomass for industry. Supply site is positive, usage is negative. Left plot shows the three different budgets with the endogenous method, right plot the three different methods endogenous (\textit{endogen}), sequential cost (\textit{seqcost}) and exogenous (\textit{exogen}). Values are also displayed in Tables \ref{tab:energy_balance_solid_biomass_for_industry_endogen}, \ref{tab:energy_balance_solid_biomass_for_industry_method}.}
	\label{fig:eb_solid_biomass_for_industry}
\end{figure*}

\begin{figure*}[!ht]
	\includegraphics[width=1\textwidth]{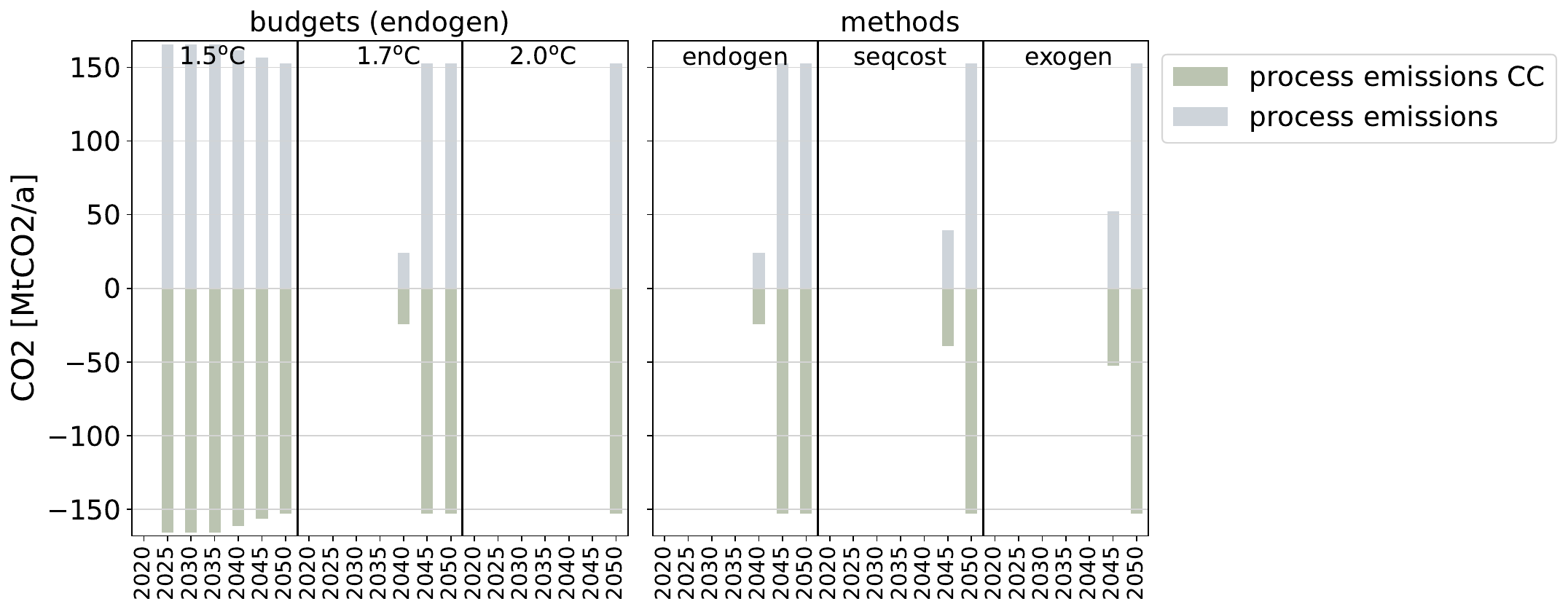}
	\caption{Energy balance for process emissions. Supply site is positive, usage is negative. Left plot shows the three different budgets with the endogenous method, right plot the three different methods endogenous (\textit{endogen}), sequential cost (\textit{seqcost}) and exogenous (\textit{exogen}). Values are also displayed in Tables \ref{tab:energy_balance_process_emissions_endogen}, \ref{tab:energy_balance_process_emissions_method}.}
	\label{fig:eb_process_emissions}
\end{figure*}

\begin{figure*}[!ht]
	\includegraphics[width=1\textwidth]{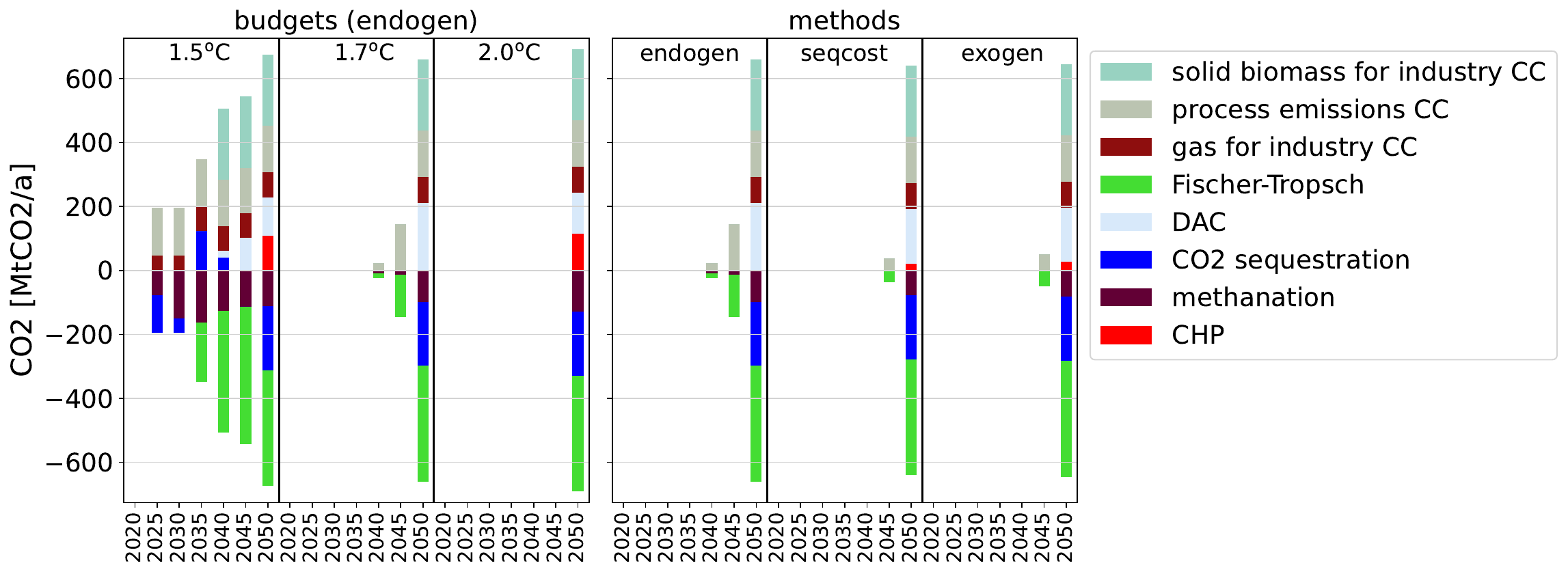}
	\caption{Energy balance for carbon dioxide storage. Supply site is positive, usage is negative. Left plot shows the three different budgets with the endogenous method, right plot the three different methods endogenous (\textit{endogen}), sequential cost (\textit{seqcost}) and exogenous (\textit{exogen}). Values are also displayed in Tables \ref{tab:energy_balance_co2_stored_endogen}, \ref{tab:energy_balance_co2_stored_method}.}
	\label{fig:eb_co2_stored}
\end{figure*}

\paragraph{Electricity}
A larger amount of electricity is produced with a tighter budget (see Figure \ref{fig:eb_AC}). For example, in 2040 about 10700 TWh of electricity are produced with the \budget{1.5}, 6600 TWh with the \budget{1.7} and 6100 TWh with the \budget{2.0}. The larger amount of electricity produced in scenarios with a \budget{1.5} compared to the \budget{2.0} is produced in renewable energy (for example +3000 TWh in 2030), while the production of conventional energy sources is lower (for example -1500 TWh in 2030). This additional electricity is mainly used for the production of green hydrogen via electrolysis and heat pumps which are employed earlier compared to the \budget{2.0} (see Figure \ref{fig:eb_low_voltage}). \\
\\ In scenarios with the endogenous method up to 1400 TWh more electricity are produced compared to the exogenous and sequential method. More electricity is generated by PV with the endogenous learning compared to the other two methods, as the investment costs decrease to a greater extent. The electricity demand is higher with the endogeouns method since the endogenous method produces hydrogen via electrolysis at higher volumes and at an earlier point in time.
\paragraph{Carbon dioxide}
At the beginning of the modelling period in 2020, CO$_2$ emissions are comparable between the three budgets (see Figure \ref{fig:eb_co2}). Large parts of the emissions are generated in the transport sector (31\%), in the combustion of fossil fuels (coal 21\%, gas 15\% and oil 11\%), in the heating sector (oil and gas boilers 12\%) and in industrial processes (7\%). 1\% of total CO$_2$ emissions are caused by the production of grey hydrogen. The stricter the available CO$_2$ budget, the faster annual emissions are reduced. In all scenarios, coal and oil are replaced by other energy sources. In the \ce{1.5} scenario coal and oil are substituted by renewable energies, in the \ce{1.7} and \ce{2.0} scenarios in a transition phase with gas (see Figure \ref{fig:eb_gas} and Figure \ref{fig:eb_gas_for_industry}). In 2030, coal-fired power plants are largely decommissioned in the \ce{1.5} and \ce{1.7} scenarios (share of total emissions is <1\%), while in the \ce{2.0} scenario they continue to contribute about 6\% to total CO$_2$ emissions. Emissions from land transport are reduced as internal combustion engines are continuously replaced by fuel and electric vehicles according to the exogenously defined transformation path (see Figure \ref{fig:eb_Li_ion}). Fossil oil is replaced by synthetic liquid hydrocarbons via the Fischer-Tropsch process. In the \ce{1.5} scenario this replacements start in 2035, in the \ce{1.7} scenario in 2045 and in the \ce{2.0} scenario in 2050. The produced synthetic liquid hydro carbons are mainly used as fuel for aircraft and naphtha production in industry (see Figure \ref{fig:eb_oil}). In the \ce{1.5} scenario, it is also used for synthetic fuels in land transport. In 2050, CO$_2$ emissions are net-zero for all budgets. Remaining emissions from industrial processes and oil are offset by bioenergy with carbon capture and storage (\gls{beccs}, see Figure \ref{fig:eb_solid_biomass} and Figure \ref{fig:eb_solid_biomass_for_industry}), carbon capture directly from the industrial process (see Figure \ref{fig:eb_process_emissions}) and \gls{dac}. This leads to even slightly net-negative emissions in scenarios with a \ce{1.5} budget.
The CO$_2$ sequestration potential of 200 Mt/CO$_2$ per year is completely exploited for all budgets in 2050 (see Figure \ref{fig:eb_co2_stored}). \\
\\ The total annual CO$_2$ emissions for the \ce{1.7} budget are similar between the three methods.  With the endogenous method, fossil oil for the production of naphtha and aviation fuels is produced to a larger extent already in 2045 via the Fischer Tropsch process compared to the other two methods (see Figure \ref{fig:eb_oil}). This is due to the cheaper production of green hydrogen. In addition, with the endogenous method, some industrial processes are already converted to capture the generated CO$_2$ in 2040. Whereas with the sequential and exogenous method, carbon capture of process emissions is used from 2045 onwards and to a lesser extent (see Figure \ref{fig:eb_process_emissions}).

\begin{figure*}[!ht]
	\includegraphics[width=1\textwidth]{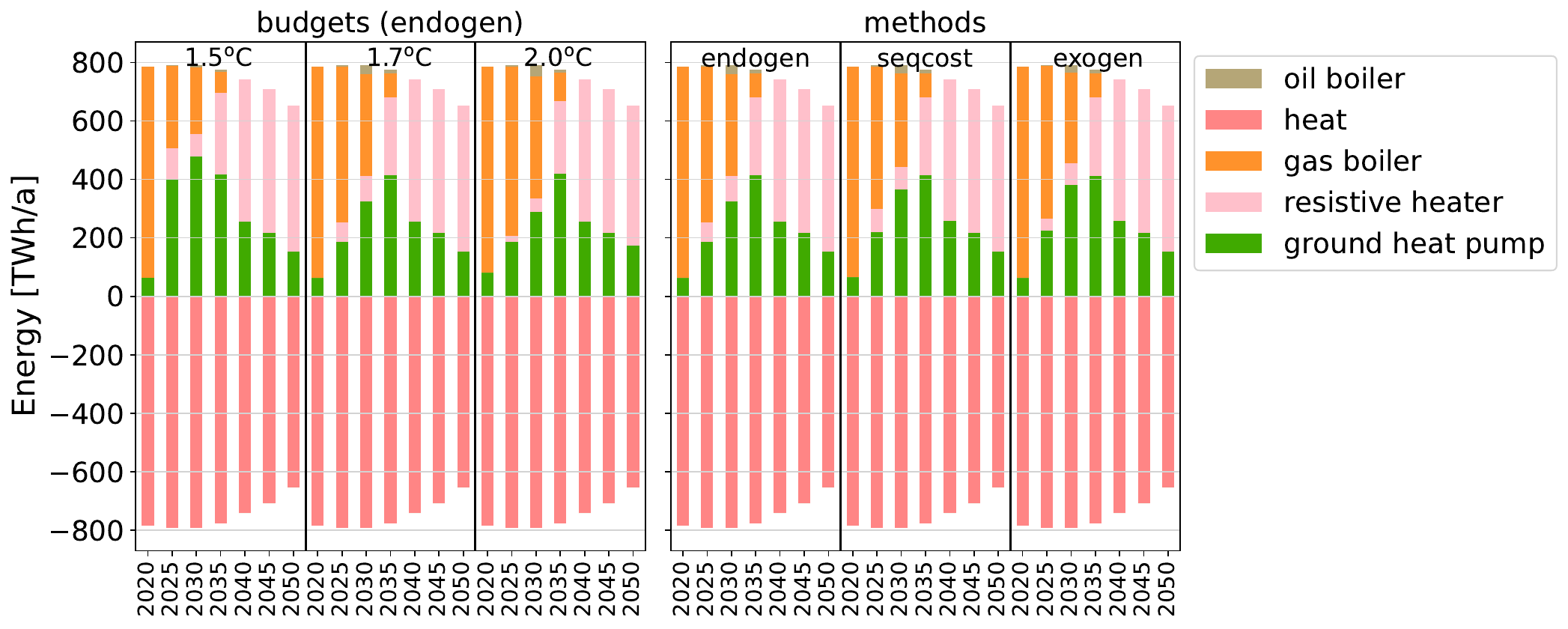}
	\caption{Energy balance for rural heat. Supply site is positive, usage is negative. Left plot shows the three different budgets with the endogenous method, right plot the three different methods endogenous (\textit{endogen}), sequential cost (\textit{seqcost}) and exogenous (\textit{exogen}). Values are also displayed in Tables \ref{tab:energy_balance_rural_heat_endogen}, \ref{tab:energy_balance_rural_heat_method}.}
	\label{fig:eb_rural_heat}
\end{figure*}

\begin{figure*}[!ht]
	\includegraphics[width=1\textwidth]{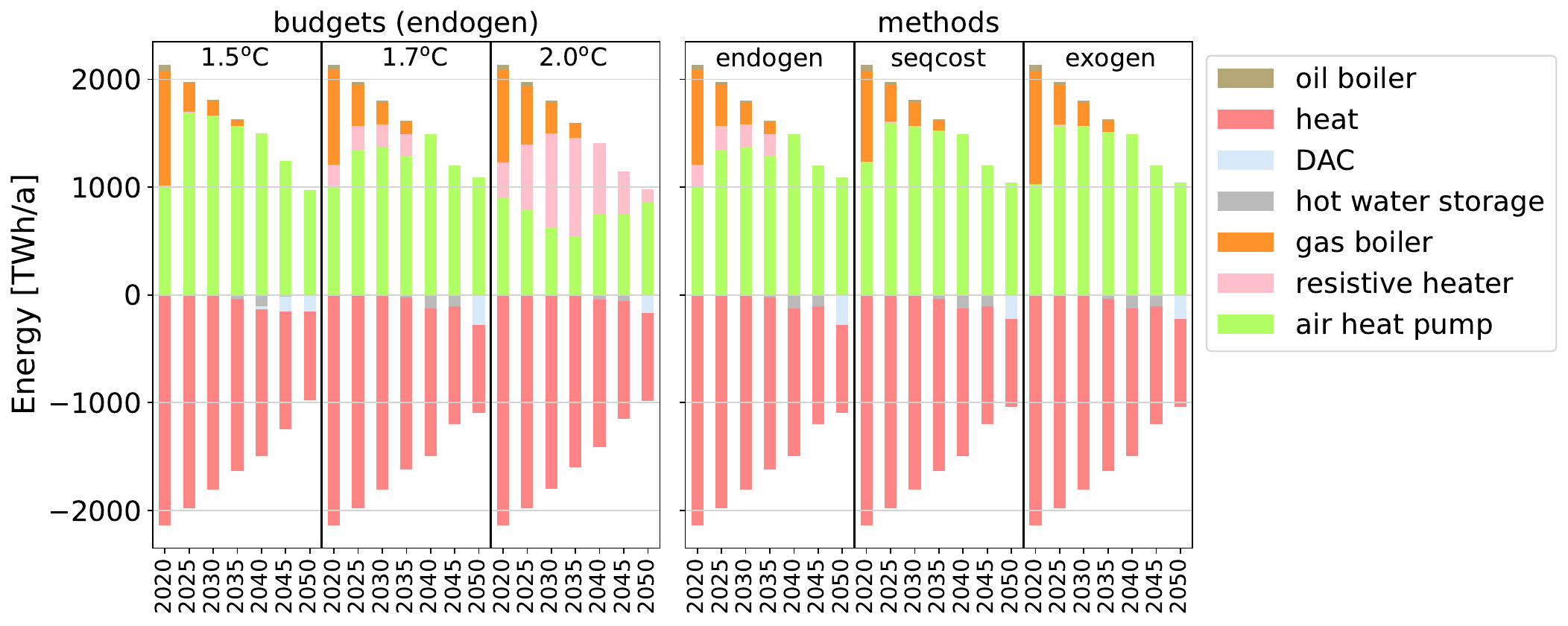}
	\caption{Energy balance for urban individual heating. Supply site is positive, usage is negative. Left plot shows the three different budgets with the endogenous method, right plot the three different methods endogenous (\textit{endogen}), sequential cost (\textit{seqcost}) and exogenous (\textit{exogen}). Values are also displayed in Tables \ref{tab:energy_balance_urban_decentral_heat_endogen}, \ref{tab:energy_balance_urban_decentral_heat_method}.}
	\label{fig:eb_urban_decentral_heat}
\end{figure*}

\begin{figure*}[!ht]
	\includegraphics[width=1\textwidth]{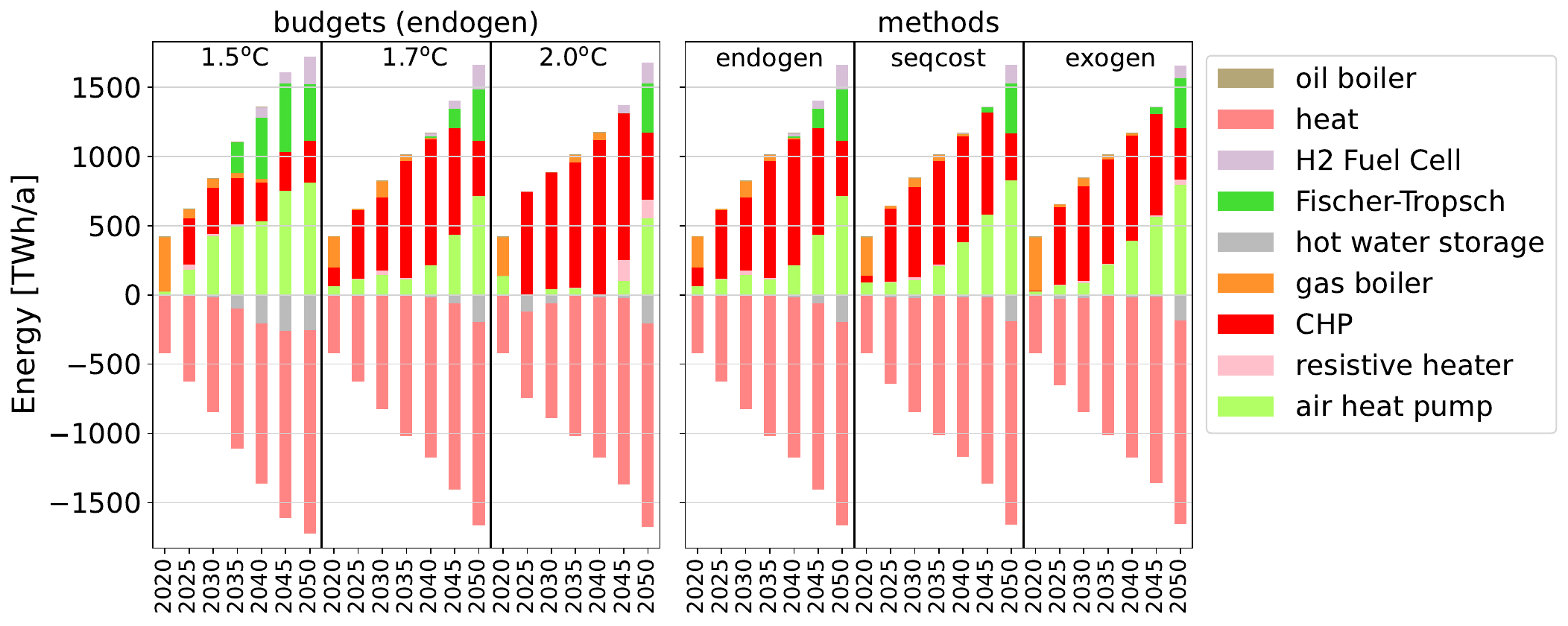}
	\caption{Energy balance for urban district heating. Supply site is positive, usage is negative. Left plot shows the three different budgets with the endogenous method, right plot the three different methods endogenous (\textit{endogen}), sequential cost (\textit{seqcost}) and exogenous (\textit{exogen}). Values are also displayed in Tables \ref{tab:energy_balance_urban_central_heat_endogen}, \ref{tab:energy_balance_urban_central_heat_method}.}
	\label{fig:eb_urban_central_heat}
\end{figure*}

\paragraph{Heat}
The heating sector is divided into rural areas (see Figure \ref{fig:eb_rural_heat}), urban areas with individual heating (see Figure \ref{fig:eb_urban_decentral_heat}) and urban areas with district heating (see Figure \ref{fig:eb_urban_central_heat}). In all scenarios, the heat sector is mostly electrified by 2050. In the district heating network, \gls{chp}s which run with biomass, biogas or synthetic gas, as well as hydrogen are also used. The stricter the CO$_2$ budget, the faster the shift from fossil fuels such as gas boilers, to heat pumps and resistive heaters. In 2030, in scenarios with a \budget{1.5}, more than half of the demand in rural areas is already met by heat pumps or resistive heaters, while in the \ce{1.7} and \ce{2.0} budget most of the demand is met by gas boilers. In district heating networks, the share of \gls{chp}s is higher in \ce{1.7} and \ce{2.0} scenarios compared to the \ce{1.5} budget. With a \ce{1.5} budget, heat pumps are used to a larger extent which do not emit additional CO$_2$.





\begin{figure*}[!ht]
	\includegraphics[width=1\textwidth]{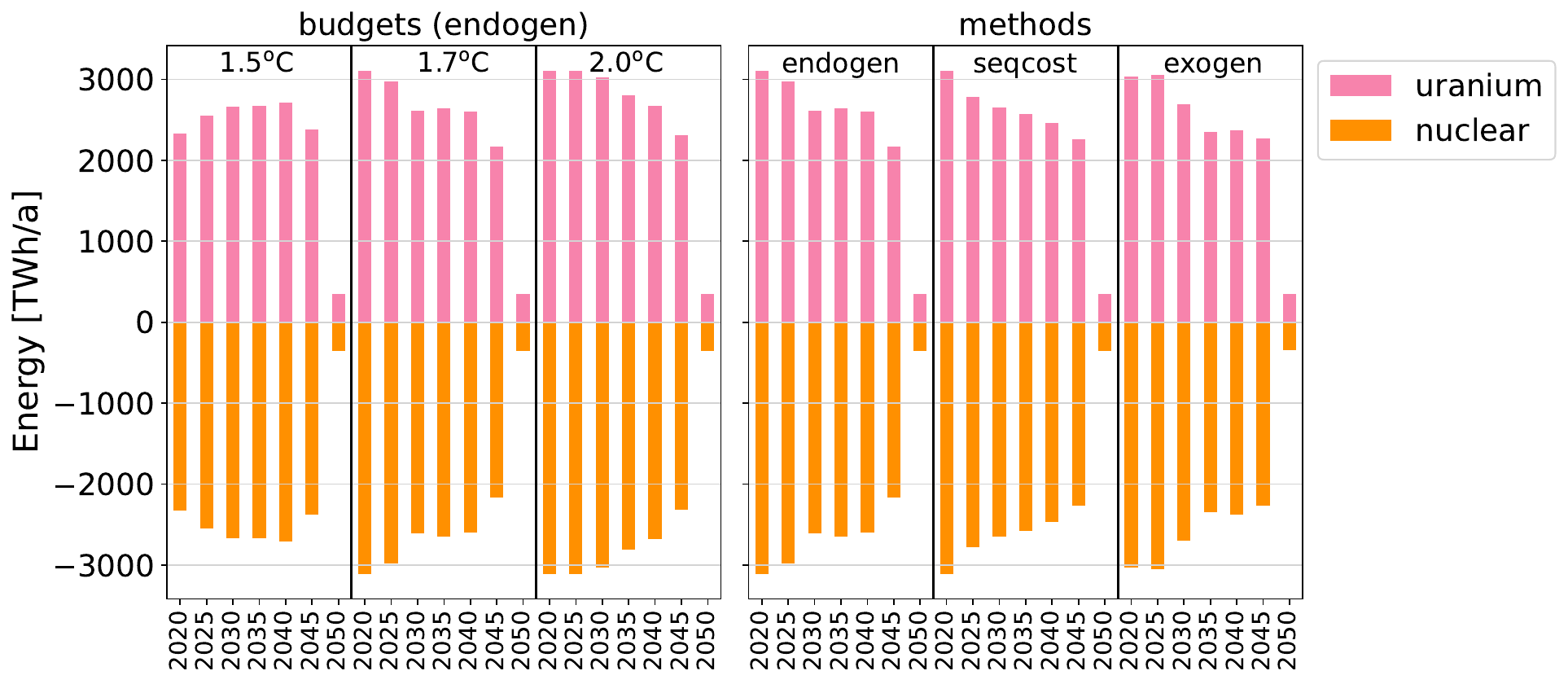}
	\caption{Energy balance for uranium. Supply site is positive, usage is negative. Left plot shows the three different budgets with the endogenous method, right plot the three different methods endogenous (\textit{endogen}), sequential cost (\textit{seqcost}) and exogenous (\textit{exogen}). Values are also displayed in Tables \ref{tab:energy_balance_uranium_endogen}, \ref{tab:energy_balance_uranium_method}.}
	\label{fig:eb_uranium}
\end{figure*}

\FloatBarrier
\subsection{Electrolysis duration curve}
\begin{figure*}[!ht]
	\centering
	\includegraphics[width=1\textwidth]{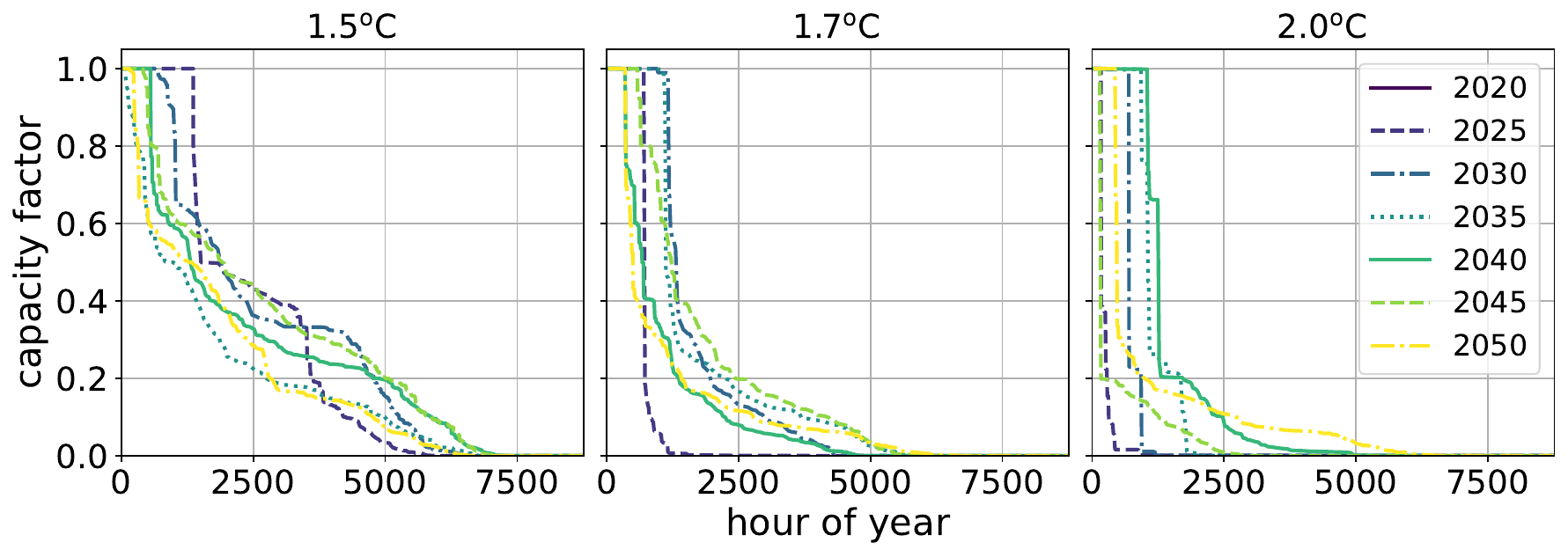}
	\caption{Duration curve of hydrogen electrolysis.}
	\label{fig:duration_curve}
\end{figure*}

The full load hours of electrolysis vary between a maximum of 1032-1334 hours per year between the \ce{2.0} and \ce{1.5} budget. In the \budget{1.5}, the electrolysis is operated on average in more hours per year (77\%) compared to the \budget{2.0} (16\%). In later years, electrolysis tends to be operated in more hours in all budgets (see Figure \ref{fig:duration_curve}).

\FloatBarrier
\section{Technology assumptions}\label{sec:tech_assumptions}
The technology assumptions are listed in the tables below. Overnight costs $c_{\text{annualised}}$ are calculated for each assets based on investment costs $inv$, fixed operational and maintenance cost (\gls{fom}) $FOM$, lifetime $n$ and discount rate $\tau$.
\begin{align}
	f_{\text{annuity}} &= (1 - (1 + \tau )^{-n})/ \tau \\
	c_{\text{annualised}} &= (f_{\text{annuity}} + FOM/100) \cdot inv
\end{align} The assumed discount rate in this study is 7\%. If variable operational and maintenance cost (\gls{vom}) or lifetime are not listed for a technology, zero costs and 25 years are assumed accordingly. The efficiencies for all thermal plants are given in units of lower heating value. The investment cost include equipment and installation cost. Infrastructure and connection expenditures within a plant are included in the investment costs. All financial data is given in Euro at the 2015-level. All our technology assumptions with a detailed description of the sources can be found openly available in the Github repository \texttt{technology-data} \cite{technologydata}, Version 0.3.0 is used for this study.
\begin{table*}[h]
	\centering
	\begin{tabular}{lrrrrrrr}
		\toprule
		build year  &  2020 &  2025 &  2030 &  2035 &  2040 &  2045 &  2050 \\
		\midrule
		electrolysis  &   650 &   550 &   450 &   375 &   300 &   275 &   250 \\
		offshore wind       &  1748 &  1661 &  1573 &  1510 &  1447 &  1432 &  1416 \\
		onshore wind       &  1119 &  1077 &  1036 &  1007 &   978 &   970 &   963 \\
		solar PV &   529 &   453 &   376 &   353 &   330 &   316 &   302 \\
		\bottomrule
	\end{tabular}
	\caption{Investment costs [\kw] in the exogenous scenarios for the learning technologies, based on \cite{cost_dea}. For wind and solar further grid connection costs are added.}
	\label{tab:exogen_costs}
\end{table*}

\begin{table*}[h]
\resizebox{0.95\textwidth}{!}{ 
\begin{threeparttable}
\begin{tabular}{llrrrrrrrll}
	\toprule
	&     &     2020 &     2025 &     2030 &     2035 &     2040 &     2045 &     2050 &            unit &             source \\
	technology & parameter &          &          &          &          &          &          &          &                 &                    \\
	\midrule
	Air-sourced heat pump central & FOM &     0.21 &     0.21 &     0.23 &     0.23 &     0.23 &     0.23 &     0.23 &          \% of investment/year &    \cite{cost_dea} \\
	& VOM &     2.19 &     2.19 &     2.51 &     2.35 &     2.19 &     2.43 &     2.67 &      EUR/MWh$_\text{th}$ &    \cite{cost_dea} \\
	& efficiency\tnote{1} &   340 &   350 &   360 &   362 &   365 &   368 &   370 &               \% &    \cite{cost_dea} \\
	& investment &   951 &   951 &   856 &   856 &   856 &   856 &   856 &       EUR/kW$_\text{th}$ &    \cite{cost_dea} \\
	& lifetime &    25 &    25 &    25 &    25 &    25 &    25 &    25 &           years &    \cite{cost_dea} \\
	Air-sourced heat pump decentral & FOM &     2.96 &     2.98 &     3 &     3.03 &     3.07 &     3.10 &     3.14 &          \% of investment/year &    \cite{cost_dea} \\
	& efficiency\tnote{1} &   340 &   350 &   360 &   365 &   370 &   375 &   380 &               \% &    \cite{cost_dea} \\
	& investment &   940 &   895 &   850 &   828 &   805 &   782 &   760 &       EUR/kW$_\text{th}$ &    \cite{cost_dea} \\
	& lifetime &    18 &    18 &    18 &    18 &    18 &    18 &    18 &           years &    \cite{cost_dea} \\
	Battery inverter & FOM &     0.20 &     0.25 &     0.34 &     0.42 &     0.54 &     0.68 &     0.90 &          \% of investment/year &    \cite{cost_dea} \\
	& efficiency &    95 &    96 &    96 &    96 &    96 &    96 &    96 &               \% &    \cite{cost_dea} \\
	& investment &   270 &   215 &   160 &   130 &   100 &    80 &    60 &          EUR/kW &    \cite{cost_dea} \\
	& lifetime &    10 &    10 &    10 &    10 &    10 &    10 &    10 &           years &    \cite{cost_dea} \\
	Battery storage & investment &   232 &   187 &   142 &   118 &    94 &    84 &    75 &         EUR/kWh &    \cite{cost_dea} \\
	& lifetime &    20 &    22 &    25 &    28 &    30 &    30 &    30 &           years &    \cite{cost_dea} \\
	Biogas upgrading & FOM &     2.51 &     2.50 &     2.49 &     2.50 &     2.50 &     2.50 &     2.51 &          \% of investment/year &    \cite{cost_dea} \\
	& VOM &     3.69 &     3.44 &     3.18 &     3.31 &     3.43 &     3.56 &     3.68 &   EUR/MWh input &    \cite{cost_dea} \\
	& investment &   423 &   402 &   381 &   372 &   362 &   352 &   343 &    EUR/kW input &    \cite{cost_dea} \\
	& lifetime &    15 &    15 &    15 &    15 &    15 &    15 &    15 &           years &    \cite{cost_dea} \\
	CCGT & FOM &     3.33 &     3.34 &     3.35 &     3.33 &     3.30 &     3.28 &     3.25 &          \% of investment/year &    \cite{cost_dea} \\
	& VOM &     4.40 &     4.30 &     4.20 &     4.15 &     4.10 &     4.05 &     4 &         EUR/MWh &    \cite{cost_dea} \\
	& efficiency &    56 &    57 &    58 &    58 &    59 &    60 &    60 &               \% &    \cite{cost_dea} \\
	& investment &   880 &   855 &   830 &   822 &   815 &   808 &   800 &          EUR/kW &    \cite{cost_dea} \\
	& lifetime &    25 &    25 &    25 &    25 &    25 &    25 &    25 &           years &    \cite{cost_dea} \\
	CHP biomass & FOM &     3.61 &     3.60 &     3.58 &     3.57 &     3.56 &     3.55 &     3.54 &          \% of investment/year &    \cite{cost_dea} \\
	& VOM &     2.11 &     2.10 &     2.10 &     2.10 &     2.10 &     2.10 &     2.10 &       EUR/MWh$_\text{el}$ &    \cite{cost_dea} \\
	& efficiency &    30 &    30 &    30 &    30 &    30 &    30 &    30 &               \% &    \cite{cost_dea} \\
	& investment &  3381 &  3296 &  3210 &  3136 &  3061 &  2987 &  2912 &        EUR/kW$_\text{el}$ &    \cite{cost_dea} \\
	& lifetime &    25 &    25 &    25 &    25 &    25 &    25 &    25 &           years &    \cite{cost_dea} \\
	CHP gas & FOM &     3.31 &     3.31 &     3.32 &     3.35 &     3.39 &     3.42 &     3.46 &          \% of investment/year &    \cite{cost_dea} \\
	& VOM &     4.40 &     4.30 &     4.20 &     4.15 &     4.10 &     4.05 &     4 &         EUR/MWh &    \cite{cost_dea} \\
	& efficiency &    40 &    40 &    41 &    42 &    42 &    42 &    43 &               \% &    \cite{cost_dea} \\
	& investment &   590 &   575 &   560 &   550 &   540 &   530 &   520 &          EUR/kW &    \cite{cost_dea} \\
	& lifetime &    25 &    25 &    25 &    25 &    25 &    25 &    25 &           years &    \cite{cost_dea} \\
	CHP solid biomass & FOM &     4.12 &     4.11 &     4.10 &     4.07 &     4.04 &     4.01 &     3.98 &          \% of investment/year &    \cite{cost_dea} \\
	& VOM &     1.86 &     1.86 &     1.85 &     1.86 &     1.87 &     1.88 &     1.88 &       EUR/MWh$_\text{el}$ &    \cite{cost_dea} \\
	& efficiency &    29 &    29 &    29 &    29 &    29 &    29 &    29 &               \% &    \cite{cost_dea} \\
	& investment &  3007 &  2929 &  2851 &  2817 &  2783 &  2749 &  2714 &        EUR/kW$_\text{el}$ &    \cite{cost_dea} \\
	& lifetime &    25 &    25 &    25 &    25 &    25 &    25 &    25 &           years &    \cite{cost_dea} \\
	CHP solid biomass with Carbon Capture & FOM &     3 &     3 &     3 &     3 &     3 &     3 &     3 &          \% of investment/year &    \cite{cost_dea} \\
	& investment &  3300 &  3000 &  2700 &  2550 &  2400 &  2200 &  2000 &  EUR/(kg$_{\text{CO}_2}$/h) &    \cite{cost_dea} \\
	& lifetime &    25 &    25 &    25 &    25 &    25 &    25 &    25 &           years &    \cite{cost_dea} \\
	CO$_2$ storage tank & FOM &     1 &     1 &     1 &     1 &     1 &     1 &     1 &          \% of investment/year &   \cite{lauri2014} \\
	& investment &  2528 &  2528 &  2528 &  2528 &  2528 &  2528 &  2528 &       EUR/t$_{\text{CO}_2}$ &   \cite{lauri2014} \\
	& lifetime &    25 &    25 &    25 &    25 &    25 &    25 &    25 &           years &   \cite{lauri2014} \\
	Cement capture & FOM &     3 &     3 &     3 &     3 &     3 &     3 &     3 &          \% of investment/year &    \cite{cost_dea} \\
	& investment &  3000 &  2800 &  2600 &  2400 &  2200 &  2000 &  1800 &  EUR/(kg$_{\text{CO}_2}$/h) &    \cite{cost_dea} \\
	& lifetime &    25 &    25 &    25 &    25 &    25 &    25 &    25 &           years &    \cite{cost_dea} \\
	Coal power plant & FOM &     1.60 &     1.60 &     1.60 &     1.60 &     1.60 &     1.60 &     1.60 &          \% of investment/year &      \cite{lazard} \\
	& VOM &     3.50 &     3.50 &     3.50 &     3.50 &     3.50 &     3.50 &     3.50 &       EUR/MWh$_\text{el}$ &      \cite{lazard} \\
	& efficiency &    33 &    33 &    33 &    33 &    33 &    33 &    33 &               \% &      \cite{lazard} \\
	& investment &  3846 &  3846 &  3846 &  3846 &  3846 &  3846 &  3846 &        EUR/kW$_\text{el}$ &      \cite{lazard} \\
	& lifetime &    40 &    40 &    40 &    40 &    40 &    40 &    40 &           years &      \cite{lazard} \\
	DAC (direct-air capture) & FOM &     4.95 &     4.95 &     4.95 &     4.95 &     4.95 &     4.95 &     4.95 &          \% of investment/year &    \cite{cost_dea} \\
	& investment &  7000 &  7000 &  6000 &  5500 &  5000 &  4500 &  4000 &  EUR/(kg$_{\text{CO}_2}$/h) &    \cite{cost_dea} \\
	& lifetime &    20 &    20 &    20 &    20 &    20 &    20 &    20 &           years &    \cite{cost_dea} \\
	Electricity distribution grid & FOM &     2 &     2 &     2 &     2 &     2 &     2 &     2 &          \% of investment/year &    \cite{cost_dea} \\
	& investment &   500 &   500 &   500 &   500 &   500 &   500 &   500 &          EUR/kW &    \cite{cost_dea} \\
	& lifetime &    40 &    40 &    40 &    40 &    40 &    40 &    40 &           years &    \cite{cost_dea} \\
	Electricity grid connection \tnote{2} & FOM &     2 &     2 &     2 &     2 &     2 &     2 &     2 &          \% of investment/year &    \cite{cost_dea} \\
	& investment &   140 &   140 &   140 &   140 &   140 &   140 &   140 &          EUR/kW &    \cite{cost_dea} \\
	& lifetime &    40 &    40 &    40 &    40 &    40 &    40 &    40 &           years &    \cite{cost_dea} \\
	Electrolysis  \tnote{3} & FOM &     2 &     2 &     2 &     2 &     2 &     2 &     2 &          \% of investment/year &    \cite{cost_dea} \\
	& efficiency &    66 &    67 &    68 &    70 &    72 &    73 &    75 &               \% &    \cite{cost_dea} \\
	& investment &   650 &   550 &   450 &   375 &   300 &   275 &   250 &        EUR/kW$_\text{el}$ &    \cite{cost_dea} \\
	& lifetime &    25 &    28 &    30 &    31 &    32 &    34 &    35 &           years &    \cite{cost_dea} \\
	Fischer-Tropsch & FOM &     3 &     3 &     3 &     3 &     3 &     3 &     3 &          \% of investment/year &  \cite{fasihi2017} \\
	& VOM &     5.30 &     4.75 &     4.20 &     3.70 &     3.20 &     2.65 &     2.10 &      EUR/MWh$_\text{FT}$ &    \cite{cost_dea} \\
	\bottomrule
\end{tabular}
\begin{tablenotes}
	\item[1] Efficiencies of heat pumps are time-dependent based on ambient temperature.
	\item[2] Grid connection costs for solar and onshore wind. For offshore wind these are calculated separately depending on the connection type. Electricity grid connection costs do not underlay any learning.
	\item[3] Technology assumptions for Alkaline Electrolysis 100 MW plant.
\end{tablenotes}
\end{threeparttable}
}
\end{table*}

\begin{table*}[h]
	\resizebox{1.\textwidth}{!}{ 
		\begin{threeparttable}
		
\begin{tabular}{llrrrrrrrll}
			\toprule
			&          &     2020 &     2025 &     2030 &     2035 &     2040 &     2045 &     2050 &            unit &                source \\
			technology & parameter &          &          &          &          &          &          &          &                 &                       \\
			\midrule
			Fischer-Tropsch & efficiency &    65 &    68 &    70 &    72 &    73 &    74 &    75 &               \% &       \cite{cost_dea} \\
			& investment &  2100 &  1850 &  1600 &  1350 &  1100 &  1000 &   900 &  EUR/kW$_\text{FT}$/year &       \cite{cost_dea} \\
			& lifetime &    25 &    25 &    25 &    25 &    25 &    25 &    25 &           years &       \cite{cost_dea} \\
			Fuel cell & FOM &     5 &     5 &     5 &     5 &     5 &     5 &     5 &          \% of investment/year &       \cite{cost_dea} \\
			& efficiency &    50 &    50 &    50 &    50 &    50 &    50 &    50 &               \% &       \cite{cost_dea} \\
			& investment &  1300 &  1200 &  1100 &  1025 &   950 &   875 &   800 &        EUR/kW$_\text{el}$ &       \cite{cost_dea} \\
\bottomrule
		\end{tabular}

\begin{tablenotes}
	\item[1] Efficiencies of heat pumps are time-dependent based on ground temperature.
\end{tablenotes}
\end{threeparttable}
	}
\end{table*}

\begin{table*}[h]
	\resizebox{1\textwidth}{!}{ 
\begin{tabular}{llrrrrrrrll}
	\toprule
	&          &       2020 &       2025 &       2030 &       2035 &       2040 &       2045 &       2050 &             unit &                 source \\
	technology & parameter &            &            &            &            &            &            &            &                  &                        \\
	\midrule
	Offshore Wind & FOM &       2.29 &       2.29 &       2.29 &       2.29 &       2.29 &       2.29 &       2.29 &           \% of investment/year &        \cite{cost_dea} \\
	& VOM &       2.97 &       2.82 &       2.67 &       2.56 &       2.46 &       2.43 &       2.40 &          EUR/MWh &        \cite{cost_dea} \\
	& investment &    1748 &    1661 &    1573 &    1510 &    1447 &    1432 &    1416 &           EUR/kW &        \cite{cost_dea} \\
	& lifetime &      27 &      28 &      30 &      30 &      30 &      30 &      30 &            years &        \cite{cost_dea} \\
	Offshore wind-ac-connection-submarine & investment &    2685 &    2685 &    2685 &    2685 &    2685 &    2685 &    2685 &        EUR/MW/km &        \cite{cost_dea} \\
	Offshore wind-ac-connection-underground & investment &    1342 &    1342 &    1342 &    1342 &    1342 &    1342 &    1342 &        EUR/MW/km &        \cite{cost_dea} \\
\bottomrule
\end{tabular}

	}
\caption{Technology assumptions. Fixed operational and maintenance cost (\gls{fom}) are given as annual percentage of the investment costs. If variable operational and maintenance cost (\gls{vom}) are not listed for a technology, zero costs are assumed.}
\end{table*}
\FloatBarrier
\section{Energy balances tables}
\begin{table*}[h]
	\resizebox{1\textwidth}{!}{ 
\begin{tabular}{lrrrrrrrrrrrrrrrrrrrrr}
	\toprule
	budget & \multicolumn{7}{l}{\ce{1.5}} & \multicolumn{7}{l}{\ce{1.7}} & \multicolumn{7}{l}{\ce{2.0}}\\
	year & 2020 & 2025 &  2030 &  2035 &  2040 &  2045 &  2050 & 2020 & 2025 & 2030 & 2035 & 2040 &  2045 &  2050 & 2020 & 2025 & 2030 & 2035 & 2040 & 2045 &  2050 \\
	\midrule
	methanation              &    0 & -478 &  -942 & -1012 &  -786 &  -708 &  -701 &    0 &    0 &    0 &    0 &  -58 &   -84 &  -611 &    0 &    0 &    0 &    0 &    0 &    0 &  -807 \\
	Fischer-Tropsch          &    0 &    0 &     0 &  -961 & -1946 & -2192 & -1844 &    0 &    0 &    0 &    0 &  -70 &  -661 & -1799 &    0 &    0 &    0 &    0 &    0 &    0 & -1788 \\
	H2 Electrolysis          &    0 &  590 &  1188 &  2385 &  3466 &  3833 &  3962 &    0 &    7 &  127 &  316 &  701 &  1565 &  3786 &    0 &    0 &    3 &   13 &  127 &  580 &  3900 \\
	H2 Fuel Cell             &    0 &    0 &     0 &     0 &  -172 &  -185 &  -444 &    0 &    0 &    0 &    0 &  -56 &  -131 &  -402 &    0 &    0 &    0 &    0 &    0 & -129 &  -332 \\
	H2 for industry          & -107 & -107 &  -114 &  -134 &  -154 &  -180 &  -196 & -107 & -107 & -114 & -134 & -154 &  -180 &  -196 & -107 & -107 & -114 & -134 & -154 & -180 &  -196 \\
	H2 for shipping          &    0 &    0 &   -25 &   -75 &  -151 &  -302 &  -503 &    0 &    0 &  -25 &  -75 & -151 &  -302 &  -503 &    0 &    0 &  -25 &  -75 & -151 & -302 &  -503 \\
	SMR                      &  107 &   50 &    22 &    16 &    19 &     9 &     2 &  107 &  155 &  141 &  113 &   63 &    67 &     0 &  107 &  162 &  265 &  416 &  454 &  306 &     0 \\
	land transport fuel cell &    0 &  -55 &  -128 &  -220 &  -275 &  -275 &  -275 &    0 &  -55 & -128 & -220 & -275 &  -275 &  -275 &    0 &  -55 & -128 & -220 & -275 & -275 &  -275 \\
	\bottomrule
\end{tabular}

	}
	\caption{Energy balance [TWh/a] for hydrogen for the three different budgets and the endogenous method. See also Figure \ref{fig:eb_H2}.}
	\label{tab:energy_balance_H2_endogen}
\end{table*}

\begin{table*}[h]
	\resizebox{1\textwidth}{!}{ 
\begin{tabular}{lrrrrrrrrrrrrrrrrrrrrr}
	\toprule
	method & \multicolumn{7}{l}{endogen} & \multicolumn{7}{l}{exogen} & \multicolumn{7}{l}{seqcost} \\
	year &    2020 & 2025 & 2030 & 2035 & 2040 &  2045 &  2050 &   2020 & 2025 & 2030 & 2035 & 2040 & 2045 &  2050 &    2020 & 2025 & 2030 & 2035 & 2040 & 2045 &  2050 \\
	\midrule
	methanation              &       0 &    0 &    0 &    0 &  -58 &   -84 &  -611 &      0 &    0 &    0 &    0 &    0 &    0 &  -487 &       0 &    0 &    0 &    0 &    0 &  -11 &  -517 \\
	Fischer-Tropsch          &       0 &    0 &    0 &    0 &  -70 &  -661 & -1799 &      0 &    0 &    0 &    0 &    0 & -187 & -1790 &       0 &    0 &    0 &    0 &    0 & -240 & -1791 \\
	H2 Electrolysis          &       0 &    7 &  127 &  316 &  701 &  1565 &  3786 &      0 &    0 &   38 &  211 &  447 &  854 &  3545 &       0 &    1 &    8 &   25 &  375 &  906 &  3497 \\
	H2 Fuel Cell             &       0 &    0 &    0 &    0 &  -56 &  -131 &  -402 &      0 &    0 &    0 &    0 &  -16 &  -17 &  -294 &       0 &    0 &    0 &    0 &   -7 &   -6 &  -204 \\
	H2 for industry          &    -107 & -107 & -114 & -134 & -154 &  -180 &  -196 &   -107 & -107 & -114 & -134 & -154 & -180 &  -196 &    -107 & -107 & -114 & -134 & -154 & -180 &  -196 \\
	H2 for shipping          &       0 &    0 &  -25 &  -75 & -151 &  -302 &  -503 &      0 &    0 &  -25 &  -75 & -151 & -302 &  -503 &       0 &    0 &  -25 &  -75 & -151 & -302 &  -503 \\
\bottomrule
\end{tabular}
		
	}
	\caption{Energy balance [TWh/a] for hydrogen for the \ce{1.7} scenario and the three methods endogenous (\textit{endogen}), exogenous (\textit{exogen}) and sequential cost (\textit{seqcost}). See also Figure \ref{fig:eb_H2}.}
	\label{tab:energy_balance_H2_method}
\end{table*}
\begin{table*}[h]
	\resizebox{1\textwidth}{!}{ 
\begin{tabular}{lrrrrrrrrrrrrrrrrrrrrr}
	\toprule
	{} & \multicolumn{7}{l}{\ce{1.5}} & \multicolumn{7}{l}{\ce{1.7}} & \multicolumn{7}{l}{\ce{2.0}} \\
	year &  2020 &  2025 &  2030 &  2035 &  2040 &  2045 &  2050 &  2020 &  2025 &  2030 &  2035 &  2040 &  2045 &  2050 &  2020 &  2025 &  2030 &  2035 &  2040 &  2045 &  2050 \\
	\midrule
	hydroelectricity              &   466 &   460 &   458 &   459 &   457 &   457 &   455 &   465 &   463 &   452 &   455 &   453 &   453 &   456 &   467 &   464 &   460 &   455 &   454 &   451 &   457 \\
	onshore wind                  &   561 &  2737 &  3181 &  3928 &  4820 &  4759 &  4943 &   436 &  1470 &  1709 &  1848 &  1961 &  2559 &  4637 &   611 &  1117 &  1528 &  1686 &  1677 &  1579 &  4853 \\
	offshore wind                 &   142 &   737 &   744 &   734 &   747 &   723 &   725 &   106 &   340 &   382 &   376 &   372 &   331 &   800 &    70 &    85 &    85 &    76 &    74 &   116 &   787 \\
	solar PV                      &   167 &   532 &  1222 &  2611 &  3483 &  4155 &  4469 &   166 &   514 &   982 &  1264 &  1918 &  2887 &  4249 &   161 &   407 &   512 &   961 &  1140 &  2129 &  4281 \\
	ground heat pump              &   -19 &    -9 &    -5 &    -2 &     0 &     0 &     0 &   -19 &   -16 &    -7 &    -2 &     0 &     0 &     0 &   -24 &   -16 &    -6 &    -2 &     0 &     0 &     0 \\
	air heat pump                 &   -29 &  -123 &   -61 &   -19 &     0 &     0 &     0 &   -57 &   -88 &   -78 &   -27 &     0 &     0 &     0 &  -117 &   -49 &   -42 &   -29 &     0 &     0 &     0 \\
\bottomrule
\end{tabular}
	}
\caption{Energy balance [TWh/a] for the electricity transmission grid for the three different budgets and the endogenous method. See also Figure \ref{fig:eb_AC}.}
\label{tab:energy_balance_AC_endogen}
\end{table*}

\begin{table*}[h]
	\resizebox{1\textwidth}{!}{ 
\begin{tabular}{lrrrrrrrrrrrrrrrrrrrrr}
	\toprule
	method & \multicolumn{7}{l}{endogen} & \multicolumn{7}{l}{exogen} & \multicolumn{7}{l}{seqcost} \\
	year &    2020 &  2025 &  2030 &  2035 &  2040 &  2045 &  2050 &   2020 &  2025 &  2030 &  2035 &  2040 &  2045 &  2050 &    2020 &  2025 &  2030 &  2035 &  2040 &  2045 &  2050 \\
	\midrule
	hydroelectricity              &     465 &   463 &   452 &   455 &   453 &   453 &   456 &    466 &   464 &   452 &   453 &   455 &   451 &   454 &     466 &   465 &   460 &   457 &   455 &   457 &   458 \\
	onshore wind                  &     436 &  1470 &  1709 &  1848 &  1961 &  2559 &  4637 &    358 &  1747 &  1767 &  2102 &  2240 &  2395 &  4730 &     348 &  1445 &  1651 &  1886 &  2440 &  2418 &  4455 \\
	offshore wind                 &     106 &   340 &   382 &   376 &   372 &   331 &   800 &     70 &    70 &    57 &    55 &    48 &    24 &     0 &     164 &   305 &   389 &   471 &   540 &   938 &  1502 \\
	offshore wind (AC)            &       0 &     0 &     0 &     0 &     0 &     0 &     0 &      0 &     0 &     0 &     0 &     0 &     0 &   134 &       0 &     0 &     0 &     0 &     0 &     0 &     0 \\
	offshore wind (DC)            &       0 &     0 &     0 &     0 &     0 &     0 &     0 &      0 &    94 &    94 &   125 &   391 &   688 &   889 &       0 &     0 &     0 &     0 &     0 &     0 &     0 \\
	solar PV                      &     166 &   514 &   982 &  1264 &  1918 &  2887 &  4249 &    141 &   409 &   810 &  1045 &  1298 &  1582 &  3650 &     141 &   358 &   453 &   701 &   991 &  1474 &  3437 \\
\bottomrule
\end{tabular}

	}
	\caption{Energy balance [TWh/a] for the electricity transmission grid for the \ce{1.7} scenario and the three methods endogenous (\textit{endogen}), exogenous (\textit{exogen}) and sequential cost (\textit{seqcost}). See also Figure \ref{fig:eb_AC}.}
	\label{tab:energy_balance_AC_method}
\end{table*}

\begin{table*}[h]
	\resizebox{1\textwidth}{!}{ 
		\begin{tabular}{lrrrrrrrrrrrrrrrrrrrrr}
			\toprule
			budget & \multicolumn{7}{l}{\ce{1.5}} & \multicolumn{7}{l}{\ce{1.7}} & \multicolumn{7}{l}{\ce{2.0}}\\
			year &  2020 &  2025 &  2030 &  2035 &  2040 &  2045 &  2050 &  2020 &  2025 &  2030 &  2035 &  2040 &  2045 &  2050 &  2020 &  2025 &  2030 &  2035 &  2040 &  2045 &  2050 \\
			\midrule
			ground heat pump              &     0 &  -101 &  -126 &  -113 &   -71 &   -60 &   -42 &     0 &   -36 &   -82 &  -112 &   -71 &   -60 &   -42 &     0 &   -36 &   -72 &  -113 &   -71 &   -60 &   -48 \\
			air heat pump                 &  -368 &  -639 &  -791 &  -826 &  -824 &  -810 &  -719 &  -345 &  -499 &  -540 &  -539 &  -683 &  -658 &  -722 &  -280 &  -262 &  -220 &  -205 &  -305 &  -343 &  -568 \\
			resistive heater              &     0 &     0 &     0 &  -275 &  -539 &  -544 &  -555 &  -223 &  -228 &  -223 &  -494 &  -539 &  -544 &  -555 &  -356 &  -649 &  -981 & -1284 & -1269 & -1131 &  -805 \\
			BEV charger                   &     0 &  -204 &  -408 &  -663 &  -870 &  -868 &  -868 &     0 &  -204 &  -410 &  -666 &  -870 &  -869 &  -869 &     0 &  -204 &  -408 &  -663 &  -866 &  -869 &  -868 \\
			V2G                           &     0 &     0 &     0 &     0 &     3 &     1 &     1 &     0 &     0 &     2 &     3 &     3 &     2 &     2 &     0 &     0 &     0 &     0 &     0 &     2 &     1 \\
			electricity distribution grid &  3836 &  4411 &  4792 &  5347 &  5752 &  5736 &  5639 &  4037 &  4434 &  4720 &  5279 &  5612 &  5584 &  5587 &  4104 &  4617 &  5148 &  5736 &  5964 &  5855 &  5741 \\
\bottomrule
		\end{tabular}
		
	}
	\caption{Energy balance [TWh/a] for electricity distribution level for the three different budgets and the endogenous method. See also Figure \ref{fig:eb_low_voltage}.}
	\label{tab:energy_balance_low_voltage_endogen}
\end{table*}

\begin{table*}[h]
	\resizebox{1\textwidth}{!}{ 
\begin{tabular}{lrrrrrrrrrrrrrrrrrrrrr}
	\toprule
	method & \multicolumn{7}{l}{endogen} & \multicolumn{7}{l}{exogen} & \multicolumn{7}{l}{seqcost} \\
	year &    2020 &  2025 &  2030 &  2035 &  2040 &  2045 &  2050 &   2020 &  2025 &  2030 &  2035 &  2040 &  2045 &  2050 &    2020 &  2025 &  2030 &  2035 &  2040 &  2045 &  2050 \\
	\midrule
	ground heat pump              &       0 &   -36 &   -82 &  -112 &   -71 &   -60 &   -42 &      0 &   -47 &   -93 &  -112 &   -71 &   -60 &   -43 &       0 &   -49 &   -98 &  -112 &   -71 &   -60 &   -43 \\
	air heat pump                 &    -345 &  -499 &  -540 &  -539 &  -683 &  -658 &  -722 &   -447 &  -607 &  -617 &  -679 &  -749 &  -713 &  -750 &    -369 &  -596 &  -622 &  -678 &  -753 &  -709 &  -736 \\
	resistive heater              &    -223 &  -228 &  -223 &  -494 &  -539 &  -544 &  -555 &      0 &     0 &     0 &  -270 &  -539 &  -544 &  -555 &       0 &     0 &     0 &  -270 &  -539 &  -551 &  -596 \\
	BEV charger                   &       0 &  -204 &  -410 &  -666 &  -870 &  -869 &  -869 &      0 &  -204 &  -409 &  -663 &  -867 &  -866 &  -870 &       0 &  -204 &  -408 &  -663 &  -866 &  -866 &  -870 \\
	V2G                           &       0 &     0 &     2 &     3 &     3 &     2 &     2 &      0 &     0 &     1 &     0 &     1 &     0 &     3 &       0 &     0 &     0 &     0 &     0 &     0 &     3 \\
	electricity distribution grid &    4037 &  4434 &  4720 &  5279 &  5612 &  5584 &  5587 &   3915 &  4326 &  4585 &  5149 &  5569 &  5530 &  5557 &    3837 &  4316 &  4464 &  5062 &  5573 &  5532 &  5588 \\
\bottomrule
\end{tabular}
		
	}
	\caption{Energy balance [TWh/a] for electricity distribution grid for the \ce{1.7} scenario and the three methods endogenous (\textit{endogen}), exogenous (\textit{exogen}) and sequential cost (\textit{seqcost}). See also Figure \ref{fig:eb_low_voltage}.}
	\label{tab:energy_balance_low_voltage_method}
\end{table*}

\begin{table*}[h]
	\resizebox{1\textwidth}{!}{ 
\begin{tabular}{lrrrrrrrrrrrrrrrrrrrrr}
	\toprule
	budget & \multicolumn{7}{l}{\ce{1.5}} & \multicolumn{7}{l}{\ce{1.7}} & \multicolumn{7}{l}{\ce{2.0}}\\
	year &  2020 &  2025 &  2030 & 2035 & 2040 & 2045 & 2050 &  2020 &  2025 &  2030 &  2035 &  2040 & 2045 & 2050 &  2020 &  2025 &  2030 &  2035 &  2040 &  2045 & 2050 \\
	\midrule
	CHP                           &     0 &     0 &     0 &    0 &    0 &   -1 & -109 &    63 &   241 &   258 &   297 &   283 &  234 &   58 &     0 &   365 &   416 &   441 &   544 &   517 &  -29 \\
	OCGT                          &    37 &     3 &     3 &   32 &   30 &   26 &   27 &    54 &     8 &    13 &   109 &   134 &   67 &   28 &   108 &    50 &   137 &   268 &   337 &   210 &   31 \\
	gas boiler                    &   442 &   127 &    85 &   35 &    5 &    0 &    0 &   375 &   191 &   138 &    50 &     1 &    0 &    0 &   377 &   231 &   145 &    59 &    11 &     0 &    0 \\
	CCGT                          &   394 &    52 &    31 &    5 &    1 &    0 &    0 &   389 &   136 &    66 &    18 &     3 &    0 &    0 &   395 &   170 &   115 &    30 &     4 &     0 &    0 \\
	DAC                           &     0 &     0 &     0 &    0 &  -22 & -102 & -118 &     0 &     0 &     0 &     0 &     0 &    0 & -212 &     0 &     0 &     0 &     0 &     0 &     0 & -127 \\
	SMR                           &    28 &    13 &     6 &    4 &    5 &    2 &    0 &    28 &    41 &    37 &    30 &    17 &   18 &    0 &    28 &    43 &    70 &   110 &   119 &    80 &    0 \\
\bottomrule
\end{tabular}

	}
	\caption{Energy balance [Mt CO$_2$/a] for carbon dioxide for the three different budgets and the endogenous method. See also Figure \ref{fig:eb_co2}.}
	\label{tab:energy_balance_co2_endogen}
\end{table*}

\begin{table*}[h]
	\resizebox{1\textwidth}{!}{ 
\begin{tabular}{lrrrrrrrrrrrrrrrrrrrrr}
	\toprule
	method & \multicolumn{7}{l}{endogen} & \multicolumn{7}{l}{exogen} & \multicolumn{7}{l}{seqcost} \\
	year &    2020 &  2025 &  2030 &  2035 &  2040 & 2045 & 2050 &   2020 &  2025 &  2030 &  2035 &  2040 & 2045 & 2050 &    2020 &  2025 &  2030 &  2035 &  2040 & 2045 & 2050 \\
	\midrule
	CHP                           &      63 &   241 &   258 &   297 &   283 &  234 &   58 &     23 &   182 &   186 &   202 &   211 &  195 &    8 &       2 &   169 &   172 &   204 &   208 &  195 &   13 \\
	OCGT                          &      54 &     8 &    13 &   109 &   134 &   67 &   28 &     58 &    12 &    30 &   134 &   175 &  133 &   38 &      59 &    16 &    29 &   122 &   148 &   96 &   31 \\
	gas boiler                    &     375 &   191 &   138 &    50 &     1 &    0 &    0 &    376 &   176 &   124 &    46 &     3 &    0 &    0 &     441 &   189 &   121 &    46 &     3 &    0 &    0 \\
	CCGT                          &     389 &   136 &    66 &    18 &     3 &    0 &    0 &    419 &   132 &    77 &    18 &     3 &    0 &    0 &     374 &   150 &    86 &    20 &     3 &    0 &    0 \\
	DAC                           &       0 &     0 &     0 &     0 &     0 &    0 & -212 &      0 &     0 &     0 &     0 &     0 &    0 & -171 &       0 &     0 &     0 &     0 &     0 &    0 & -170 \\
	SMR                           &      28 &    41 &    37 &    30 &    17 &   18 &    0 &     28 &    43 &    61 &    57 &    39 &   28 &    0 &      28 &    42 &    69 &   107 &    56 &   29 &    0 \\
\bottomrule
\end{tabular}
		
	}
	\caption{Energy balance [Mt CO$_2$/a] for carbon dioxide for the \ce{1.7} scenario and the three methods endogenous (\textit{endogen}), exogenous (\textit{exogen}) and sequential cost (\textit{seqcost}). See also Figure \ref{fig:eb_co2}.}
	\label{tab:energy_balance_co2_method}
\end{table*}

\begin{table*}[h]
	\resizebox{1\textwidth}{!}{ 
		\begin{tabular}{lrrrrrrrrrrrrrrrrrrrrr}
			\toprule
			budget & \multicolumn{7}{l}{\ce{1.5}} & \multicolumn{7}{l}{\ce{1.7}} & \multicolumn{7}{l}{\ce{2.0}}\\
			year &  2020 & 2025 & 2030 & 2035 & 2040 & 2045 & 2050 &  2020 &  2025 &  2030 &  2035 &  2040 &  2045 & 2050 &  2020 &  2025 &  2030 &  2035 &  2040 &  2045 & 2050 \\
			\midrule
			CHP                 &     0 &    0 &    0 &    0 &    0 &    0 &    0 &  -315 & -1205 & -1290 & -1483 & -1413 & -1171 & -288 &     0 & -1823 & -2080 & -2203 & -2718 & -2583 & -432 \\
			OCGT                &  -185 &  -13 &  -17 & -162 & -150 & -131 & -135 &  -272 &   -40 &   -67 &  -546 &  -671 &  -333 & -139 &  -541 &  -249 &  -687 & -1338 & -1687 & -1050 & -153 \\
			gas boiler          & -2210 & -633 & -427 & -176 &  -27 &    0 &    0 & -1875 &  -953 &  -691 &  -252 &    -7 &     0 &    0 & -1884 & -1157 &  -725 &  -295 &   -53 &     0 &    0 \\
			gas                 &  4888 &  999 &  279 &    0 &    0 &    0 &    0 &  4930 &  3461 &  2947 &  2902 &  2523 &  1906 &  362 &  4920 &  4673 &  4797 &  4917 &  5457 &  4417 &  363 \\
			methanation         &     0 &  383 &  754 &  809 &  629 &  566 &  561 &     0 &     0 &     0 &     0 &    46 &    67 &  489 &     0 &     0 &     0 &     0 &     0 &     0 &  645 \\
			CCGT                & -1971 & -262 & -154 &  -27 &   -3 &    0 &    0 & -1946 &  -679 &  -331 &   -91 &   -13 &     0 &    0 & -1973 &  -849 &  -575 &  -151 &   -21 &     0 &    0 \\
\bottomrule
		\end{tabular}

	}
	\caption{Energy balance [TWh/a] for gas for the three different budgets and the endogenous method. See also Figure \ref{fig:eb_gas}.}
	\label{tab:energy_balance_gas_endogen}
\end{table*}

\begin{table*}[h]
	\resizebox{1\textwidth}{!}{ 
		\begin{tabular}{lrrrrrrrrrrrrrrrrrrrrr}
			\toprule
			method & \multicolumn{7}{l}{endogen} & \multicolumn{7}{l}{exogen} & \multicolumn{7}{l}{seqcost} \\
			year &    2020 &  2025 &  2030 &  2035 &  2040 &  2045 & 2050 &   2020 &  2025 &  2030 &  2035 &  2040 &  2045 & 2050 &    2020 &  2025 &  2030 &  2035 &  2040 &  2045 & 2050 \\
			\midrule
			CHP                 &    -315 & -1205 & -1290 & -1483 & -1413 & -1171 & -288 &   -115 &  -908 &  -930 & -1012 & -1053 &  -977 & -141 &     -10 &  -843 &  -861 & -1020 & -1040 &  -974 & -199 \\
			OCGT                &    -272 &   -40 &   -67 &  -546 &  -671 &  -333 & -139 &   -289 &   -60 &  -149 &  -669 &  -873 &  -666 & -188 &    -295 &   -79 &  -146 &  -610 &  -741 &  -481 & -154 \\
			gas boiler          &   -1875 &  -953 &  -691 &  -252 &    -7 &     0 &    0 &  -1879 &  -880 &  -621 &  -232 &   -15 &     0 &    0 &   -2207 &  -945 &  -605 &  -232 &   -15 &     0 &    0 \\
			gas                 &    4930 &  3461 &  2947 &  2902 &  2523 &  1906 &  362 &   4900 &  3105 &  2767 &  2671 &  2535 &  2163 &  363 &    4903 &  3212 &  2765 &  2877 &  2472 &  1970 &  363 \\
			methanation         &       0 &     0 &     0 &     0 &    46 &    67 &  489 &      0 &     0 &     0 &     0 &     0 &     0 &  390 &       0 &     0 &     0 &     0 &     0 &     9 &  413 \\
			CCGT                &   -1946 &  -679 &  -331 &   -91 &   -13 &     0 &    0 &  -2094 &  -662 &  -383 &   -90 &   -15 &     0 &    0 &   -1869 &  -752 &  -429 &  -101 &   -14 &     0 &    0 \\
\bottomrule
		\end{tabular}

	}
	\caption{Energy balance [TWh/a] for gas for the \ce{1.7} scenario and the three methods endogenous (\textit{endogen}), exogenous (\textit{exogen}) and sequential cost (\textit{seqcost}). See also Figure \ref{fig:eb_gas}.}
	\label{tab:energy_balance_gas_method}
\end{table*}
\begin{table*}[h]
	\resizebox{1\textwidth}{!}{ 
\begin{tabular}{lrrrrrrrrrrrrrrrrrrrrr}
	\toprule
	budget & \multicolumn{7}{l}{\ce{1.5}} & \multicolumn{7}{l}{\ce{1.7}} & \multicolumn{7}{l}{\ce{2.0}}\\
	year & 2020 & 2025 & 2030 & 2035 & 2040 & 2045 & 2050 & 2020 & 2025 & 2030 & 2035 & 2040 & 2045 & 2050 & 2020 & 2025 & 2030 & 2035 & 2040 & 2045 & 2050 \\
	\midrule
	gas for industry    &    0 & -229 & -229 & -381 & -381 & -381 & -381 &    0 &    0 &    0 &    0 &    0 &    0 & -381 &    0 &    0 &    0 &    0 &    0 &    0 & -381 \\
	gas for industry CC &    0 &  229 &  229 &  381 &  381 &  381 &  381 &    0 &    0 &    0 &    0 &    0 &    0 &  381 &    0 &    0 &    0 &    0 &    0 &    0 &  381 \\
	\bottomrule
\end{tabular}
	
	}
	\caption{Energy balance [TWh/a] for gas for industry for the three different budgets and the endogenous method. See also Figure \ref{fig:eb_gas_for_industry}.}
	\label{tab:energy_balance_gas_for_industry_endogen}
\end{table*}

\begin{table*}[h]
	\resizebox{1\textwidth}{!}{ 
\begin{tabular}{lrrrrrrrrrrrrrrrrrrrrr}
	\toprule
	method & \multicolumn{7}{l}{endogen} & \multicolumn{7}{l}{exogen} & \multicolumn{7}{l}{seqcost} \\
	year &    2020 & 2025 & 2030 & 2035 & 2040 & 2045 & 2050 &   2020 & 2025 & 2030 & 2035 & 2040 & 2045 & 2050 &    2020 & 2025 & 2030 & 2035 & 2040 & 2045 & 2050 \\
	\midrule
	gas for industry    &       0 &    0 &    0 &    0 &    0 &    0 & -381 &      0 &    0 &    0 &    0 &    0 &    0 & -381 &       0 &    0 &    0 &    0 &    0 &    0 & -381 \\
	gas for industry CC &       0 &    0 &    0 &    0 &    0 &    0 &  381 &      0 &    0 &    0 &    0 &    0 &    0 &  381 &       0 &    0 &    0 &    0 &    0 &    0 &  381 \\
	\bottomrule
\end{tabular}

	}
	\caption{Energy balance [TWh/a] for gas for industry for the \ce{1.7} scenario and the three methods endogenous (\textit{endogen}), exogenous (\textit{exogen}) and sequential cost (\textit{seqcost}). See also Figure \ref{fig:eb_gas_for_industry}.}
	\label{tab:energy_balance_gas_for_industry_method}
\end{table*}
\begin{table*}[h]
	\resizebox{1\textwidth}{!}{ 
		\begin{tabular}{lrrrrrrrrrrrrrrrrrrrrr}
			\toprule
			budget & \multicolumn{7}{l}{\ce{1.5}} & \multicolumn{7}{l}{\ce{1.7}} & \multicolumn{7}{l}{\ce{2.0}}\\
			year & 2020 & 2025 & 2030 & 2035 & 2040 & 2045 & 2050 & 2020 & 2025 & 2030 & 2035 & 2040 & 2045 & 2050 & 2020 & 2025 & 2030 & 2035 & 2040 & 2045 & 2050 \\
			\midrule
			BEV charger &    0 &  184 &  367 &  596 &  783 &  781 &  781 &    0 &  184 &  369 &  599 &  783 &  782 &  782 &    0 &  183 &  367 &  597 &  780 &  782 &  781 \\
			Li ion      &    0 & -183 & -367 & -596 & -780 & -780 & -780 &    0 & -183 & -367 & -596 & -780 & -780 & -780 &    0 & -183 & -367 & -596 & -780 & -780 & -780 \\
			V2G         &    0 &   -1 &    0 &    0 &   -3 &   -1 &   -1 &    0 &   -1 &   -2 &   -3 &   -3 &   -3 &   -2 &    0 &    0 &    0 &    0 &    0 &   -2 &   -2 \\
			\bottomrule
		\end{tabular}

	}
	\caption{Energy balance [TWh/a] for electric vehicles for the three different budgets and the endogenous method. See also Figure \ref{fig:eb_Li_ion}.}
	\label{tab:energy_balance_Li_ion_endogen}
\end{table*}

\begin{table*}[h]
	\resizebox{1\textwidth}{!}{ 
		\begin{tabular}{lrrrrrrrrrrrrrrrrrrrrr}
			\toprule
			method & \multicolumn{7}{l}{endogen} & \multicolumn{7}{l}{exogen} & \multicolumn{7}{l}{seqcost} \\
			year &    2020 & 2025 & 2030 & 2035 & 2040 & 2045 & 2050 &   2020 & 2025 & 2030 & 2035 & 2040 & 2045 & 2050 &    2020 & 2025 & 2030 & 2035 & 2040 & 2045 & 2050 \\
			\midrule
			BEV charger &       0 &  184 &  369 &  599 &  783 &  782 &  782 &      0 &  184 &  368 &  596 &  781 &  780 &  783 &       0 &  184 &  367 &  596 &  780 &  780 &  783 \\
			Li ion      &       0 & -183 & -367 & -596 & -780 & -780 & -780 &      0 & -183 & -367 & -596 & -780 & -780 & -780 &       0 & -183 & -367 & -596 & -780 & -780 & -780 \\
			V2G         &       0 &   -1 &   -2 &   -3 &   -3 &   -3 &   -2 &      0 &   -1 &   -2 &    0 &   -1 &    0 &   -3 &       0 &    0 &    0 &    0 &    0 &    0 &   -3 \\
			\bottomrule
		\end{tabular}

	}
	\caption{Energy balance [TWh/a] for electric vehicles for the \ce{1.7} scenario and the three methods endogenous (\textit{endogen}), exogenous (\textit{exogen}) and sequential cost (\textit{seqcost}). See also Figure \ref{fig:eb_Li_ion}.}
	\label{tab:energy_balance_Li_ion_method}
\end{table*}
\begin{table*}[h]
	\resizebox{1\textwidth}{!}{ 
		\begin{tabular}{lrrrrrrrrrrrrrrrrrrrrr}
			\toprule
			budget & \multicolumn{7}{l}{\ce{1.5}} & \multicolumn{7}{l}{\ce{1.7}} & \multicolumn{7}{l}{\ce{2.0}}\\
			year &  2020 &  2025 &  2030 &  2035 &  2040 &  2045 &  2050 &  2020 &  2025 &  2030 &  2035 &  2040 &  2045 &  2050 &  2020 &  2025 &  2030 &  2035 &  2040 &  2045 &  2050 \\
			\midrule
			Fischer-Tropsch       &     0 &     0 &     0 &   692 &  1411 &  1592 &  1341 &     0 &     0 &     0 &     0 &    51 &   489 &  1341 &     0 &     0 &     0 &     0 &     0 &     0 &  1341 \\
			imported oil          &  5103 &  4344 &  3589 &  1909 &   377 &     0 &     0 &  5072 &  4362 &  3613 &  2606 &  1731 &  1103 &     0 &  5080 &  4372 &  3621 &  2601 &  1784 &  1592 &     0 \\
			kerosene for aviation &  -612 &  -612 &  -612 &  -612 &  -612 &  -612 &  -612 &  -612 &  -612 &  -612 &  -612 &  -612 &  -612 &  -612 &  -612 &  -612 &  -612 &  -612 &  -612 &  -612 &  -612 \\
			land transport oil    & -3056 & -2355 & -1621 &  -703 &     0 &     0 &     0 & -3056 & -2355 & -1621 &  -703 &     0 &     0 &     0 & -3056 & -2355 & -1621 &  -703 &     0 &     0 &     0 \\
			naphtha for industry  &  -728 &  -728 &  -728 &  -728 &  -728 &  -728 &  -728 &  -728 &  -728 &  -728 &  -728 &  -728 &  -728 &  -728 &  -728 &  -728 &  -728 &  -728 &  -728 &  -728 &  -728 \\
			oil boiler            &   -78 &   -20 &   -30 &   -22 &    -8 &     0 &     0 &   -46 &   -38 &   -54 &   -28 &    -1 &     0 &     0 &   -54 &   -48 &   -63 &   -23 &    -3 &     0 &     0 \\
			shipping oil          &  -629 &  -629 &  -597 &  -534 &  -440 &  -251 &     0 &  -629 &  -629 &  -597 &  -534 &  -440 &  -251 &     0 &  -629 &  -629 &  -597 &  -534 &  -440 &  -251 &     0 \\
			\bottomrule
		\end{tabular}
		
	}
	\caption{Energy balance [TWh/a] for oil for the three different budgets and the endogenous method. See also Figure \ref{fig:eb_oil}.}
	\label{tab:energy_balance_oil_endogen}
\end{table*}

\begin{table*}[h]
	\resizebox{1\textwidth}{!}{ 
		\begin{tabular}{lrrrrrrrrrrrrrrrrrrrrr}
			\toprule
			method & \multicolumn{7}{l}{endogen} & \multicolumn{7}{l}{exogen} & \multicolumn{7}{l}{seqcost} \\
			year &    2020 &  2025 &  2030 &  2035 &  2040 &  2045 &  2050 &   2020 &  2025 &  2030 &  2035 &  2040 &  2045 &  2050 &    2020 &  2025 &  2030 &  2035 &  2040 &  2045 &  2050 \\
			\midrule
			Fischer-Tropsch       &       0 &     0 &     0 &     0 &    51 &   489 &  1341 &      0 &     0 &     0 &     0 &     0 &   138 &  1341 &       0 &     0 &     0 &     0 &     0 &   178 &  1341 \\
			imported oil          &    5072 &  4362 &  3613 &  2606 &  1731 &  1103 &     0 &   5095 &  4352 &  3613 &  2606 &  1782 &  1454 &     0 &    5092 &  4357 &  3609 &  2605 &  1782 &  1414 &     0 \\
			kerosene for aviation &    -612 &  -612 &  -612 &  -612 &  -612 &  -612 &  -612 &   -612 &  -612 &  -612 &  -612 &  -612 &  -612 &  -612 &    -612 &  -612 &  -612 &  -612 &  -612 &  -612 &  -612 \\
			land transport oil    &   -3056 & -2355 & -1621 &  -703 &     0 &     0 &     0 &  -3056 & -2355 & -1621 &  -703 &     0 &     0 &     0 &   -3056 & -2355 & -1621 &  -703 &     0 &     0 &     0 \\
			naphtha for industry  &    -728 &  -728 &  -728 &  -728 &  -728 &  -728 &  -728 &   -728 &  -728 &  -728 &  -728 &  -728 &  -728 &  -728 &    -728 &  -728 &  -728 &  -728 &  -728 &  -728 &  -728 \\
			oil boiler            &     -46 &   -38 &   -54 &   -28 &    -1 &     0 &     0 &    -69 &   -28 &   -54 &   -27 &    -1 &     0 &     0 &     -66 &   -32 &   -50 &   -26 &    -1 &     0 &     0 \\
			shipping oil          &    -629 &  -629 &  -597 &  -534 &  -440 &  -251 &     0 &   -629 &  -629 &  -597 &  -534 &  -440 &  -251 &     0 &    -629 &  -629 &  -597 &  -534 &  -440 &  -251 &     0 \\
			\bottomrule
		\end{tabular}

	}
	\caption{Energy balance [TWh/a] for oil for the \ce{1.7} scenario and the three methods endogenous (\textit{endogen}), exogenous (\textit{exogen}) and sequential cost (\textit{seqcost}). See also Figure \ref{fig:eb_oil}.}
	\label{tab:energy_balance_oil_method}
\end{table*}
\begin{table*}[h]
	\resizebox{1\textwidth}{!}{ 
		\begin{tabular}{lrrrrrrrrrrrrrrrrrrrrr}
			\toprule
			budget & \multicolumn{7}{l}{\ce{1.5}} & \multicolumn{7}{l}{\ce{1.7}} & \multicolumn{7}{l}{\ce{2.0}}\\
			year & 2020 &  2025 &  2030 &  2035 &  2040 &  2045 &  2050 & 2020 & 2025 & 2030 &  2035 &  2040 &  2045 &  2050 & 2020 & 2025 & 2030 & 2035 & 2040 & 2045 &  2050 \\
			\midrule
			solid biomass                 &  702 &  1186 &  1186 &  1186 &  1186 &  1186 &  1186 &  702 &  702 &  702 &  1039 &  1182 &  1122 &  1186 &  702 &  702 &  702 &  702 &  702 &  702 &  1186 \\
			CHP                           &    0 &  -483 &  -483 &  -483 &  -405 &  -405 &  -405 &    0 &    0 &    0 &  -337 &  -479 &  -420 &  -405 &    0 &    0 &    0 &    0 &    0 &    0 &  -405 \\
			solid biomass for industry    & -702 &  -702 &  -702 &  -702 &     0 &     0 &     0 & -702 & -702 & -702 &  -702 &  -702 &  -702 &     0 & -702 & -702 & -702 & -702 & -702 & -702 &     0 \\
			solid biomass for industry CC &    0 &     0 &     0 &     0 &  -780 &  -780 &  -780 &    0 &    0 &    0 &     0 &     0 &     0 &  -780 &    0 &    0 &    0 &    0 &    0 &    0 &  -780 \\
			\bottomrule
		\end{tabular}

	}
	\caption{Energy balance [TWh/a] for solid biomass for the three different budgets and the endogenous method. See also Figure \ref{fig:eb_solid_biomass}.}
	\label{tab:energy_balance_solid_biomass_endogen}
\end{table*}

\begin{table*}[h]
	\resizebox{1\textwidth}{!}{ 
		\begin{tabular}{lrrrrrrrrrrrrrrrrrrrrr}
			\toprule
			method & \multicolumn{7}{l}{endogen} & \multicolumn{7}{l}{exogen} & \multicolumn{7}{l}{seqcost} \\
			year &    2020 & 2025 & 2030 &  2035 &  2040 &  2045 &  2050 &   2020 & 2025 &  2030 &  2035 &  2040 &  2045 &  2050 &    2020 &  2025 &  2030 &  2035 &  2040 &  2045 &  2050 \\
			\midrule
			solid biomass                 &     702 &  702 &  702 &  1039 &  1182 &  1122 &  1186 &    702 &  929 &  1097 &  1186 &  1186 &  1186 &  1186 &     702 &  1010 &  1184 &  1186 &  1186 &  1186 &  1186 \\
			CHP                           &       0 &    0 &    0 &  -337 &  -479 &  -420 &  -405 &      0 & -227 &  -395 &  -483 &  -483 &  -483 &  -405 &       0 &  -308 &  -481 &  -483 &  -483 &  -483 &  -405 \\
			solid biomass for industry    &    -702 & -702 & -702 &  -702 &  -702 &  -702 &     0 &   -702 & -702 &  -702 &  -702 &  -702 &  -702 &     0 &    -702 &  -702 &  -702 &  -702 &  -702 &  -702 &     0 \\
			solid biomass for industry CC &       0 &    0 &    0 &     0 &     0 &     0 &  -780 &      0 &    0 &     0 &     0 &     0 &     0 &  -780 &       0 &     0 &     0 &     0 &     0 &     0 &  -780 \\
			\bottomrule
		\end{tabular}

	}
	\caption{Energy balance [TWh/a] for solid biomass for industry for the \ce{1.7} scenario and the three methods endogenous (\textit{endogen}), exogenous (\textit{exogen}) and sequential cost (\textit{seqcost}). See also Figure \ref{fig:eb_solid_biomass}.}
	\label{tab:energy_balance_solid_biomass_method}
\end{table*}
\begin{table*}[h]
	\resizebox{1\textwidth}{!}{ 
		\begin{tabular}{lrrrrrrrrrrrrrrrrrrrrr}
			\toprule
			budget & \multicolumn{7}{l}{\ce{1.5}} & \multicolumn{7}{l}{\ce{1.7}} & \multicolumn{7}{l}{\ce{2.0}}\\
			year & 2020 & 2025 & 2030 & 2035 & 2040 & 2045 & 2050 & 2020 & 2025 & 2030 & 2035 & 2040 & 2045 & 2050 & 2020 & 2025 & 2030 & 2035 & 2040 & 2045 & 2050 \\
			\midrule
			solid biomass for industry    &    0 &    0 &    0 &    0 & -702 & -702 & -702 &    0 &    0 &    0 &    0 &    0 &    0 & -702 &    0 &    0 &    0 &    0 &    0 &    0 & -702 \\
			solid biomass for industry CC &    0 &    0 &    0 &    0 &  702 &  702 &  702 &    0 &    0 &    0 &    0 &    0 &    0 &  702 &    0 &    0 &    0 &    0 &    0 &    0 &  702 \\
			\bottomrule
		\end{tabular}

	}
	\caption{Energy balance [TWh/a] for solid biomass for industry for the three different budgets and the endogenous method. See also Figure \ref{fig:eb_solid_biomass_for_industry}.}
	\label{tab:energy_balance_solid_biomass_for_industry_endogen}
\end{table*}

\begin{table*}[h]
	\resizebox{1\textwidth}{!}{ 
		\begin{tabular}{lrrrrrrrrrrrrrrrrrrrrr}
			\toprule
			method & \multicolumn{7}{l}{endogen} & \multicolumn{7}{l}{exogen} & \multicolumn{7}{l}{seqcost} \\
			year &    2020 & 2025 & 2030 & 2035 & 2040 & 2045 & 2050 &   2020 & 2025 & 2030 & 2035 & 2040 & 2045 & 2050 &    2020 & 2025 & 2030 & 2035 & 2040 & 2045 & 2050 \\
			\midrule
			solid biomass for industry    &       0 &    0 &    0 &    0 &    0 &    0 & -702 &      0 &    0 &    0 &    0 &    0 &    0 & -702 &       0 &    0 &    0 &    0 &    0 &    0 & -702 \\
			solid biomass for industry CC &       0 &    0 &    0 &    0 &    0 &    0 &  702 &      0 &    0 &    0 &    0 &    0 &    0 &  702 &       0 &    0 &    0 &    0 &    0 &    0 &  702 \\
			\bottomrule
		\end{tabular}

	}
	\caption{Energy balance [TWh/a] for solid biomass for the \ce{1.7} scenario and the three methods endogenous (\textit{endogen}), exogenous (\textit{exogen}) and sequential cost (\textit{seqcost}). See also Figure \ref{fig:eb_solid_biomass_for_industry}.}
	\label{tab:energy_balance_solid_biomass_for_industry_method}
\end{table*}
\begin{table*}[h]
	\resizebox{1\textwidth}{!}{ 
		\begin{tabular}{lrrrrrrrrrrrrrrrrrrrrr}
			\toprule
			budget & \multicolumn{7}{l}{\ce{1.5}} & \multicolumn{7}{l}{\ce{1.7}} & \multicolumn{7}{l}{\ce{2.0}}\\
			year & 2020 & 2025 & 2030 & 2035 & 2040 & 2045 & 2050 & 2020 & 2025 & 2030 & 2035 & 2040 & 2045 & 2050 & 2020 & 2025 & 2030 & 2035 & 2040 & 2045 & 2050 \\
			\midrule
			process emissions    &    0 &  166 &  166 &  166 &  162 &  157 &  153 &    0 &    0 &    0 &    0 &   24 &  153 &  153 &    0 &    0 &    0 &    0 &    0 &    0 &  153 \\
			process emissions CC &    0 & -166 & -166 & -166 & -162 & -157 & -153 &    0 &    0 &    0 &    0 &  -24 & -153 & -153 &    0 &    0 &    0 &    0 &    0 &    0 & -153 \\
			\bottomrule
		\end{tabular}

	}
	\caption{Energy balance [TWh/a] for process emissions for the three different budgets and the endogenous method. See also Figure \ref{fig:eb_process_emissions}.}
	\label{tab:energy_balance_process_emissions_endogen}
\end{table*}

\begin{table*}[h]
	\resizebox{1\textwidth}{!}{ 
		\begin{tabular}{lrrrrrrrrrrrrrrrrrrrrr}
			\toprule
			method & \multicolumn{7}{l}{endogen} & \multicolumn{7}{l}{exogen} & \multicolumn{7}{l}{seqcost} \\
			year &    2020 & 2025 & 2030 & 2035 & 2040 & 2045 & 2050 &   2020 & 2025 & 2030 & 2035 & 2040 & 2045 & 2050 &    2020 & 2025 & 2030 & 2035 & 2040 & 2045 & 2050 \\
			\midrule
			process emissions    &       0 &    0 &    0 &    0 &   24 &  153 &  153 &      0 &    0 &    0 &    0 &    0 &   39 &  153 &       0 &    0 &    0 &    0 &    0 &   52 &  153 \\
			process emissions CC &       0 &    0 &    0 &    0 &  -24 & -153 & -153 &      0 &    0 &    0 &    0 &    0 &  -39 & -153 &       0 &    0 &    0 &    0 &    0 &  -52 & -153 \\
			\bottomrule
		\end{tabular}

	}
	\caption{Energy balance [TWh/a] for process emissions for the \ce{1.7} scenario and the three methods endogenous (\textit{endogen}), exogenous (\textit{exogen}) and sequential cost (\textit{seqcost}). See also Figure \ref{fig:eb_process_emissions}.}
	\label{tab:energy_balance_process_emissions_method}
\end{table*}
\begin{table*}[h]
	\resizebox{1\textwidth}{!}{ 
		\begin{tabular}{lrrrrrrrrrrrrrrrrrrrrr}
			\toprule
			budget & \multicolumn{7}{l}{\ce{1.5}} & \multicolumn{7}{l}{\ce{1.7}} & \multicolumn{7}{l}{\ce{2.0}}\\
			year & 2020 & 2025 & 2030 & 2035 & 2040 & 2045 & 2050 & 2020 & 2025 & 2030 & 2035 & 2040 & 2045 & 2050 & 2020 & 2025 & 2030 & 2035 & 2040 & 2045 & 2050 \\
			\midrule
			CHP                           &    0 &    0 &    0 &    0 &    0 &    1 &  109 &    0 &    0 &    0 &    0 &    0 &    0 &    0 &    0 &    0 &    0 &    0 &    0 &    0 &  115 \\
			methanation                   &    0 &  -77 & -151 & -162 & -126 & -113 & -112 &    0 &    0 &    0 &    0 &   -9 &  -13 &  -98 &    0 &    0 &    0 &    0 &    0 &    0 & -129 \\
			CO2 sequestration             &    0 & -119 &  -44 &  123 &   40 &    0 & -200 &    0 &    0 &    0 &    0 &    0 &    0 & -200 &    0 &    0 &    0 &    0 &    0 &    0 & -200 \\
			DAC                           &    0 &    0 &    0 &    0 &   22 &  102 &  118 &    0 &    0 &    0 &    0 &    0 &    0 &  212 &    0 &    0 &    0 &    0 &    0 &    0 &  127 \\
			Fischer-Tropsch               &    0 &    0 &    0 & -187 & -381 & -430 & -362 &    0 &    0 &    0 &    0 &  -14 & -132 & -362 &    0 &    0 &    0 &    0 &    0 &    0 & -362 \\
			gas for industry CC           &    0 &   46 &   46 &   77 &   77 &   77 &   79 &    0 &    0 &    0 &    0 &    0 &    0 &   80 &    0 &    0 &    0 &    0 &    0 &    0 &   80 \\
			process emissions CC          &    0 &  149 &  149 &  149 &  145 &  141 &  145 &    0 &    0 &    0 &    0 &   23 &  145 &  145 &    0 &    0 &    0 &    0 &    0 &    0 &  145 \\
			solid biomass for industry CC &    0 &    0 &    0 &    0 &  222 &  222 &  222 &    0 &    0 &    0 &    0 &    0 &    0 &  222 &    0 &    0 &    0 &    0 &    0 &    0 &  222 \\
			\bottomrule
		\end{tabular}

	}
	\caption{Energy balance [Mt CO$_2$/a] for carbon dioxide storage for the three different budgets and the endogenous method. See also Figure \ref{fig:eb_co2_stored}.}
	\label{tab:energy_balance_co2_stored_endogen}
\end{table*}

\begin{table*}[h]
	\resizebox{1\textwidth}{!}{ 
		\begin{tabular}{lrrrrrrrrrrrrrrrrrrrrr}
			\toprule
			method & \multicolumn{7}{l}{endogen} & \multicolumn{7}{l}{exogen} & \multicolumn{7}{l}{seqcost} \\
			year &    2020 & 2025 & 2030 & 2035 & 2040 & 2045 & 2050 &   2020 & 2025 & 2030 & 2035 & 2040 & 2045 & 2050 &    2020 & 2025 & 2030 & 2035 & 2040 & 2045 & 2050 \\
			\midrule
			CHP                           &       0 &    0 &    0 &    0 &    0 &    0 &    0 &      0 &    0 &    0 &    0 &    0 &    0 &   21 &       0 &    0 &    0 &    0 &    0 &    0 &   27 \\
			methanation                   &       0 &    0 &    0 &    0 &   -9 &  -13 &  -98 &      0 &    0 &    0 &    0 &    0 &    0 &  -78 &       0 &    0 &    0 &    0 &    0 &   -2 &  -83 \\
			CO2 sequestration             &       0 &    0 &    0 &    0 &    0 &    0 & -200 &      0 &    0 &    0 &    0 &    0 &    0 & -200 &       0 &    0 &    0 &    0 &    0 &    0 & -200 \\
			DAC                           &       0 &    0 &    0 &    0 &    0 &    0 &  212 &      0 &    0 &    0 &    0 &    0 &    0 &  171 &       0 &    0 &    0 &    0 &    0 &    0 &  170 \\
			Fischer-Tropsch               &       0 &    0 &    0 &    0 &  -14 & -132 & -362 &      0 &    0 &    0 &    0 &    0 &  -37 & -362 &       0 &    0 &    0 &    0 &    0 &  -48 & -362 \\
			gas for industry CC           &       0 &    0 &    0 &    0 &    0 &    0 &   80 &      0 &    0 &    0 &    0 &    0 &    0 &   80 &       0 &    0 &    0 &    0 &    0 &    0 &   80 \\
			process emissions CC          &       0 &    0 &    0 &    0 &   23 &  145 &  145 &      0 &    0 &    0 &    0 &    0 &   37 &  145 &       0 &    0 &    0 &    0 &    0 &   50 &  145 \\
			solid biomass for industry CC &       0 &    0 &    0 &    0 &    0 &    0 &  222 &      0 &    0 &    0 &    0 &    0 &    0 &  222 &       0 &    0 &    0 &    0 &    0 &    0 &  222 \\
			\bottomrule
		\end{tabular}
		
	}
	\caption{Energy balance [Mt CO$_2$/a] for carbon dioxide storage for the \ce{1.7} scenario and the three methods endogenous (\textit{endogen}), exogenous (\textit{exogen}) and sequential cost (\textit{seqcost}). See also Figure \ref{fig:eb_co2_stored}.}
	\label{tab:energy_balance_co2_stored_method}
\end{table*}
\begin{table*}[h]
	\resizebox{1\textwidth}{!}{ 
\begin{tabular}{lrrrrrrrrrrrrrrrrrrrrr}
	\toprule
	budget & \multicolumn{7}{l}{\ce{1.5}} & \multicolumn{7}{l}{\ce{1.7}} & \multicolumn{7}{l}{\ce{2.0}}\\
	year & 2020 & 2025 & 2030 & 2035 & 2040 & 2045 & 2050 & 2020 & 2025 & 2030 & 2035 & 2040 & 2045 & 2050 & 2020 & 2025 & 2030 & 2035 & 2040 & 2045 & 2050 \\
	\midrule
	ground heat pump &   64 &  401 &  479 &  417 &  257 &  218 &  153 &   64 &  187 &  325 &  414 &  257 &  218 &  153 &   80 &  187 &  288 &  418 &  257 &  218 &  173 \\
	resistive heater &    0 &  105 &   77 &  280 &  485 &  490 &  500 &    0 &   65 &   87 &  267 &  485 &  490 &  500 &    0 &   20 &   48 &  249 &  485 &  490 &  480 \\
	gas boiler       &  721 &  285 &  226 &   72 &    0 &    0 &    0 &  721 &  533 &  349 &   83 &    0 &    0 &    0 &  705 &  578 &  415 &   99 &    0 &    0 &    0 \\
	heat             & -785 & -791 & -791 & -776 & -742 & -708 & -653 & -785 & -791 & -791 & -776 & -742 & -708 & -653 & -785 & -791 & -791 & -776 & -742 & -708 & -653 \\
	oil boiler       &    0 &    0 &    8 &    7 &    0 &    0 &    0 &    0 &    6 &   30 &   12 &    0 &    0 &    0 &    0 &    6 &   40 &    9 &    0 &    0 &    0 \\
	\bottomrule
\end{tabular}		
	}
	\caption{Energy balance [TWh/a] for rural heat for the three different budgets and the endogenous method. See also Figure \ref{fig:eb_rural_heat}.}
	\label{tab:energy_balance_rural_heat_endogen}
\end{table*}

\begin{table*}[h]
	\resizebox{1\textwidth}{!}{ 
\begin{tabular}{lrrrrrrrrrrrrrrrrrrrrr}
	\toprule
	method & \multicolumn{7}{l}{endogen} & \multicolumn{7}{l}{exogen} & \multicolumn{7}{l}{seqcost} \\
	year &    2020 & 2025 & 2030 & 2035 & 2040 & 2045 & 2050 &   2020 & 2025 & 2030 & 2035 & 2040 & 2045 & 2050 &    2020 & 2025 & 2030 & 2035 & 2040 & 2045 & 2050 \\
	\midrule
	ground heat pump &      64 &  187 &  325 &  414 &  257 &  218 &  153 &     65 &  220 &  366 &  414 &  257 &  218 &  154 &      64 &  225 &  381 &  413 &  257 &  218 &  154 \\
	resistive heater &       0 &   65 &   87 &  267 &  485 &  490 &  500 &      0 &   80 &   78 &  267 &  485 &  490 &  500 &       0 &   40 &   74 &  267 &  485 &  490 &  500 \\
	gas boiler       &     721 &  533 &  349 &   83 &    0 &    0 &    0 &    720 &  487 &  320 &   83 &    0 &    0 &    0 &     721 &  522 &  310 &   83 &    0 &    0 &    0 \\
	heat             &    -785 & -791 & -791 & -776 & -742 & -708 & -653 &   -785 & -791 & -791 & -776 & -742 & -708 & -653 &    -785 & -791 & -791 & -776 & -742 & -708 & -653 \\
	oil boiler       &       0 &    6 &   30 &   12 &    0 &    0 &    0 &      0 &    4 &   28 &   12 &    0 &    0 &    0 &       0 &    4 &   26 &   13 &    0 &    0 &    0 \\
	\bottomrule
\end{tabular}	
		
	}
	\caption{Energy balance [TWh/a] for rural heat for the \ce{1.7} scenario and the three methods endogenous (\textit{endogen}), exogenous (\textit{exogen}) and sequential cost (\textit{seqcost}). See also Figure \ref{fig:eb_rural_heat}.}
	\label{tab:energy_balance_rural_heat_method}
\end{table*}

\begin{table*}[h]
	\resizebox{1\textwidth}{!}{ 
\begin{tabular}{lrrrrrrrrrrrrrrrrrrrrr}
	\toprule
	budget & \multicolumn{7}{l}{\ce{1.5}} & \multicolumn{7}{l}{\ce{1.7}} & \multicolumn{7}{l}{\ce{2.0}}\\
	year &  2020 &  2025 &  2030 &  2035 &  2040 &  2045 & 2050 &  2020 &  2025 &  2030 &  2035 &  2040 &  2045 &  2050 &  2020 &  2025 &  2030 &  2035 &  2040 &  2045 & 2050 \\
	\midrule
	air heat pump     &  1011 &  1683 &  1667 &  1562 &  1499 &  1245 &  974 &   993 &  1342 &  1375 &  1289 &  1493 &  1199 &  1093 &   895 &   793 &   621 &   536 &   754 &   755 &  865 \\
	resistive heater  &     6 &    19 &     1 &     4 &     0 &     0 &    0 &   217 &   227 &   207 &   202 &     0 &     0 &     0 &   337 &   604 &   878 &   922 &   657 &   393 &  119 \\
	gas boiler        &  1054 &   259 &   127 &    60 &     0 &     0 &    0 &   885 &   380 &   208 &   117 &     0 &     0 &     0 &   857 &   544 &   288 &   129 &     0 &     0 &    0 \\
	hot water storage &     0 &     0 &    -5 &   -36 &  -107 &   -18 &   -3 &     0 &     0 &    -1 &   -21 &  -128 &  -105 &     0 &     0 &     0 &     0 &     0 &   -46 &   -55 &    0 \\
	DAC               &     0 &     0 &     0 &     0 &   -28 &  -133 & -153 &     0 &     0 &     0 &     0 &     0 &     0 &  -275 &     0 &     0 &     0 &     0 &     0 &     0 & -166 \\
	heat              & -2136 & -1978 & -1804 & -1597 & -1365 & -1094 & -818 & -2136 & -1978 & -1804 & -1597 & -1365 & -1094 &  -818 & -2136 & -1978 & -1804 & -1597 & -1365 & -1094 & -818 \\
	oil boiler        &    65 &    17 &    14 &     7 &     0 &     0 &    0 &    42 &    28 &    15 &    11 &     0 &     0 &     0 &    47 &    37 &    17 &    10 &     0 &     0 &    0 \\
	\bottomrule
\end{tabular}
	
	}
	\caption{Energy balance [TWh/a] for urban individual heat for the three different budgets and the endogenous method. See also Figure \ref{fig:eb_urban_decentral_heat}.}
	\label{tab:energy_balance_urban_decentral_heat_endogen}
\end{table*}

\begin{table*}[h]
	\resizebox{1\textwidth}{!}{ 
\begin{tabular}{lrrrrrrrrrrrrrrrrrrrrr}
	\toprule
	method & \multicolumn{7}{l}{endogen} & \multicolumn{7}{l}{exogen} & \multicolumn{7}{l}{seqcost} \\
	year &    2020 &  2025 &  2030 &  2035 &  2040 &  2045 &  2050 &   2020 &  2025 &  2030 &  2035 &  2040 &  2045 &  2050 &    2020 &  2025 &  2030 &  2035 &  2040 &  2045 &  2050 \\
	\midrule
	air heat pump     &     993 &  1342 &  1375 &  1289 &  1493 &  1199 &  1093 &   1232 &  1592 &  1565 &  1524 &  1493 &  1199 &  1041 &    1021 &  1566 &  1566 &  1513 &  1493 &  1199 &  1040 \\
	resistive heater  &     217 &   227 &   207 &   202 &     0 &     0 &     0 &      6 &    17 &     2 &     1 &     0 &     0 &     0 &       6 &    14 &     3 &     2 &     0 &     0 &     0 \\
	gas boiler        &     885 &   380 &   208 &   117 &     0 &     0 &     0 &    837 &   349 &   222 &    99 &     0 &     0 &     0 &    1054 &   374 &   222 &   111 &     0 &     0 &     0 \\
	hot water storage &       0 &     0 &    -1 &   -21 &  -128 &  -105 &     0 &      0 &     0 &    -4 &   -36 &  -128 &  -105 &     0 &       0 &     0 &    -3 &   -36 &  -128 &  -105 &    -1 \\
	DAC               &       0 &     0 &     0 &     0 &     0 &     0 &  -275 &      0 &     0 &     0 &     0 &     0 &     0 &  -223 &       0 &     0 &     0 &     0 &     0 &     0 &  -220 \\
	heat              &   -2136 & -1978 & -1804 & -1597 & -1365 & -1094 &  -818 &  -2136 & -1978 & -1804 & -1597 & -1365 & -1094 &  -818 &   -2136 & -1978 & -1804 & -1597 & -1365 & -1094 &  -818 \\
	oil boiler        &      42 &    28 &    15 &    11 &     0 &     0 &     0 &     61 &    21 &    18 &    10 &     0 &     0 &     0 &      56 &    25 &    17 &     8 &     0 &     0 &     0 \\
	\bottomrule
\end{tabular}

	}
	\caption{Energy balance [TWh/a] for urban individual heat for the \ce{1.7} scenario and the three methods endogenous (\textit{endogen}), exogenous (\textit{exogen}) and sequential cost (\textit{seqcost}). See also Figure \ref{fig:eb_urban_decentral_heat}.}
	\label{tab:energy_balance_urban_decentral_heat_method}
\end{table*}
\begin{table*}[h]
	\resizebox{1\textwidth}{!}{ 
\begin{tabular}{lrrrrrrrrrrrrrrrrrrrrr}
	\toprule
	budget & \multicolumn{7}{l}{\ce{1.5}} & \multicolumn{7}{l}{\ce{1.7}} & \multicolumn{7}{l}{\ce{2.0}}\\
	year & 2020 & 2025 & 2030 &  2035 &  2040 &  2045 &  2050 & 2020 & 2025 & 2030 &  2035 &  2040 &  2045 &  2050 & 2020 & 2025 & 2030 &  2035 &  2040 &  2045 &  2050 \\
	\midrule
	air heat pump     &   25 &  181 &  421 &   505 &   532 &   754 &   813 &   64 &  118 &  146 &   111 &   211 &   436 &   713 &  139 &    1 &   39 &    47 &     0 &   102 &   552 \\
	resistive heater  &    0 &   36 &   21 &     2 &     0 &     0 &     0 &    0 &    1 &   28 &    12 &     0 &     0 &     0 &    0 &    0 &    0 &     8 &     0 &   148 &   138 \\
	CHP               &    0 &  333 &  333 &   333 &   280 &   280 &   298 &  131 &  495 &  529 &   844 &   913 &   771 &   400 &    0 &  744 &  849 &   900 &  1118 &  1061 &   481 \\
	gas boiler        &  392 &   74 &   64 &    41 &    28 &     0 &     0 &  226 &   12 &  121 &    48 &     7 &     0 &     0 &  281 &    0 &    0 &    62 &    54 &     0 &     0 \\
	hot water storage &    0 &    0 &  -18 &  -101 &  -206 &  -262 &  -253 &    0 &    0 &   -1 &    -9 &   -18 &   -59 &  -193 &    0 & -120 &  -62 &    -9 &   -20 &   -24 &  -208 \\
	Fischer-Tropsch   &    0 &    0 &    0 &   221 &   437 &   491 &   411 &    0 &    0 &    0 &     0 &    15 &   139 &   368 &    0 &    0 &    0 &     0 &     0 &     0 &   358 \\
	H2 Fuel Cell      &    0 &    0 &    0 &     0 &    78 &    83 &   200 &    0 &    0 &    0 &     0 &    25 &    59 &   181 &    0 &    0 &    0 &     0 &     0 &    58 &   149 \\
	heat              & -422 & -626 & -826 & -1009 & -1156 & -1346 & -1470 & -422 & -626 & -826 & -1009 & -1156 & -1346 & -1470 & -422 & -626 & -826 & -1009 & -1156 & -1346 & -1470 \\
	oil boiler        &    5 &    1 &    5 &     6 &     7 &     0 &     0 &    0 &    0 &    3 &     2 &     1 &     0 &     0 &    1 &    0 &    0 &     1 &     3 &     0 &     0 \\
	\bottomrule
\end{tabular}

	}
	\caption{Energy balance [TWh/a] for urban district heat for the three different budgets and the endogenous method. See also Figure \ref{fig:eb_urban_central_heat}.}
	\label{tab:energy_balance_urban_central_heat_endogen}
\end{table*}

\begin{table*}[h]
	\resizebox{1\textwidth}{!}{ 
\begin{tabular}{lrrrrrrrrrrrrrrrrrrrrr}
	\toprule
	method & \multicolumn{7}{l}{endogen} & \multicolumn{7}{l}{exogen} & \multicolumn{7}{l}{seqcost} \\
	year &    2020 & 2025 & 2030 &  2035 &  2040 &  2045 &  2050 &   2020 & 2025 & 2030 &  2035 &  2040 &  2045 &  2050 &    2020 & 2025 & 2030 &  2035 &  2040 &  2045 &  2050 \\
	\midrule
	air heat pump     &      64 &  118 &  146 &   111 &   211 &   436 &   713 &     91 &   93 &  105 &   210 &   380 &   580 &   825 &      25 &   70 &   83 &   217 &   391 &   566 &   793 \\
	resistive heater  &       0 &    1 &   28 &    12 &     0 &     0 &     0 &      0 &    4 &   24 &     8 &     0 &     0 &     0 &       0 &    4 &   19 &    11 &     0 &     7 &    41 \\
	CHP               &     131 &  495 &  529 &   844 &   913 &   771 &   400 &     48 &  528 &  653 &   749 &   766 &   735 &   343 &       4 &  557 &  684 &   753 &   761 &   734 &   368 \\
	gas boiler        &     226 &   12 &  121 &    48 &     7 &     0 &     0 &    282 &   20 &   64 &    46 &    15 &     0 &     0 &     389 &   23 &   58 &    33 &    16 &     0 &     0 \\
	hot water storage &       0 &    0 &   -1 &    -9 &   -18 &   -59 &  -193 &      0 &  -18 &  -22 &    -7 &   -15 &   -16 &  -190 &       0 &  -29 &  -21 &    -6 &   -17 &   -13 &  -185 \\
	Fischer-Tropsch   &       0 &    0 &    0 &     0 &    15 &   139 &   368 &      0 &    0 &    0 &     0 &     0 &    39 &   360 &       0 &    0 &    0 &     0 &     0 &    50 &   361 \\
\bottomrule
\end{tabular}

	}
	\caption{Energy balance [TWh/a] for urban district heat for the \ce{1.7} scenario and the three methods endogenous (\textit{endogen}), exogenous (\textit{exogen}) and sequential cost (\textit{seqcost}). See also Figure \ref{fig:eb_urban_central_heat}.}
	\label{tab:energy_balance_urban_central_heat_method}
\end{table*}

\begin{table*}[h]
	\resizebox{1\textwidth}{!}{ 
\begin{tabular}{lrrrrrrrrrrrrrrrrrrrrr}
	\toprule
	budget & \multicolumn{7}{l}{\ce{1.5}} & \multicolumn{7}{l}{\ce{1.7}} & \multicolumn{7}{l}{\ce{2.0}} \\
	year &  2020 &  2025 &  2030 &  2035 &  2040 &  2045 & 2050 &  2020 &  2025 &  2030 &  2035 &  2040 &  2045 & 2050 &  2020 &  2025 &  2030 &  2035 &  2040 &  2045 & 2050 \\
	\midrule
	nuclear & -2328 & -2547 & -2664 & -2672 & -2712 & -2379 & -352 & -3105 & -2975 & -2609 & -2643 & -2601 & -2166 & -350 & -3105 & -3105 & -3028 & -2807 & -2677 & -2313 & -350 \\
	uranium &  2328 &  2547 &  2664 &  2672 &  2712 &  2379 &  352 &  3105 &  2975 &  2609 &  2643 &  2601 &  2166 &  350 &  3105 &  3105 &  3028 &  2807 &  2677 &  2313 &  350 \\
	\bottomrule
\end{tabular}

	}
	\caption{Energy balance [TWh/a] for uranium for the three different budgets and the endogenous method. See also Figure \ref{fig:eb_uranium}.}
	\label{tab:energy_balance_uranium_endogen}
\end{table*}

\begin{table*}[h]
	\resizebox{1\textwidth}{!}{ 
\begin{tabular}{lrrrrrrrrrrrrrrrrrrrrr}
	\toprule
	method & \multicolumn{7}{l}{endogen} & \multicolumn{7}{l}{exogen} & \multicolumn{7}{l}{seqcost} \\
	year &    2020 &  2025 &  2030 &  2035 &  2040 &  2045 & 2050 &   2020 &  2025 &  2030 &  2035 &  2040 &  2045 & 2050 &    2020 &  2025 &  2030 &  2035 &  2040 &  2045 & 2050 \\
	\midrule
	nuclear &   -3105 & -2975 & -2609 & -2643 & -2601 & -2166 & -350 &  -3105 & -2780 & -2650 & -2572 & -2463 & -2261 & -350 &   -3030 & -3053 & -2695 & -2345 & -2371 & -2266 & -344 \\
	uranium &    3105 &  2975 &  2609 &  2643 &  2601 &  2166 &  350 &   3105 &  2780 &  2650 &  2572 &  2463 &  2261 &  350 &    3030 &  3053 &  2695 &  2345 &  2371 &  2266 &  344 \\
	\bottomrule
\end{tabular}

	}
	\caption{Energy balance [TWh/a] for uranium for the \ce{1.7} scenario and the three methods endogenous (\textit{endogen}), exogenous (\textit{exogen}) and sequential cost (\textit{seqcost}). See also Figure \ref{fig:eb_uranium}.}
	\label{tab:energy_balance_uranium_method}
\end{table*}

\begin{table*}[h]
	\resizebox{1\textwidth}{!}{ 
\begin{tabular}{lrrrrrrrrrrrrrrrrrrrrr}
	\toprule
	method & \multicolumn{7}{l}{endogen} & \multicolumn{7}{l}{seqcost} & \multicolumn{7}{l}{exogen} \\
	year &    2020 &  2025 &  2030 &  2035 &  2040 &  2045 &  2050 &    2020 & 2025 &  2030 &  2035 &  2040 &  2045 &  2050 &   2020 & 2025 &  2030 &  2035 &  2040 &  2045 &  2050 \\
	\midrule
	Offshore wind [GW]                 &      35 &   187 &   193 &   193 &   190 &   182 &   175 &      41 &  162 &   184 &   316 &   506 &   498 &   478 &     18 &   65 &   155 &   207 &   255 &   250 &   263 \\
	Solar rooftop [GW]                 &       0 &     0 &     0 &     0 &    16 &    16 &    16 &       0 &   58 &    96 &    96 &    96 &    96 &    97 &      0 &    0 &     5 &    80 &    96 &    96 &    96 \\
	Solar PV [GW]                      &     172 &   436 &   940 &  1970 &  2616 &  3106 &  3335 &     153 &  337 &   516 &  1488 &  1509 &  1818 &  2611 &    153 &  577 &   791 &  1168 &  2217 &  2203 &  2419 \\
	Onshore wind [GW]                  &     257 &  1031 &  1222 &  1530 &  1811 &  1774 &  1785 &     199 &  715 &  1235 &  1315 &  1671 &  1634 &  1599 &    199 &  986 &  1240 &  1540 &  1958 &  1921 &  1890 \\
	OCGT [GW]                          &      96 &    60 &    55 &   371 &   449 &   429 &   444 &     155 &  120 &   107 &   331 &   363 &   328 &   405 &    171 &  136 &   123 &   379 &   470 &   427 &   464 \\
	CCGT [GW]                          &     241 &   153 &   103 &    24 &     4 &     0 &     0 &     241 &  153 &   103 &    24 &     4 &     0 &     0 &    241 &  153 &   103 &    24 &     4 &     0 &     0 \\
\bottomrule
\end{tabular}
	}
	\caption{Installed capacities in the \ce{1.5} scenarios.}
	\label{tab:capacities_1p5}
\end{table*}

\begin{table*}[h]
	\resizebox{1\textwidth}{!}{ 
\begin{tabular}{lrrrrrrrrrrrrrrrrrrrrr}
	\toprule
	method & \multicolumn{7}{l}{endogen} & \multicolumn{7}{l}{seqcost} & \multicolumn{7}{l}{exogen} \\
	year &    2020 & 2025 & 2030 & 2035 &  2040 &  2045 &  2050 &    2020 & 2025 & 2030 & 2035 & 2040 &  2045 &  2050 &   2020 & 2025 & 2030 & 2035 & 2040 &  2045 &  2050 \\
	\midrule
	Offshore wind [GW]                 &      27 &   82 &   97 &   96 &    94 &    86 &   190 &      41 &   74 &   96 &  118 &  138 &   230 &   367 &     18 &   41 &   41 &   47 &  108 &   171 &   257 \\
	Solar rooftop [GW]                 &       0 &    0 &    0 &    0 &    16 &    16 &    56 &       0 &    0 &   96 &   96 &   96 &    96 &    96 &      0 &    0 &    1 &   33 &   96 &    96 &    99 \\
	Solar PV [GW]                      &     172 &  423 &  764 &  972 &  1455 &  2165 &  3152 &     153 &  310 &  379 &  559 &  774 &  1118 &  2550 &    153 &  347 &  640 &  811 &  997 &  1197 &  2707 \\
	Onshore wind [GW]                  &     218 &  572 &  678 &  728 &   761 &  1016 &  1702 &     181 &  557 &  637 &  748 &  961 &   924 &  1651 &    184 &  661 &  693 &  835 &  882 &   923 &  1802 \\
	OCGT [GW]                          &     142 &  106 &   94 &  315 &   489 &   422 &   422 &     152 &  117 &  104 &  393 &  626 &   549 &   549 &    149 &  114 &  101 &  397 &  614 &   540 &   540 \\
	CCGT [GW]                          &     241 &  153 &  103 &   24 &     4 &     0 &     0 &     241 &  153 &  103 &   24 &    4 &     0 &     0 &    241 &  153 &  103 &   24 &    4 &     0 &     0 \\
\bottomrule
\end{tabular}

	}
	\caption{Installed capacities in the \ce{1.7} scenarios.}
	\label{tab:capacities_1p7}
\end{table*}

\begin{table*}[h]
	\resizebox{1\textwidth}{!}{ 
\begin{tabular}{lrrrrrrrrrrrrrrrrrrrrr}
	\toprule
	method & \multicolumn{7}{l}{endogen} & \multicolumn{7}{l}{seqcost} & \multicolumn{7}{l}{exogen} \\
	year &    2020 & 2025 & 2030 &  2035 &  2040 &  2045 &  2050 &    2020 & 2025 & 2030 & 2035 & 2040 & 2045 &  2050 &   2020 & 2025 & 2030 & 2035 &  2040 &  2045 &  2050 \\
	\midrule
	Offshore wind [GW]                 &      18 &   22 &   22 &    21 &    20 &    30 &   190 &      18 &   18 &   18 &   18 &   15 &   30 &   204 &     18 &   18 &   18 &   18 &    15 &    10 &   245 \\
	Solar rooftop [GW]                 &       0 &    0 &    0 &     0 &    16 &    16 &    17 &       0 &    0 &   62 &   62 &   97 &   99 &   218 &      0 &    0 &    0 &   25 &    81 &    81 &    99 \\
	Solar PV [GW]                      &     167 &  345 &  422 &   747 &   879 &  1602 &  3178 &     153 &  153 &  177 &  540 &  589 &  579 &  2054 &    153 &  153 &  472 &  558 &   709 &   810 &  2852 \\
	Onshore wind [GW]                  &     263 &  427 &  555 &   617 &   609 &   580 &  1839 &     277 &  373 &  513 &  570 &  586 &  644 &  2251 &    276 &  373 &  488 &  608 &   561 &   615 &  1833 \\
	OCGT [GW]                          &     320 &  285 &  272 &   668 &   662 &   519 &   519 &     325 &  289 &  277 &  464 &  634 &  629 &   629 &    325 &  289 &  277 &  578 &   622 &   581 &   581 \\
	CCGT [GW]                          &     241 &  153 &  103 &    24 &     4 &     0 &     0 &     241 &  153 &  103 &   24 &    4 &    0 &     0 &    241 &  153 &  103 &   24 &     4 &     0 &     0 \\
\bottomrule
\end{tabular}

	}
	\caption{Installed capacities in the \ce{2.0} scenarios.}
	\label{tab:capacities_2p0}
\end{table*}

\FloatBarrier

\bibliography{sn-bibliography}
\clearpage
\newpage\null\thispagestyle{empty}
\printglossary[type=\acronymtype]
\clearpage
\newpage\null\thispagestyle{empty}